\documentclass[twocolumn]{aastex631}
\usepackage{amsmath}

\newcommand\kms{km$\,$s$^{-1}$}
\newcommand\Msol{M$_{\odot}$}

\newcommand{\hi}{H\,{\sc i}}
\newcommand{\hii}{H\,{\sc ii}}

\defcitealias{Bellazzini+2022}{Paper I}

\submitjournal{AAS Journals}

\shorttitle{Isolated Blue Stellar Systems}
\shortauthors{Jones et al.}

\begin{document}

\title{Young, blue, and isolated stellar systems in the Virgo Cluster. II. A new class of stellar system}

\correspondingauthor{Michael G. Jones}
\email{jonesmg@arizona.edu}

\author[0000-0002-5434-4904]{Michael G. Jones}
\affiliation{Steward Observatory, University of Arizona, 933 North Cherry Avenue, Rm. N204, Tucson, AZ 85721-0065, USA}

\author[0000-0003-4102-380X]{David J. Sand}
\affiliation{Steward Observatory, University of Arizona, 933 North Cherry Avenue, Rm. N204, Tucson, AZ 85721-0065, USA}

\author[0000-0001-8200-810X]{Michele Bellazzini}
\affil{INAF – Osservatorio di Astrofisica e Scienza dello Spazio di Bologna, Via Gobetti 93/3, 40129 Bologna, Italy}

\author[0000-0002-0956-7949]{Kristine Spekkens}
\affiliation{Department of Physics and Space Science, Royal Military College of Canada P.O. Box 17000, Station Forces Kingston, ON K7K 7B4, Canada}
\affiliation{Department of Physics, Engineering Physics and Astronomy, Queen’s University, Kingston, ON K7L 3N6, Canada}

\author[0000-0001-8855-3635]{Ananthan Karunakaran}
\affiliation{Instituto de Astrof\'{i}sica de Andaluc\'{i}a (CSIC), Glorieta de la Astronom\'{i}a, 18008 Granada, Spain}
\affiliation{Department of Physics, Engineering Physics and Astronomy, Queen’s University, Kingston, ON K7L 3N6, Canada}

\author[0000-0002-9798-5111]{Elizabeth A. K. Adams}
\affil{ASTRON, the Netherlands Institute for Radio Astronomy, Oude Hoogeveesedijk 4,7991 PD Dwingeloo, The Netherlands}
\affil{Kapteyn Astronomical Institute, PO Box 800, 9700 AV Groningen, The Netherlands}

\author[0000-0002-6551-4294]{Giuseppina Battaglia}
\affiliation{Instituto de Astrof\'{ı}sica de Canarias, V\'{ı}a L\'{a}ctea s/n 38205 La Laguna, Spain}
\affiliation{Universidad de La Laguna, Avda. Astrofísico Fco. Sánchez , La Laguna, Tenerife E-38205,  Spain}

\author[0000-0002-3865-9906]{Giacomo Beccari}
\affil{European Southern Observatory, Karl-Schwarzschild-Strasse 2, D-85748 Garching bei M\"{u}nchen, Germany}

\author[0000-0001-8354-7279]{Paul Bennet}
\affiliation{Space Telescope Science Institute, 3700 San Martin Drive, Baltimore, MD 21218, USA}

\author[0000-0002-1821-7019]{John M. Cannon}
\affiliation{Department of Physics \& Astronomy, Macalester College, 1600 Grand Avenue, Saint Paul, MN 55105, USA}

\author[0000-0002-5281-1417]{Giovanni Cresci}
\affil{INAF - Osservatorio Astrofisico di Arcetri, Largo E. Fermi 5, 50125 Firenze, Italy}

\author[0000-0002-1763-4128]{Denija Crnojevi\'{c}}
\affil{University of Tampa, 401 West Kennedy Boulevard, Tampa, FL 33606, USA}

\author[0000-0003-2352-3202]{Nelson Caldwell}
\affiliation{Center for Astrophysics, Harvard \& Smithsonian, 60 Garden Street, Cambridge, MA 02138, USA}

\author[0000-0002-8598-439X]{Jackson Fuson}
\affiliation{Department of Physics \& Astronomy, Macalester College, 1600 Grand Avenue, Saint Paul, MN 55105, USA}

\author[0000-0001-8867-4234]{Puragra Guhathakurta}
\affiliation{UCO/Lick Observatory, University of California Santa Cruz, 1156 High Street, Santa Cruz, CA 95064, USA}

\author[0000-0001-5334-5166]{Martha P. Haynes}
\affiliation{Cornell Center for Astrophysics and Planetary Science, Space Sciences Building, Cornell University, Ithaca, NY 14853, USA}

\author[0000-0002-9724-8998]{John L. Inoue}
\affiliation{Department of Physics \& Astronomy, Macalester College, 1600 Grand Avenue, Saint Paul, MN 55105, USA}

\author[0000-0003-4486-6802]{Laura Magrini}
\affil{INAF - Osservatorio Astrofisico di Arcetri, Largo E. Fermi 5, 50125 Firenze, Italy}

\author[0000-0002-0810-5558]{Ricardo R. Mu\~{n}oz}
\affiliation{Departamento de Astronom\'{ı}a, Universidad de Chile, Camino El Observatorio 1515, Las Condes, Santiago}

\author[0000-0001-9649-4815]{Bur\c{c}in Mutlu-Pakdil}
\affil{Kavli Institute for Cosmological Physics, University of Chicago, Chicago, IL 60637, USA}
\affil{Department of Astronomy and Astrophysics, University of Chicago, Chicago IL 60637, USA}

\author[0000-0003-0248-5470]{Anil Seth}
\affiliation{Department of Physics \& Astronomy, University of Utah, Salt Lake City, UT, 84112, USA}

\author[0000-0002-1468-9668]{Jay Strader}
\affiliation{Center for Data Intensive and Time Domain Astronomy, Department of Physics and Astronomy, Michigan State University, East Lansing, MI 48824, USA}

\author[0000-0001-6443-5570]{Elisa Toloba}
\affiliation{Department of Physics, University of the Pacific, 3601 Pacific Avenue, Stockton, CA 95211, USA}

\author[0000-0002-5177-727X]{Dennis Zaritsky}
\affiliation{Steward Observatory, University of Arizona, 933 North Cherry Avenue, Rm. N204, Tucson, AZ 85721-0065, USA}



\begin{abstract}

We discuss five blue stellar systems in the direction of the Virgo cluster, analogous to the enigmatic object SECCO~1 (AGC~226067). These objects were identified based on their optical and UV morphology and followed up with \hi \ observations with the VLA (and GBT), MUSE/VLT optical spectroscopy, and HST imaging. 
These new data indicate that one system is a distant group of galaxies. The remaining four are extremely low mass ($M_\ast \sim 10^5$~\Msol), are dominated by young, blue stars, have highly irregular and clumpy morphologies, are only a few kpc across, yet host an abundance of metal-rich, $12 + \log (\mathrm{O/H}) > 8.2$, H\,{\sc ii} regions. These high metallicities indicate that these stellar systems formed from gas stripped from much more massive galaxies. 
Despite the young age of their stellar populations, only one system is detected in \hi, while the remaining three have minimal (if any) gas reservoirs. Furthermore, two systems are surprisingly isolated and have no plausible parent galaxy within $\sim$30\arcmin \ ($\sim$140~kpc). 
Although tidal stripping cannot be conclusively excluded as the formation mechanism of these objects, ram pressure stripping more naturally explains their properties, in particular their isolation, owing to the higher velocities, relative to the parent system, that can be achieved.
Therefore, we posit that most of these systems formed from ram pressure stripped gas removed from new infalling cluster members, and survived in the intracluster medium long enough to become separated from their parent galaxies by hundreds of kiloparsecs, and that they thus represent a new type of stellar system. 
\end{abstract}



\section{Introduction} 
\label{sec:intro}

Exceptionally high gas-to-stellar mass ratio systems are of particular interest in extragalactic astronomy as they represent one extreme of galaxy formation, namely some of the lowest mass objects that succeed in forming any stars. Blind radio surveys of neutral hydrogen (\hi) have uncovered a plethora of gas-rich systems that have few, or perhaps no, stars \citep{Saul+2012,Adams+2013,Taylor+2013,Cannon+2015}. 
However, distinguishing those which may be genuine extreme low-mass dwarf galaxies from other classes of objects \citep{Cannon+2015}, such as tidal debris and high velocity clouds \citep{Adams+2016}, is a challenging process owing to the faintness of any associated stellar counterpart to these objects \citep[e.g.][]{Janesh+2019}, as well as confusion with foreground Milky Way \hi \ emission, which often dominates the velocity range where candidates are expected to be detectable.

However, attempts to distinguish these objects have led to surprising discoveries, most notably SECCO~1 \citep[also called AGC~226067;][]{Adams+2013,Bellazzini+2015,Sand+2015,Adams+2015,Beccari+2017,Sand+2017,Bellazzini+2018} and AGC~226178 \citep{Cannon+2015,Junais+2021,Jones+2022}. These are two young, blue, extremely low-mass ($M_\ast \sim 10^5$~\Msol), gas-rich, metal-rich, actively star-forming stellar systems in the Virgo cluster. AGC~226178 has a gas-to-stellar mass ratio $(1.4M_\mathrm{HI}/\mathrm{M_\ast})\sim1000$, while SECCO~1 has a ratio of $\sim$150.\footnote{Here a factor of 1.4 is used to account for helium in the gas mass.} The properties of both systems imply that they formed via in situ star formation (SF) in gaseous debris stripped from a much larger object. In the case of AGC~226178, the likely parent object has been identified as the nearby galaxy VCC~2034, to which it is connected via a tenuous, low column density, 70~kpc-long \hi \ bridge \citep{Jones+2022}. However, it is unclear whether this gas was stripped by a high-speed tidal encounter, or by ram pressure from the intracluster medium (ICM). In the case of SECCO~1, despite it being relatively close to the Virgo cluster center, it is still sufficiently isolated that its origin is uncertain, and multiple possible parent objects have been suggested \citep{Sand+2017,Bellazzini+2018}.

As alluded to by \citet{Sand+2017} and \citet{Jones+2022},  these two objects are not unique but instead appear to be part of a larger population of similar objects in Virgo. SECCO~1 and AGC~226178 were originally identified through their \hi \ line emission, thereby guaranteeing gas richness. However, with the latest and deepest wide field imaging surveys it is possible to visually identify objects in the Virgo cluster with similar optical and UV properties, though not necessarily equivalently gas-rich.

In this work, we present comprehensive observations of a sample of isolated, blue stellar systems in the Virgo cluster as part of a campaign to understand their physical properties and origins. 
These additional candidate objects, along with AGC226067/SECCO1 and AGC226178, were followed up with Hubble Space Telescope (HST) F606W and F814W imaging with the Advanced Camera for Surveys (ACS), and \hi \ observations with the Jansky Very Large Array (VLA) and the Green Bank Telescope (GBT). Additional observations with the MUSE (Multi Unit Spectroscopic Explorer) integral field spectrograph on the VLT (Very Large Telescope), are presented in a companion paper, \citet{Bellazzini+2022}, hereafter \citetalias{Bellazzini+2022}.

The sample identification is described in \S\ref{sec:sample} and their follow-up observations in \S\ref{sec:obs}. The results, \hi \ and stellar masses, star formation rates (SFRs), and metallicity measurements are presented in \S\ref{sec:results}. In \S\ref{sec:discuss} we search for potential points of origin of these objects and we discuss potential formation scenarios in \S\ref{sec:formation}. Finally, in \S\ref{sec:fate} \& \S\ref{sec:future} we discuss the fate of these objects and future directions of investigation, before drawing our conclusions in \S\ref{sec:conclude}.

We adopt 16.5~Mpc \citep{Mei+2007} as the distance to the Virgo cluster throughout.

\section{Target identification}
\label{sec:sample}

We performed a visual search for isolated, blue stellar systems, similar in optical appearance to SECCO~1, using the $\sim$100~deg$^2$ of NGVS \citep[Next Generation Virgo cluster Survey,][]{Ferrarese+2012} $ugi$ imaging of the Virgo cluster, along with GALEX \citep[Galaxy Evolution Explorer,][]{Martin+2005} UV imaging when available.  Characteristic systems display an over-density of compact blue sources with strong associated UV emission.  They also lack a diffuse red component typical of Virgo dwarf galaxies, even when they have ongoing SF. 
Partial results from this search were presented in \citet{Sand+2017}.  In total, five isolated, blue stellar system candidates (or BCs), which we number 1-5, were identified. All five were followed-up with observations with HST, the VLA, and MUSE/VLT. The coordinates of these five targets are listed in Table \ref{tab:BCs}, and their locations relative to the Virgo cluster are shown in Figure \ref{fig:BC_locs}.

The object we refer to as BC3 is an independent re-identification (based on optical appearance) of the \hi-selected object AGC~226178 from the ALFALFA survey \citep{Haynes+2011,Cannon+2015}. This object has already been studied in detail \citep{Cannon+2015,Junais+2021,Jones+2022} and is the BC most similar to SECCO~1.

As discussed in the remainder of this paper, we are now confident that four of the five BCs are genuine SECCO~1 analogs.

\begin{table}
\centering
\caption{BC coordinates and \hi \ velocities}
\begin{tabular}{cccc}
\hline \hline
Object & R.A. & Dec. & $v_\mathrm{HI}/\mathrm{km\,s^{-1}}$\\
\hline
BC1    & 12:39:02.0 & +12:12:16.7 &      \\
BC2$^\ddag$    & 12:44:27.9 & +12:37:13.4 &      \\
BC3    & 12:46:42.5 & +10:22:04.8 & 1581 \\
BC4    & 12:26:25.7 & +14:23:12.2 &      \\
BC5    & 12:26:30.9 & +15:10:26.2 &      \\ 
SECCO1 & 12:21:53.9 & +13:27:37.0 & $-142^\dag$  \\ \hline
\end{tabular}
\tablenotetext{}{Columns: (1) object name; (2 \& 3) coordinates (J2000) of the main body of each object; (4) heliocentric velocity of \hi \ emission \citep{Haynes+2011}.
$^\dag$Value for the main body from \citet{Adams+2015}.
$^\ddag$BC2 is a spurious object (\S\ref{sec:morph}). 
}
\label{tab:BCs}
\end{table}

\section{Observations and reduction}\
\label{sec:obs}

After the initial identification of our target BCs using NGVS and GALEX we had little information about their properties except that they were similar to SECCO~1 in optical appearance (extremely blue, faint and clumpy) and that their UV emission indicated some recent or ongoing SF. We therefore pursued a three-pronged observational strategy to uncover their nature: 1) HST imaging to better understand their detailed morphology and stellar populations; 2) Observations with MUSE/VLT to measure their redshifts via the H$\alpha$ line and obtain metallicity measurements; 3) VLA D-array and GBT observations to search for any associated \hi \ line emission and quantify their neutral gas content.

\subsection{HST observations}

Each of the five candidates was observed with ACS in the F606W and F814W filters as part of program 15183 (PI: D.~Sand). Each target was observed for a total of 2120~s and 2180~s in the two filters respectively, except BC4, which was observed for 2000~s in each filter. \texttt{DOLPHOT}'s \citep{Dolphin2000,dolphot} ACS module was used to align the exposures and perform point source photometry of the resolved stellar population. The dust maps of \citet{Schlegel+1998} and $R_\mathrm{F606W}$ and $R_\mathrm{F814W}$ values of \citet{Schlafly+2011} were used to correct for Galactic extinction at the position of each source. Stars were selected from the resulting \texttt{DOLPHOT} catalog following a similar approach to \citet{Jones+2022}. Briefly, we select all point-like (type 1 and 2) objects with no photometry flags from the \texttt{DOLPHOT} source catalog. We removed sources with greater than 1 mag of crowding (combined, from the two filters). Finally, the combined (in quadrature) absolute sharpness value was enforced to be below $\sqrt{0.075}$ and a roundness threshold of less than 1 (in both filters) was set.
Completeness limits were also estimated as in \citet{Jones+2022}, based on artificial stars added evenly over both images in each field. The measured 90\% completeness limits were fit with the combination of a horizontal line and a one-sided parabola (e.g. Figure \ref{fig:BC1_HST_GLX}, bottom panels), and the 50\% limits were fit with straight lines.

In addition to the point source photometry in \S\ref{sec:stellarmasses} we also perform aperture photometry on the combined, drizzled images in each band to measure the integrated magnitudes and colors of the systems. This was performed using the \texttt{Astropy} package \texttt{Photutils} \citep{photutils} and a combination of manually-constructed circular and elliptical apertures enclosing the various clumps of each source. In each case the sky background was subtracted based on the median value within an annulus (circular or elliptical) surrounding the aperture.

\subsection{MUSE/VLT observations}

To robustly identify \hii \ regions, obtain optical redshifts and basic kinematics, and measure metallicities, we observed all BCs with MUSE/VLT \citep{Bacon+2014}. These observations were carried out as part of program 0101.B-0376A (P.I: R. Mu\~noz). They covered the spectral range 4650-9300~\AA \ and a $\simeq1.0\arcmin \times 1.0\arcmin$ field centered on each target. These observations are discussed in detail in \citetalias{Bellazzini+2022}, and here we present an outline of the data reduction process. 

The reduction and analysis of these data followed \citet{Beccari+2017}. The individual dithered exposures were calibrated separately and then combined into a single stacked data cube for each target. H$\alpha$ (and integrated light) peaks at least 3$\sigma$ above the background were identified using \texttt{Sextractor} \citep{Bertin+1996}. The flux of each of these detected sources was measured using a 1.5\arcsec \ (radius) aperture and a 1D spectrum (with a step size of 1.25~\AA)  of each source was produced.

Redshifts were measured for all detected H$\alpha$ clumps, and line fluxes for H$\beta$, [N{\sc ii}], and [O{\sc iii}] were measured wherever possible \citepalias[Tables 2 \& 3 of][]{Bellazzini+2022}. In Section \ref{sec:MUSE_results} we summarize the findings of these measurements and their implications for the origins of BCs.

\subsection{GALEX data}
\label{sec:galex_data}

We searched for archival NUV and FUV data from GALEX at the location of each BC (and SECCO~1).
Most of the BCs are within the footprint of the GALEX Ultraviolet Virgo Cluster Survey \citep[GUViCS,][]{Boselli+2011}, however, this is not always the deepest data available. For BCs 2, 3, 4 and SECCO~1 we use tiles from GUViCS (typically $\sim$1.6~ks in both bands), but no FUV data are available for either BC4 or SECCO~1. For BC1 we use tiles ``Virgo\_Epoque\_MOS05" ($\sim$16~ks) and ``NGA\_Virgo\_MOS04" ($\sim$1.6~ks) for NUV and FUV, respectively. For BC5 we use tiles ``NGA\_NGC4421" ($\sim$2~ks) and ``GI1\_079012\_Group5" ($\sim$1.6~ks). 

In \S\ref{sec:SFRs} we perform aperture photometry on these GALEX tiles and estimate the SFR in each object. The flux within each aperture was measured from the corresponding background subtracted GALEX tile. Uncertainties were estimated by placing 10000 circular apertures (equal in area to the target apertures) randomly across the GALEX tile after masking the brightest 1\% of pixels. Magnitudes were calculated following the conversions of \citet{Morrissey+2007} and extinction corrections used $R_\mathrm{NUV} = 8.20$ and $R_\mathrm{FUV} = 8.24$ \citep{Wyder+2007}. Finally, these magnitude measurements were converted to SFRs following \citet{Iglesias-Paramo+2006}, using 4.74 as the bolometric solar absolute magnitude. 

\subsection{VLA observations}

BC3 was observed previously as part of the ALFALFA ``Almost Dark" galaxies sample \citep[VLA program 13A-028, PI: J.~Cannon;][]{Cannon+2015}. These data were obtained in D-configuration, have a channel width of 7.81~kHz ($\sim$1.65~\kms), a total bandwidth of 8~MHz, and a total on-source integration time of approximately 1.6 h. These data were re-reduced by \citet{Jones+2022} using standard reduction methods in the Common Astronomy Software Applications package \citep[\texttt{CASA},][]{CASA}. The final imaging used Brigg's robust=0.5 weighting to provide a compromise between sensitivity and angular resolution for the detected \hi \ emission. The channels were averaged and re-binned to a velocity resolution of 5~\kms.

The remaining 4 candidates were observed in the VLA program 18A-185 (PI: K.~Spekkens). Each target was observed on-source for approximately 1.5~h in D-configuration. The initial observations of both BC1 and BC2 suffered from severe interference and were subsequently re-observed, greatly improving the data quality. As the redshift of the objects were not known prior to the observations, we used a 32~MHz bandwidth (from 1394.416 to 1426.416~MHz, or approximately $-$1250 to 5500~\kms) to search for any \hi \ emission associated with the optical candidates. This range was split up into 3072 channels of 10.42~kHz ($\sim$2.2~\kms), which during the data reduction was averaged over 4 channels resulting in a velocities resolution of 8.8~\kms.

Initially the entire bandwidth of the data were reduced to search for \hi \ emission. However, after the redshifts for all candidates were obtained from MUSE spectroscopy with the VLT, a narrow sub-band was re-reduced (spanning $\sim$1000~\kms), allowing for improved local continuum subtraction. The reduction was performed with a \texttt{Python} and \texttt{CASA}-based pipeline that will be presented in full in Jones et al. (in prep.). The most severe interference was flagged manually and the \texttt{tfcrop} flagging algorithm was run in addition. For BC1 we also used \texttt{rflag}, after initial calibrations, as there were no bright lines which might be mistaken for interference (other than Milky Way emission). Imaging used Brigg's robust=2 weighting in order to maximize our detection capabilities. Refer to Table \ref{tab:VLAobs} for details of the beam sizes and rms noise for each observation.

\begin{table}
\centering
\caption{VLA data summary}
\begin{tabular}{cccc}
\hline \hline
Object & Beam size                                                & $\sigma_\mathrm{rms}/\mathrm{mJy \, beam^{-1}}$ & $\Delta v_\mathrm{chan}/$\kms \\ \hline
BC1    & 60\arcsec$\times$51\arcsec & 1.9 & 8.8       \\
BC2    & 63\arcsec$\times$55\arcsec & 1.1 & 8.8       \\
BC3    & 56\arcsec$\times$45\arcsec & 1.2 & 5         \\
BC4    & 65\arcsec$\times$54\arcsec & 0.9 & 8.8       \\
BC5    & 65\arcsec$\times$54\arcsec & 1.0 & 8.8 \\ \hline    
\end{tabular}
\tablenotetext{}{Columns: (1) object name; (2) synthesized beam size; (3) rms noise; (4) velocity resolution.}
\label{tab:VLAobs}
\end{table}

\subsection{GBT observations}

The large surface area and low system temperature of the GBT allow it to obtain much deeper \hi \ spectra than the VLA, providing a more stringent constraint on any neutral gas content. However, after the redshifts of the candidates were known (from their H$\alpha$ emission) it was determined that only BC1 was suitable for single dish follow-up as BC4 and BC5 would be confused with Milky Way emission, BC3 had already been strongly detected with the VLA \citep{Cannon+2015}, and the HST imaging of BC2 indicated that it was a background galaxy group (\S\ref{sec:morph}). A director's discretionary time proposal (21A-433, PI: M.~Jones) was submitted to the GBT and BC1 was observed for a total of 3~h using $\mathrm{ON-OFF}$ position switching. The data were reduced using standard GBT \texttt{IDL} procedures. The resulting spectrum has an rms noise of 0.25~mJy (within $\pm300$~\kms \ of the redshift of BC1) after smoothing to a velocity resolution of 30~\kms.

\section{Results}
\label{sec:results}

\begin{figure}
    \centering
    \includegraphics[width=\columnwidth]{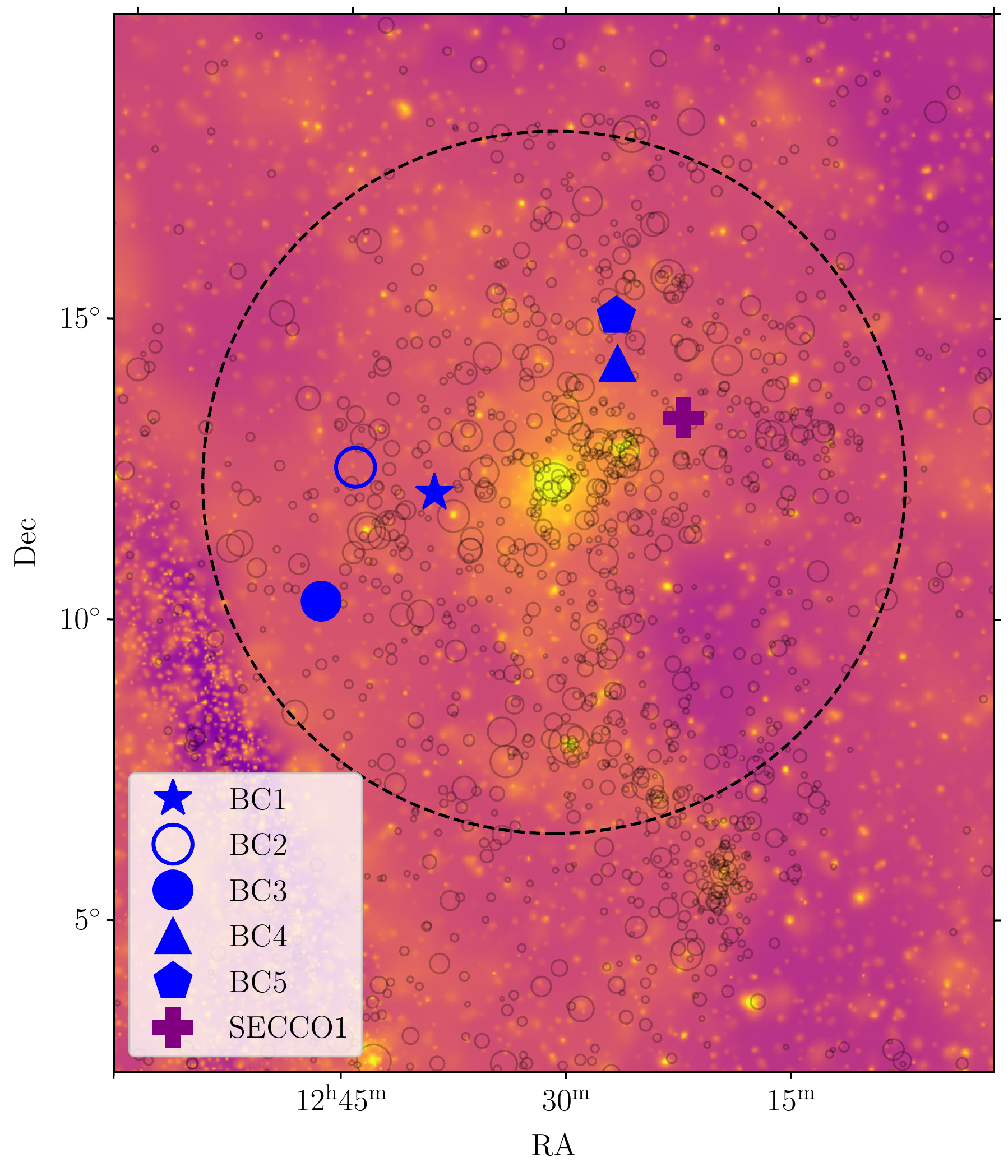}
    \caption{Locations of BCs (and SECCO~1) in the direction of Virgo overlaid on a ROSAT mosaic of hard (0.4-2.4~keV) X-ray emission \citep{Brown+2021}. Virgo members and possible members \citep[from the Extended Virgo Cluster Catalog;][]{Kim+2014} are plotted as faint black, unfilled circles. The area of each circle is proportional to the total $r$-band flux of the galaxy it represents. The BCs are shown with blue symbols (see legend) and SECCO~1 is shown as a purple cross. The symbol for BC2 is unfilled as this object is spurious (see \S\ref{sec:morph}). The approximate virial radius \citep[taken to be 1.7~Mpc,][]{Kashibadze+2020} of the cluster is shown by a large dashed black circle.}
    \label{fig:BC_locs}
\end{figure}

\begin{figure*}
    \centering
    \includegraphics[width=\columnwidth]{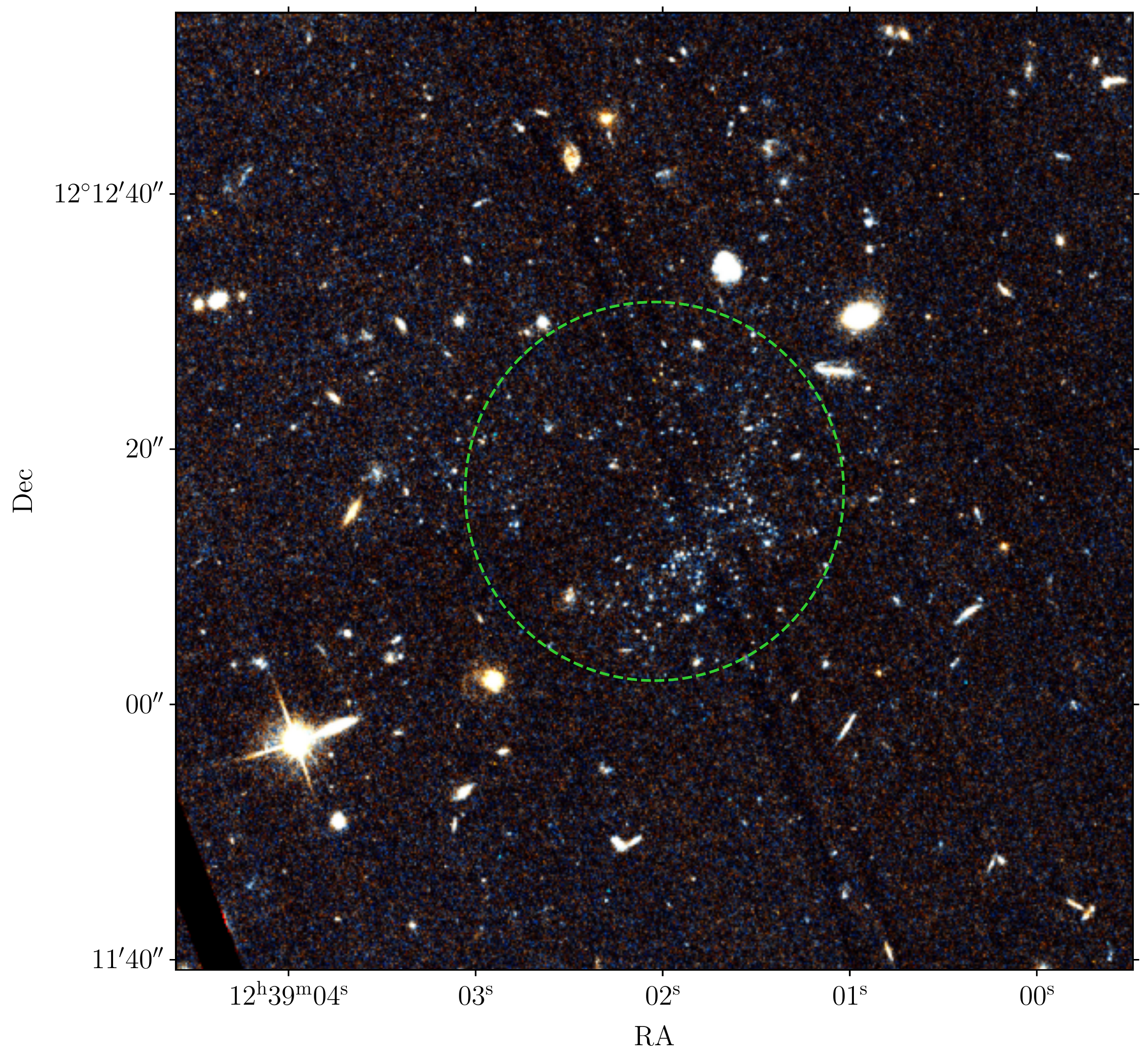}
    \includegraphics[width=\columnwidth]{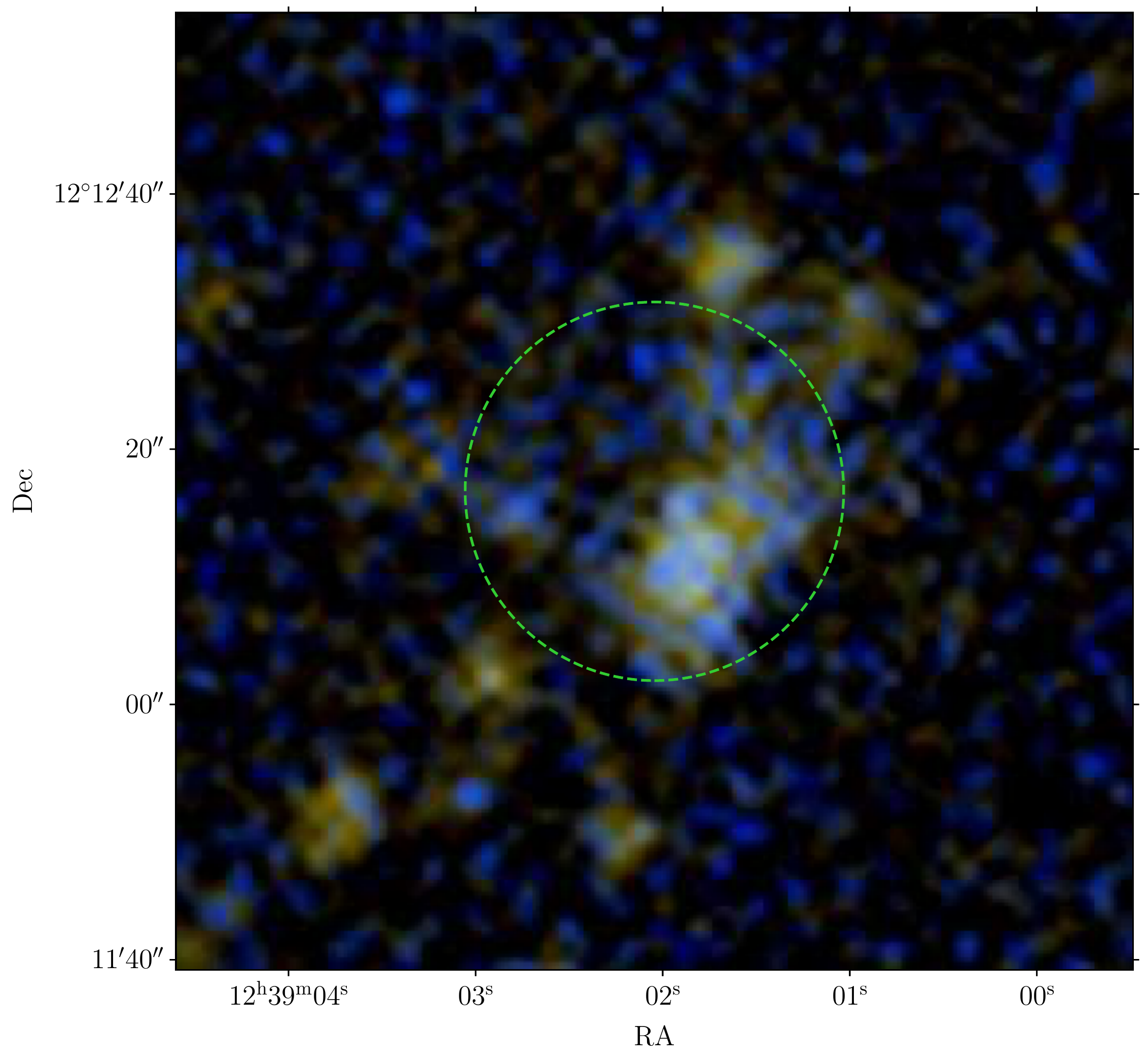}
    \includegraphics[width=0.5\columnwidth]{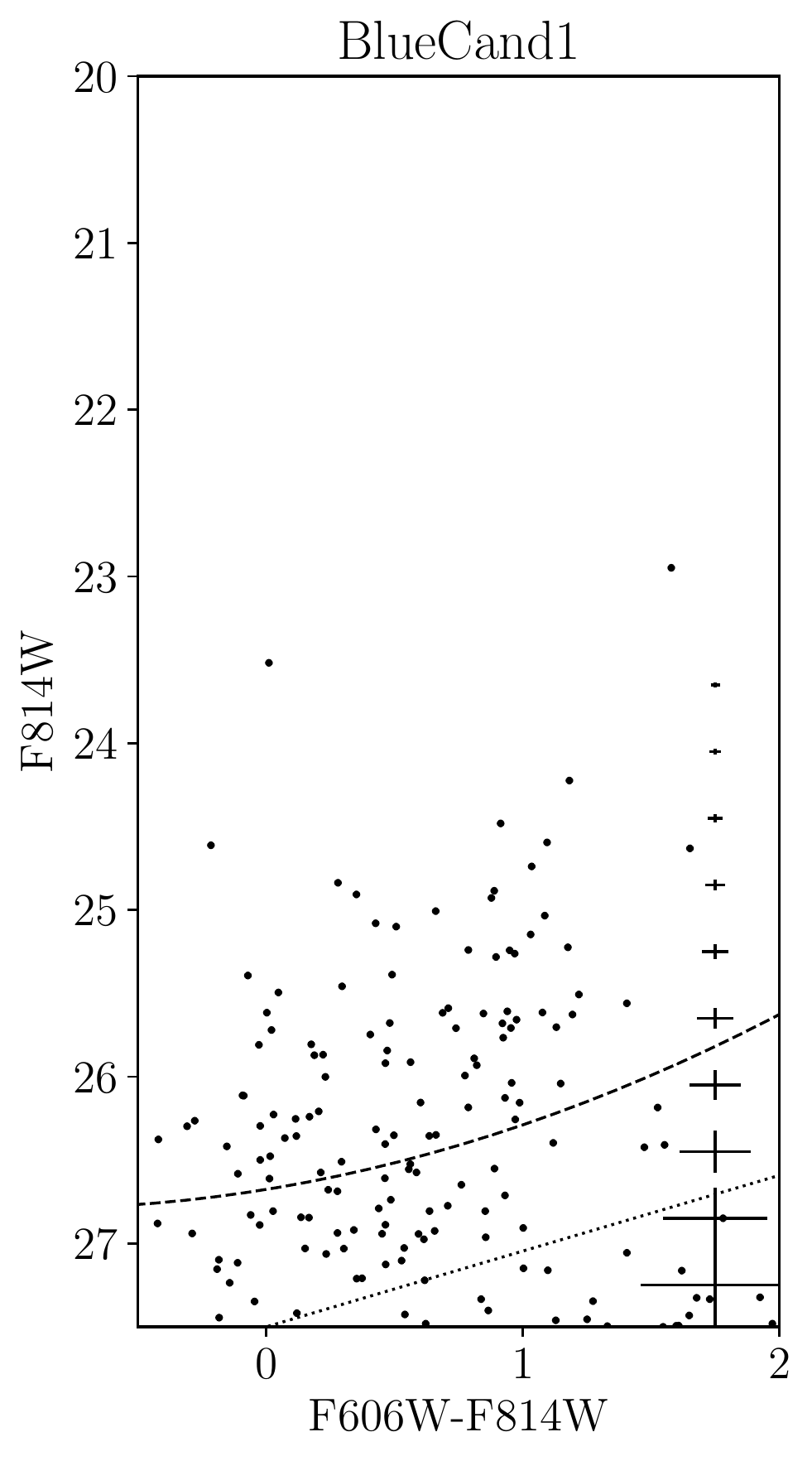}
    \includegraphics[width=0.5\columnwidth]{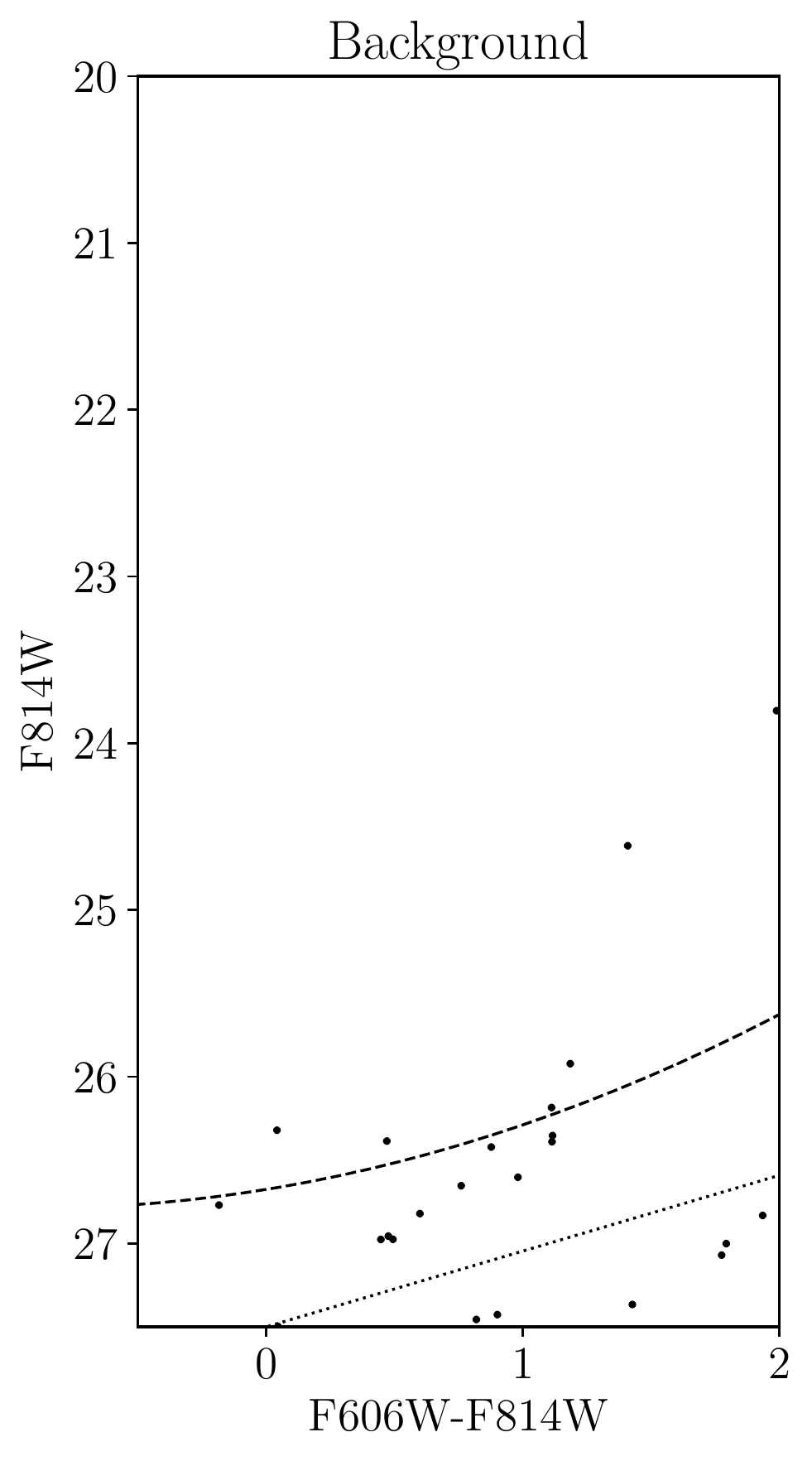}
    \caption{\textit{Top-left}: False color HST F606W+F814W image of BC1. The dashed green circle shows the region used to construct the CMD. At the distance of the Virgo cluster (16.5~Mpc) 20\arcsec \ is 1.6~kpc. \textit{Top-right}: GALEX NUV+FUV image showing the same field. \textit{Bottom-left}: CMD of the point sources within the aperture shown. The dashed lines indicates the 90\% completeness limit and the dotted line the 50\% limit. The errorbars indicate the typical uncertainties (from artificial star tests) in the F814W magnitude and F606W-F814W color, as a function of F814W magnitude. \textit{Bottom-right}: The CMD of a background region of the HST image away from bright sources. The aperture used was equal in area to the target aperture.}
    \label{fig:BC1_HST_GLX}
\end{figure*}

\begin{figure*}
    \centering
    \includegraphics[width=\columnwidth]{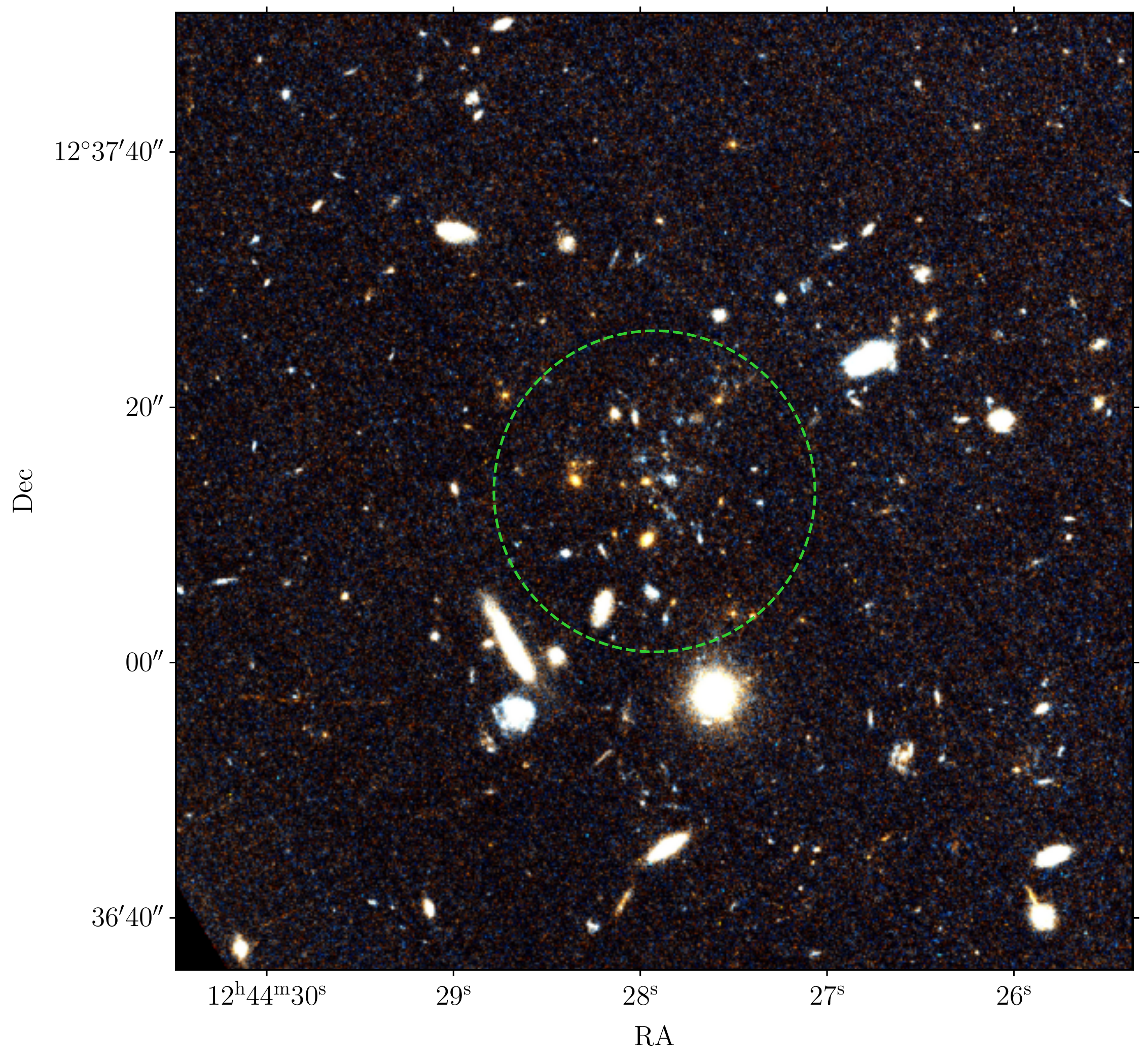}
    \includegraphics[width=\columnwidth]{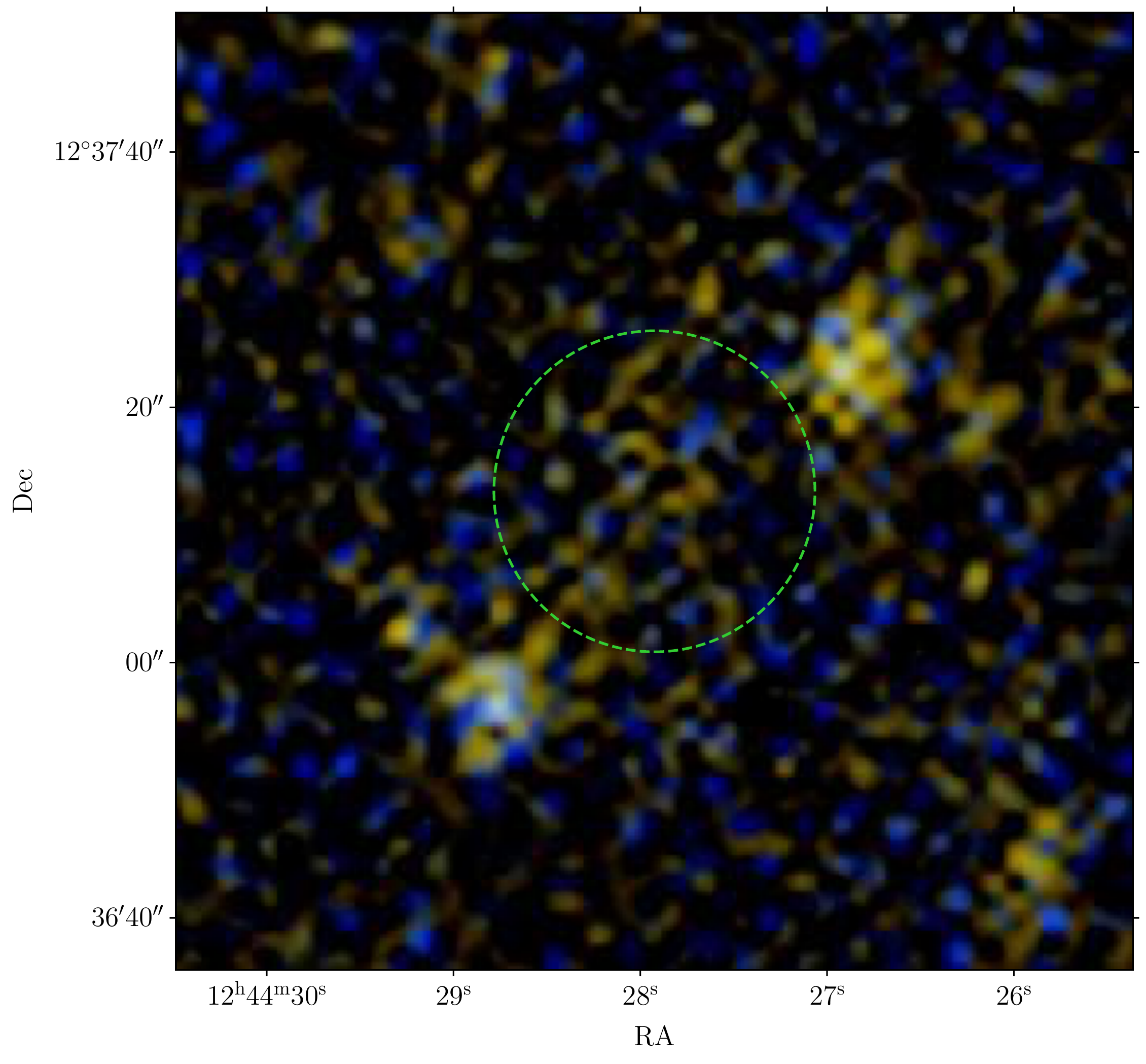}
    \includegraphics[width=0.5\columnwidth]{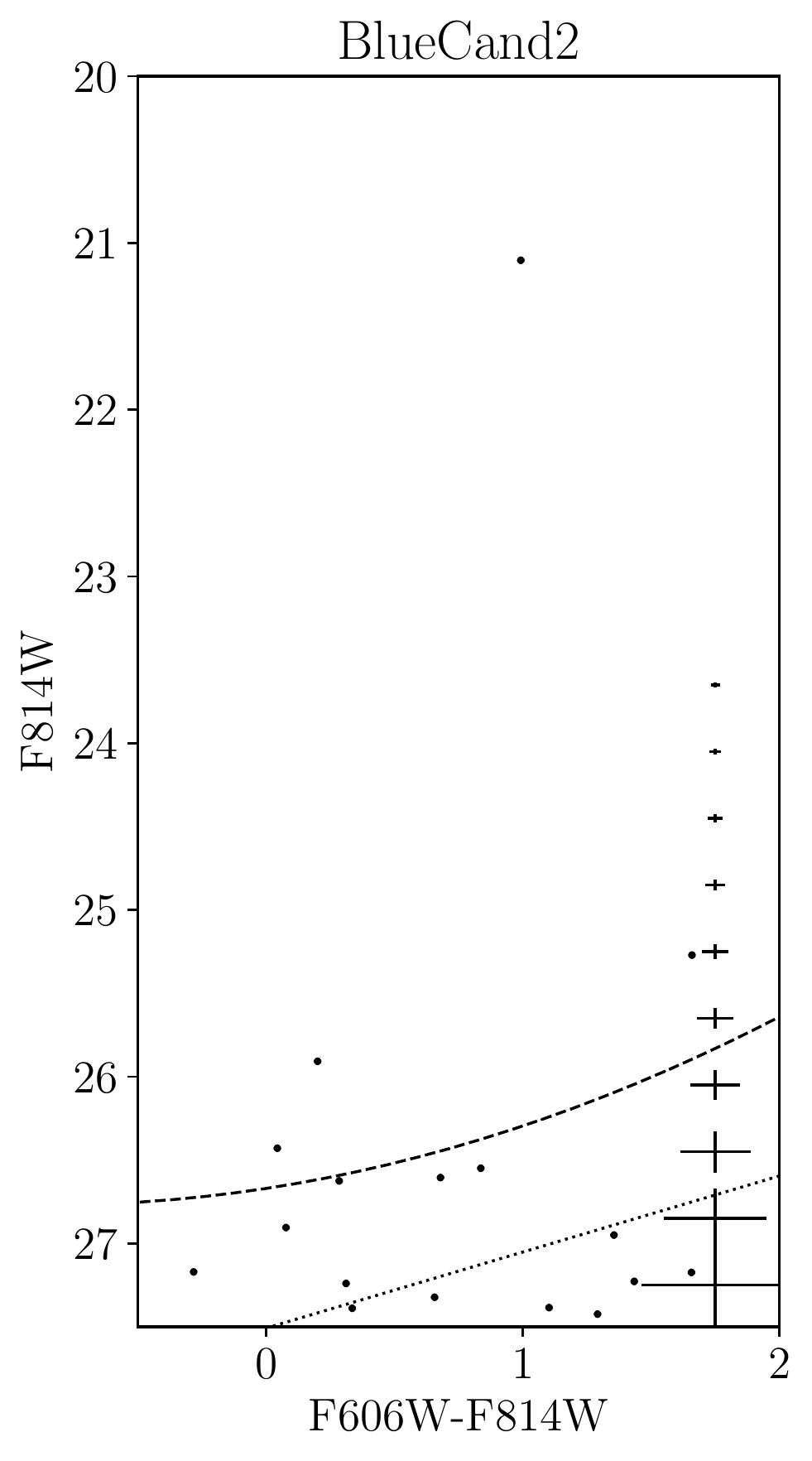}
    \includegraphics[width=0.5\columnwidth]{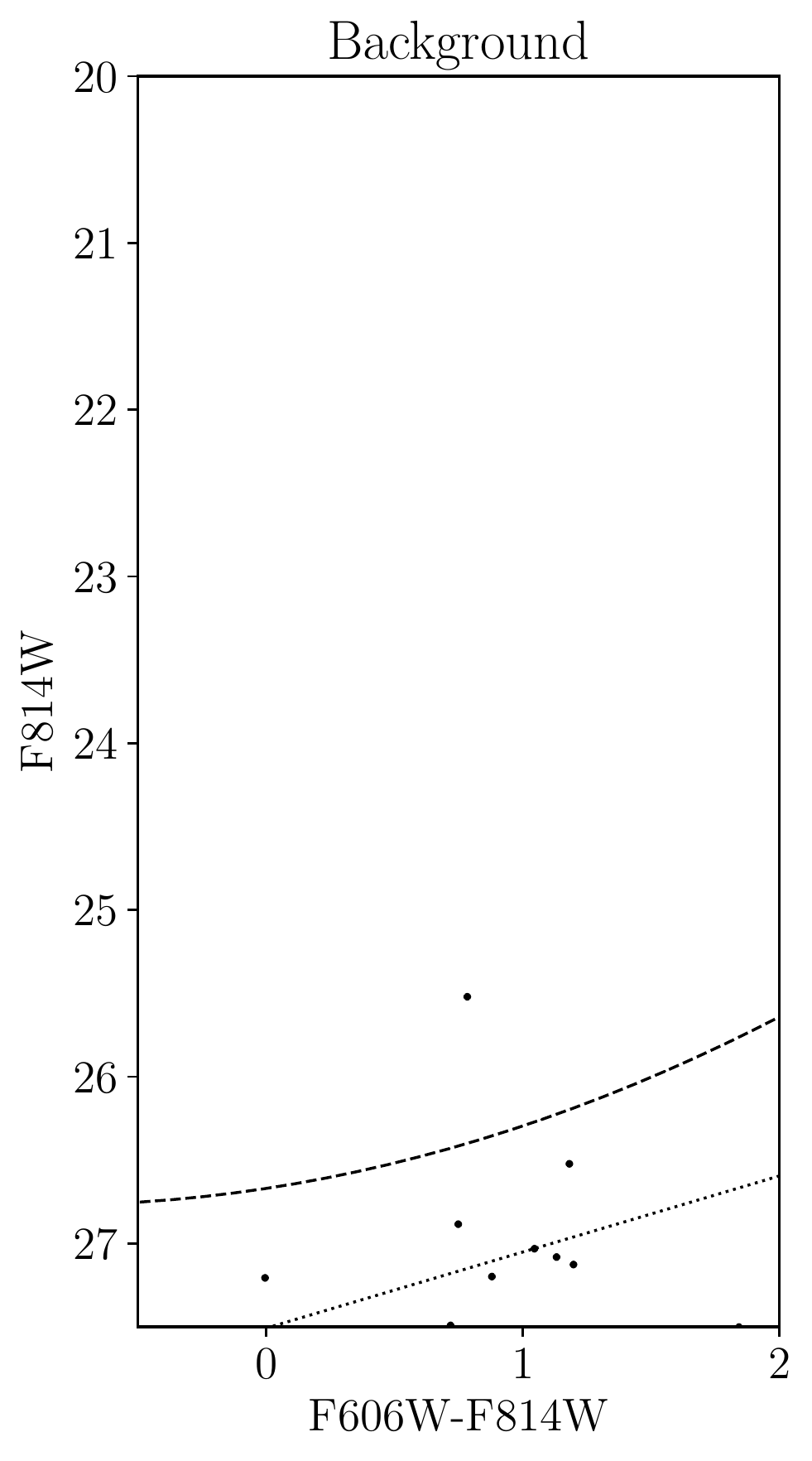}
    \caption{\textit{Top-left}: False color HST F606W+F814W image of BC2. The dashed green circle shows the region used to construct the CMD. Unlike the other BCs, this HST images appears to indicate that this is a background galaxy group. \textit{Top-right}: GALEX NUV+FUV image showing the same field. There is only very weak NUV emission associated with BC2. \textit{Bottom}: CMD within the aperture shown (left) and a blank field aperture (right).  The dotted and dashed lines, and error bars, are the same as described in Figure \ref{fig:BC1_HST_GLX}, bottom panels. This CMD appears to be consistent with background, supporting the conclusion that this is a spurious blue stellar system candidate.}
    \label{fig:BC2_HST_GLX}
\end{figure*}

\begin{figure*}
    \centering
    \includegraphics[width=\columnwidth]{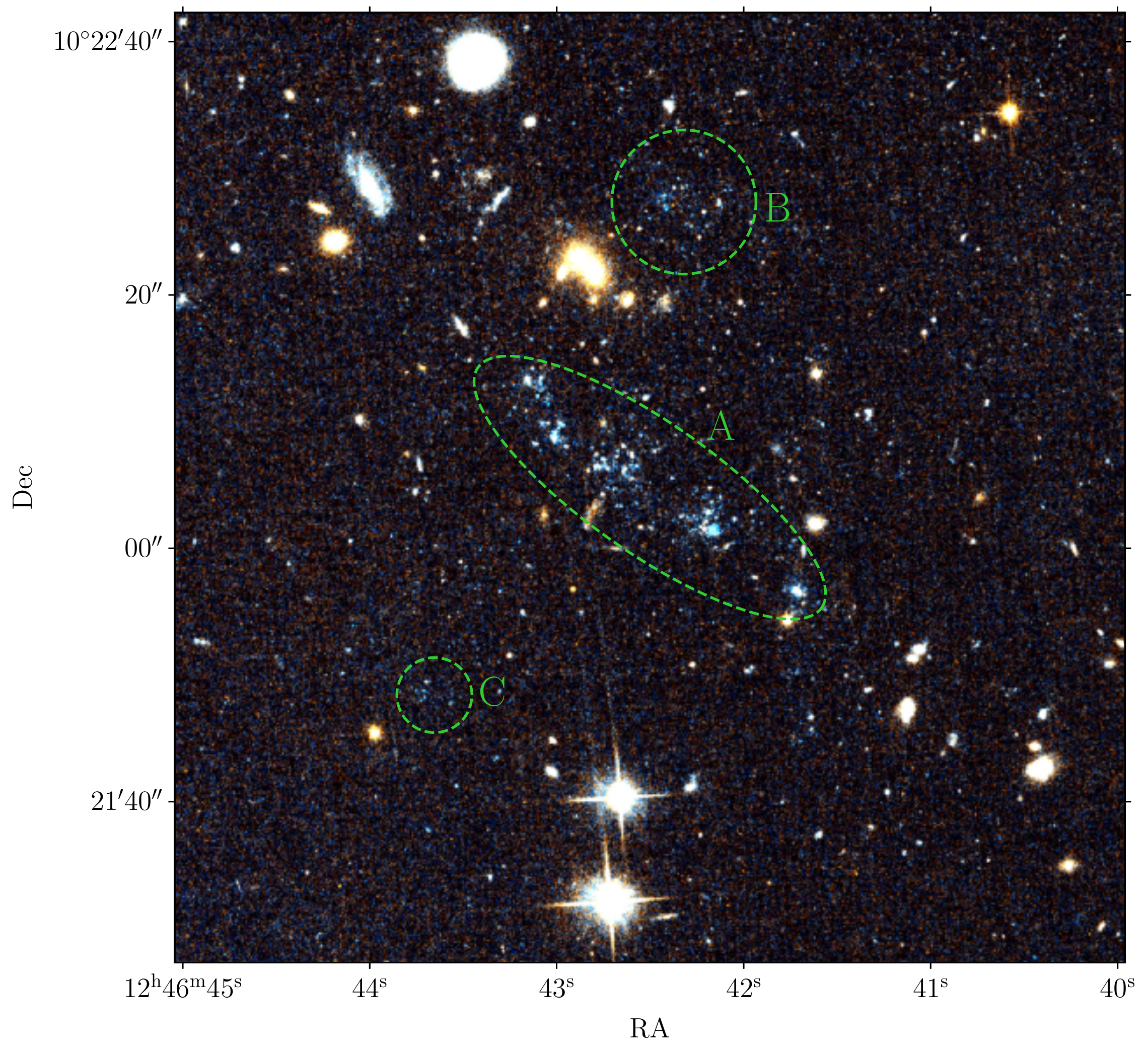}
    \includegraphics[width=\columnwidth]{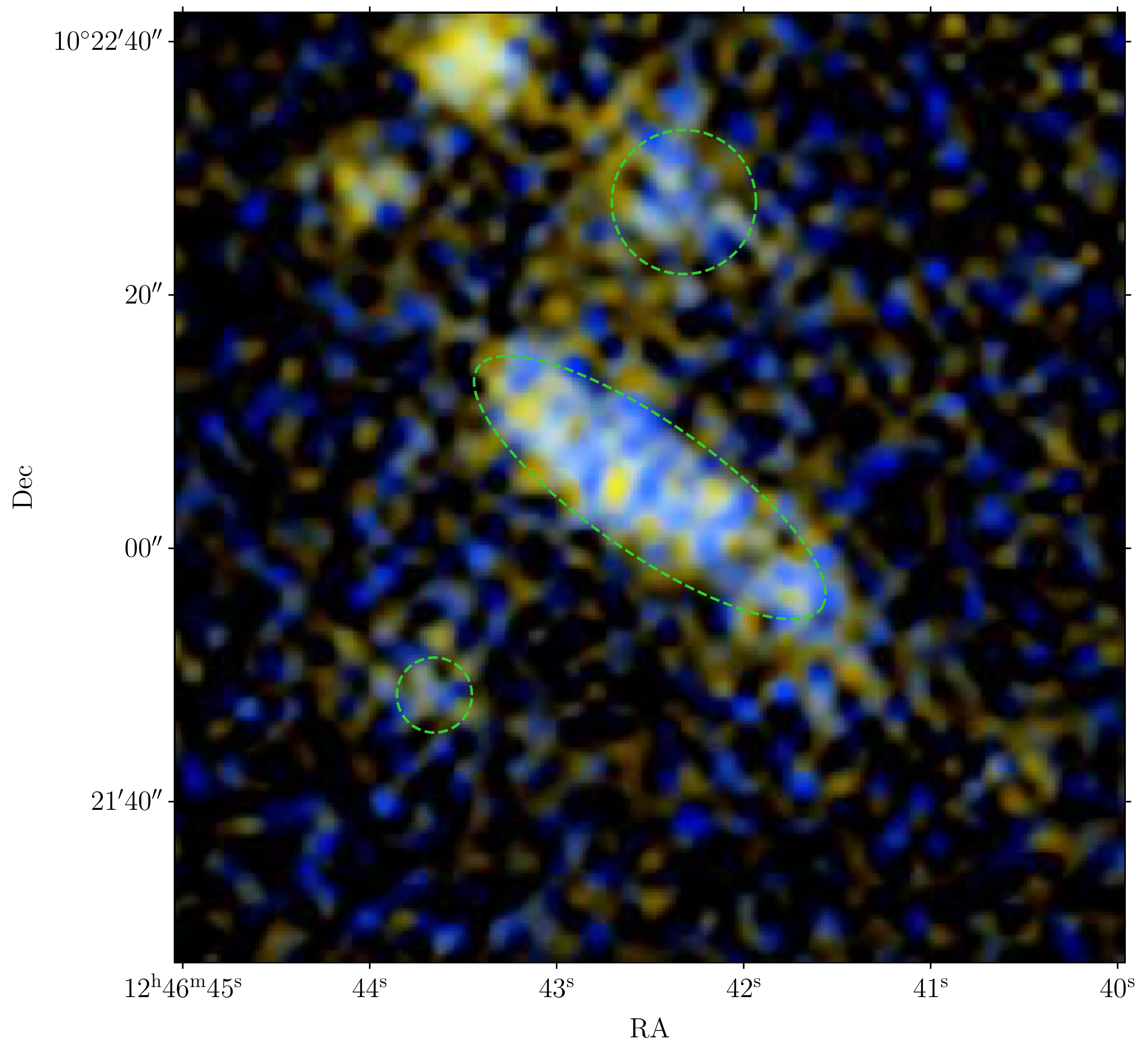}
    \includegraphics[width=0.5\columnwidth]{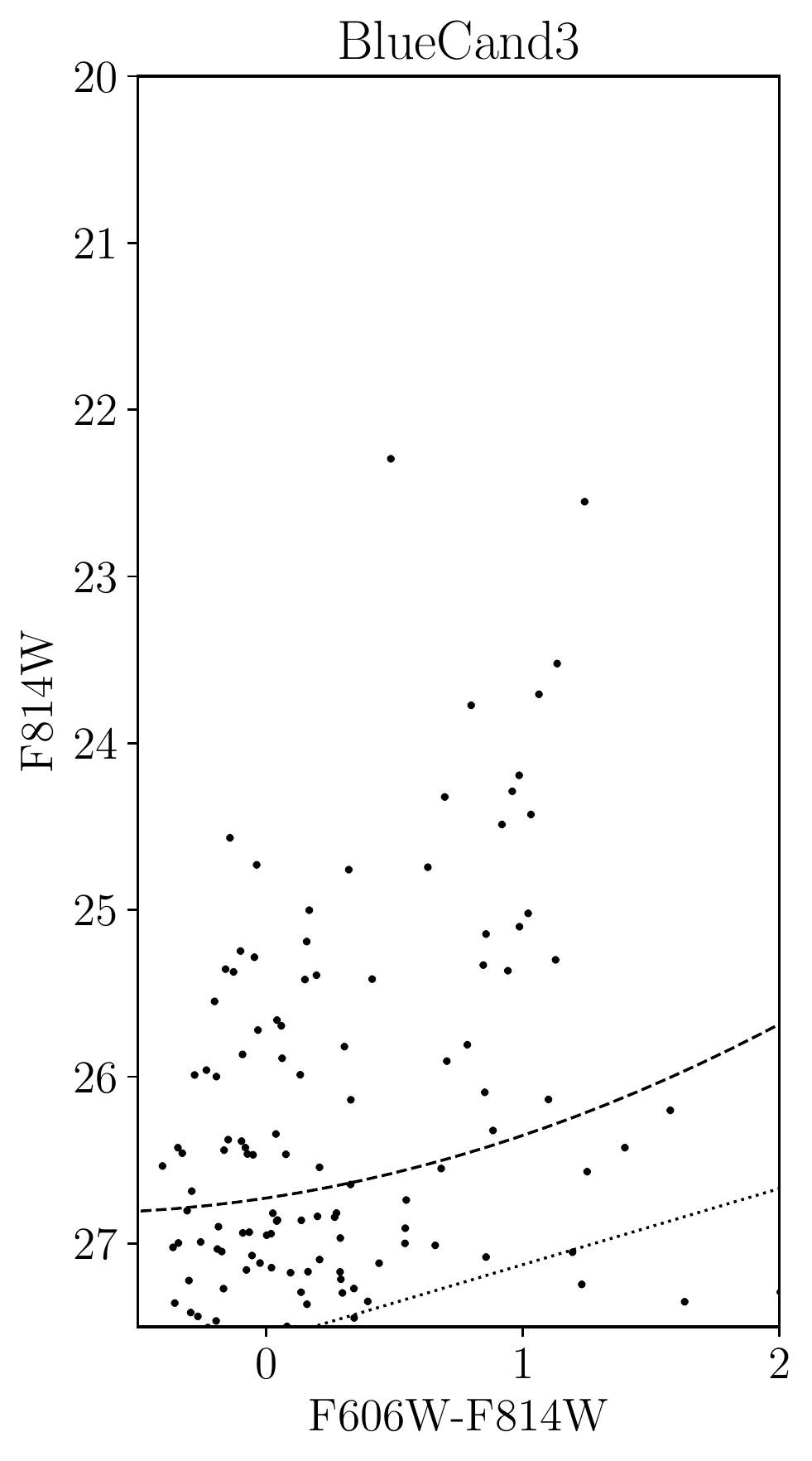}
    \includegraphics[width=0.5\columnwidth]{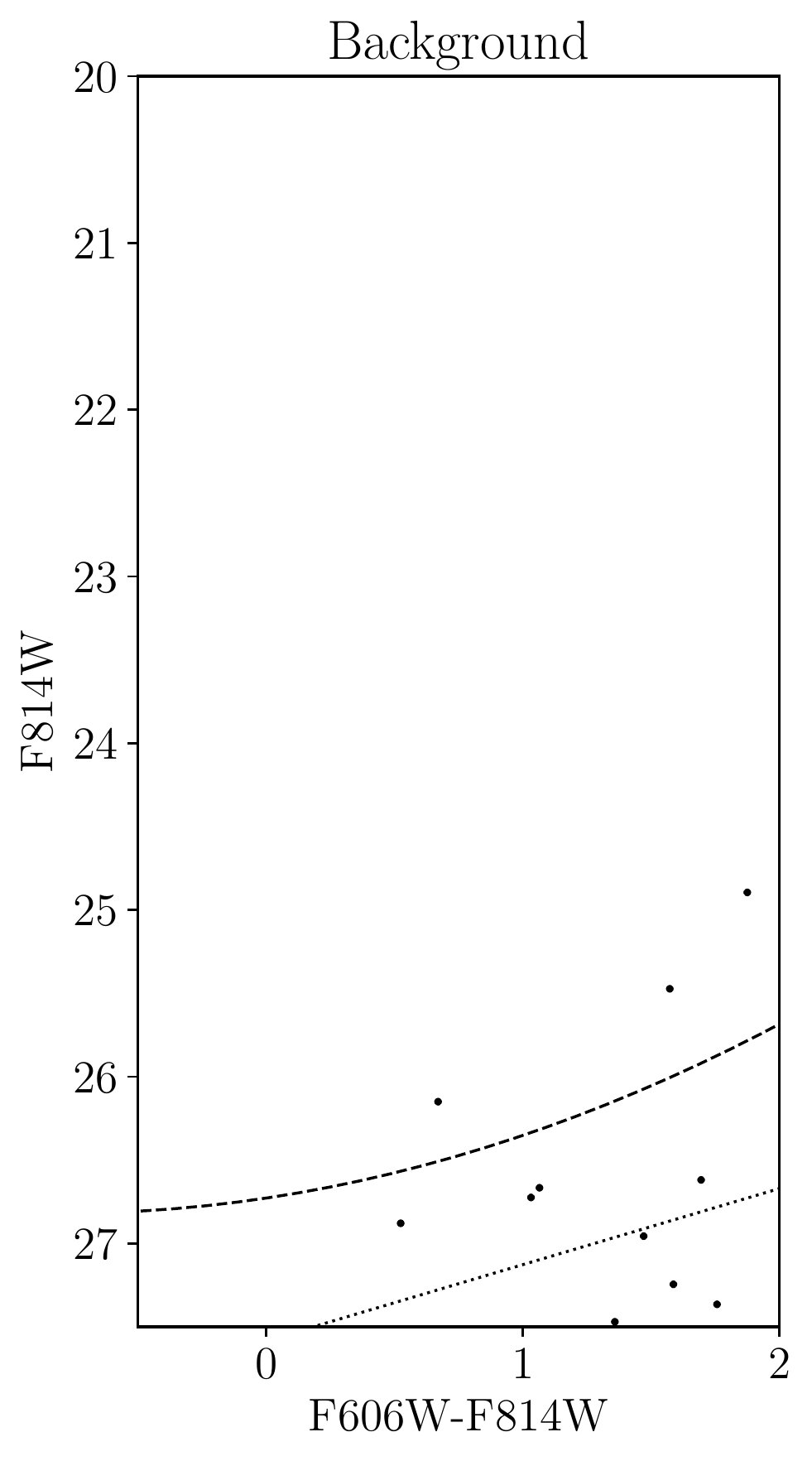}
    \caption{\textit{Top-left}: False color HST F606W+F814W image of BC3. The dashed green ellipse and circles show the regions used to construct the CMD. \textit{Top-right}: GALEX NUV+FUV image showing the same field. \textit{Bottom}: CMD within the apertures shown (left) and a blank field aperture (right). See Figure \ref{fig:BC1_HST_GLX} caption for further details.}
    \label{fig:BC3_HST_GLX}
\end{figure*}

\begin{figure*}
    \centering
    \includegraphics[width=1.5\columnwidth]{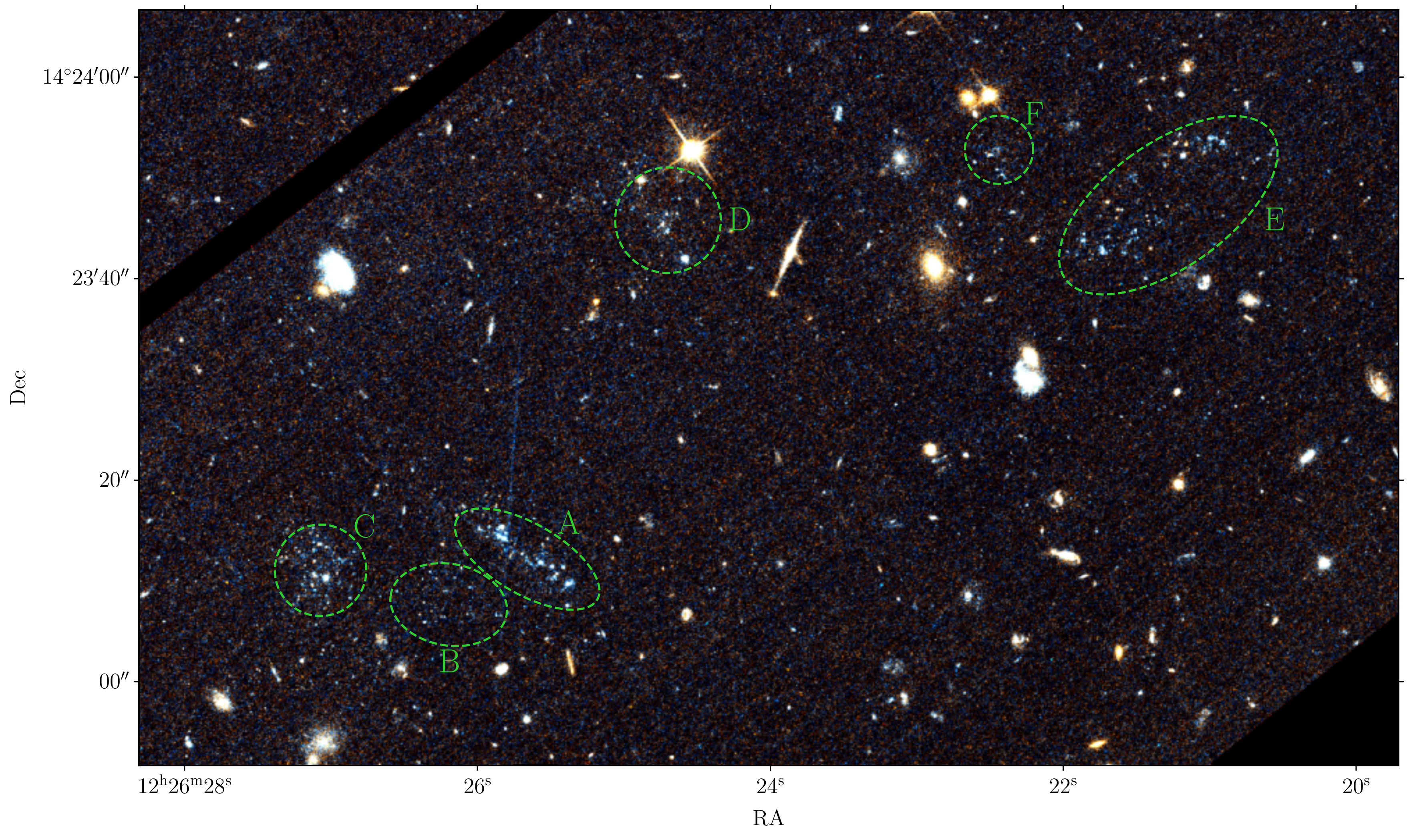}
    \includegraphics[width=1.5\columnwidth]{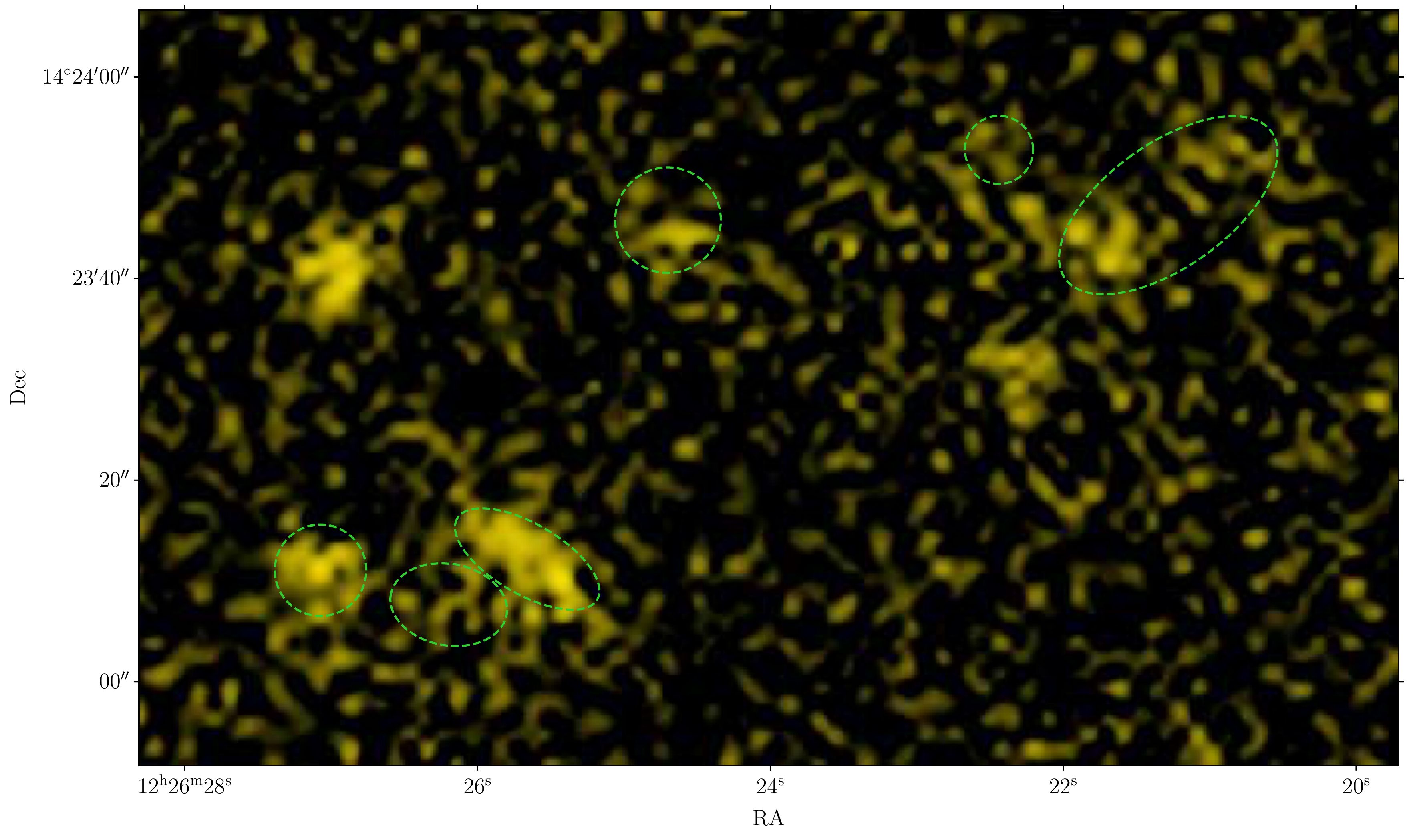} \\
    \includegraphics[width=0.5\columnwidth]{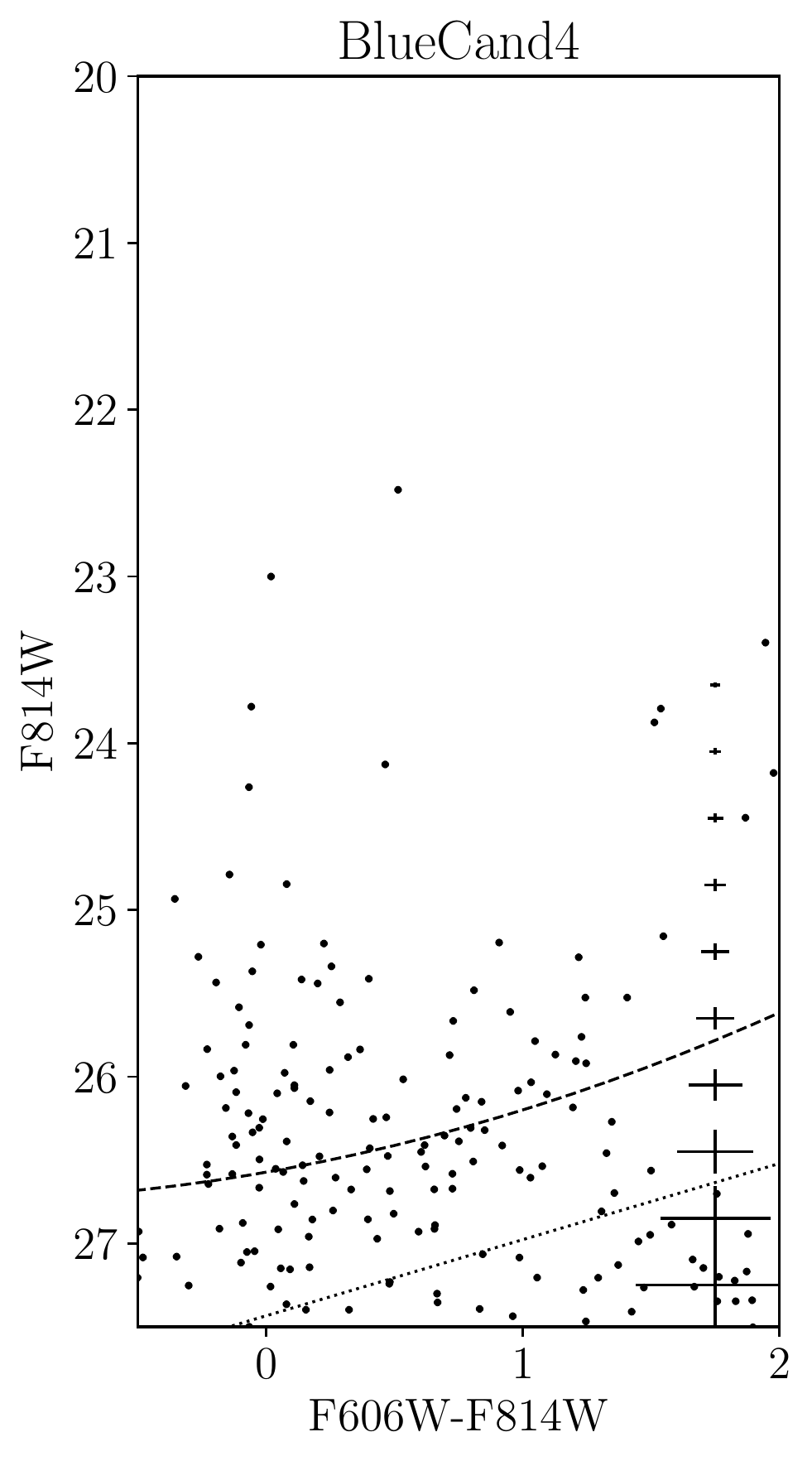}
    \includegraphics[width=0.5\columnwidth]{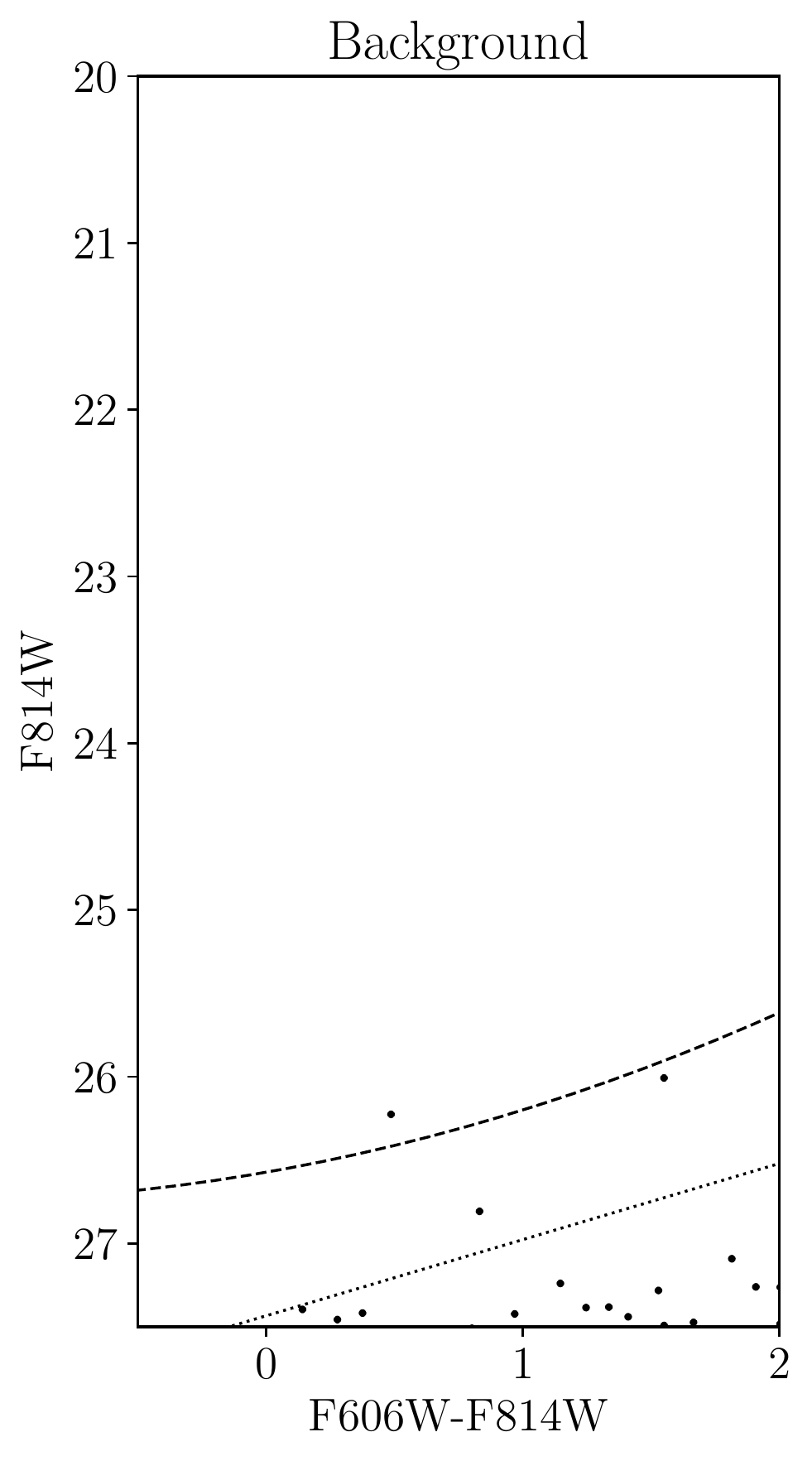}
    \caption{\textit{Top}: False color HST F606W+F814W image of BC4. The dashed green ellipses and circles show the regions used to construct the CMD. \textit{Middle}: GALEX NUV image showing the same field. \textit{Bottom}: CMD within the apertures shown (left) and a blank field aperture (right). See Figure \ref{fig:BC1_HST_GLX} caption for further details.}
    \label{fig:BC4_HST_GLX}
\end{figure*}

\begin{figure*}
    \centering
    \includegraphics[width=\columnwidth]{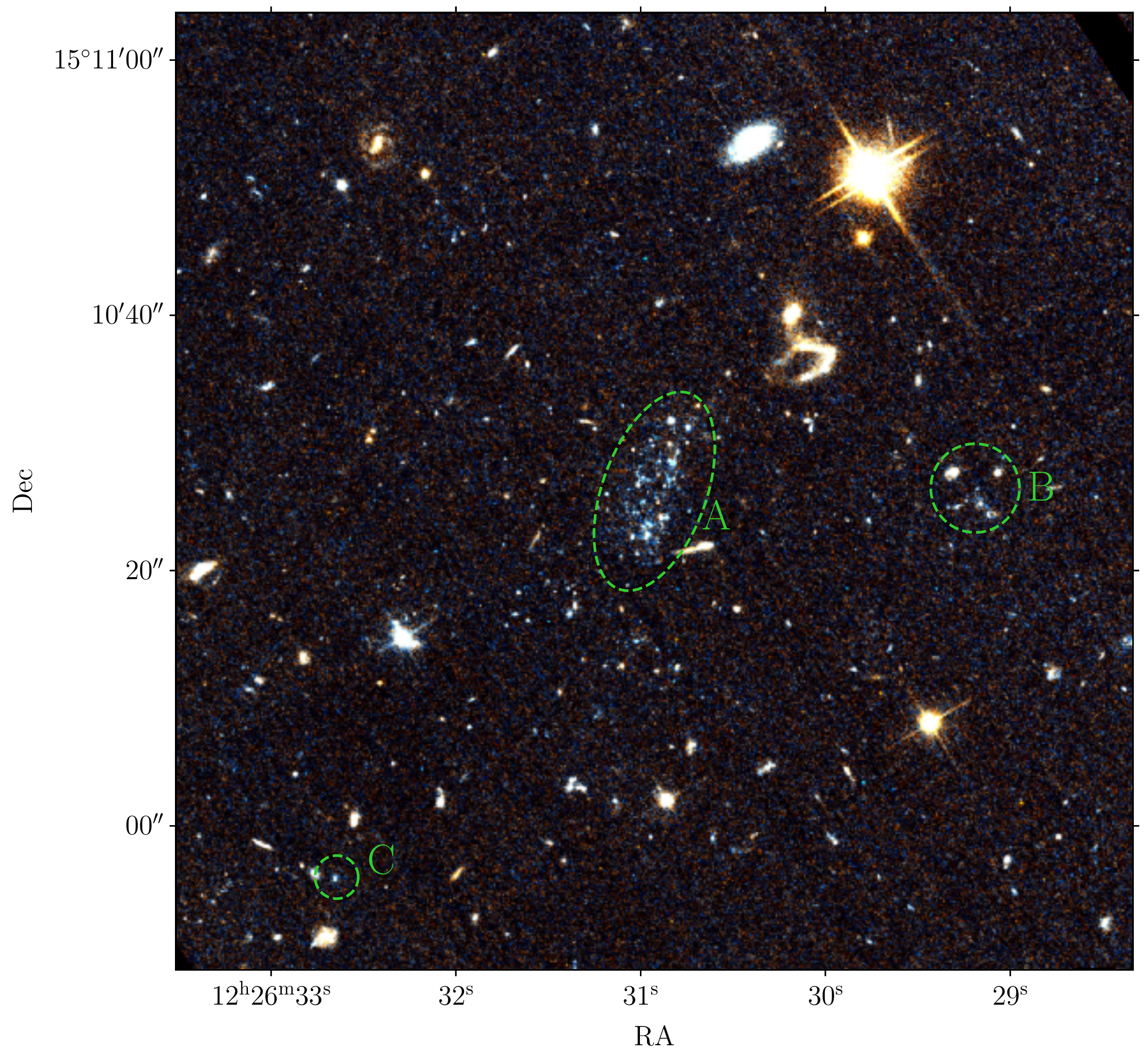}
    \includegraphics[width=\columnwidth]{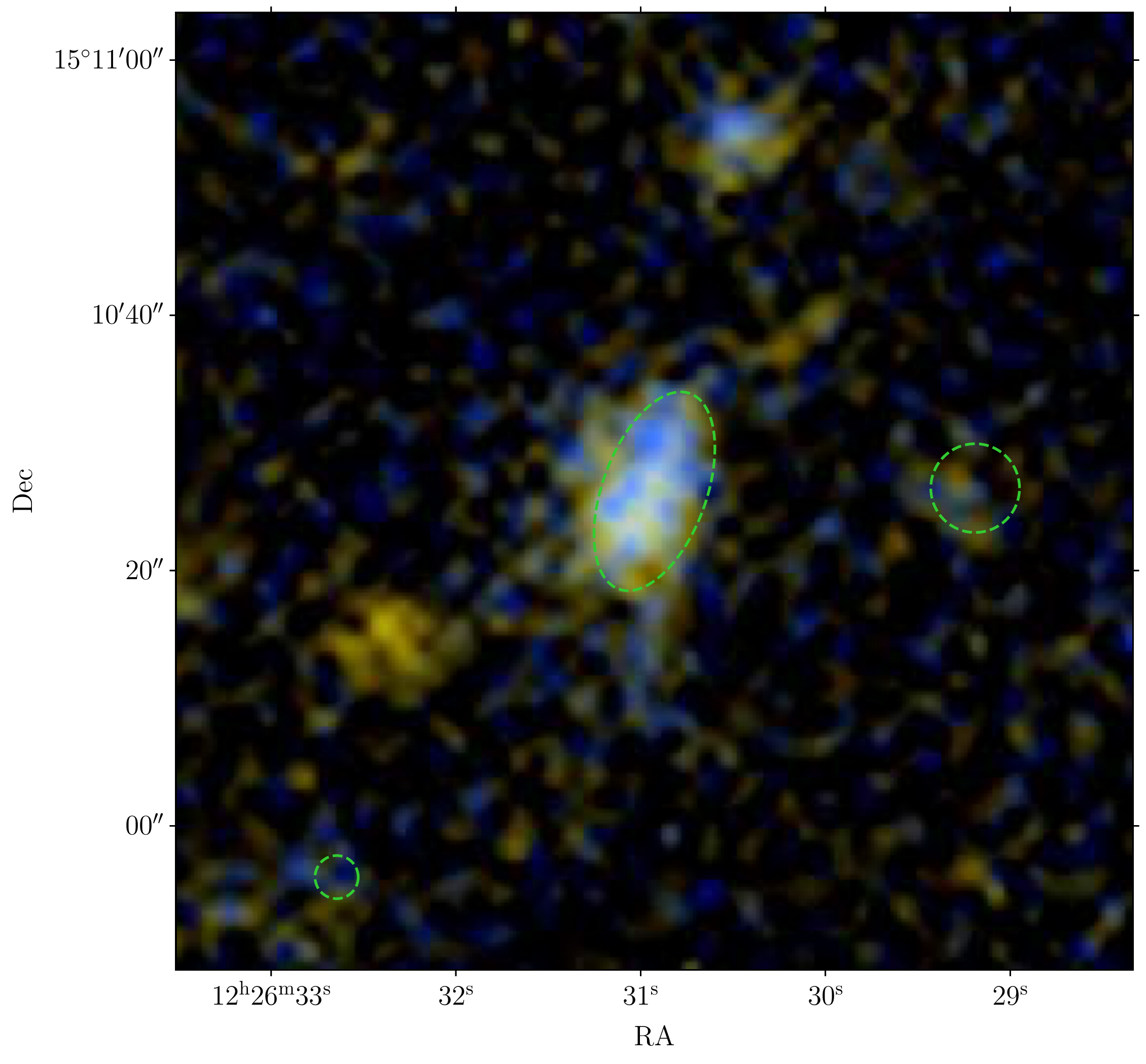}
    \includegraphics[width=0.49\columnwidth]{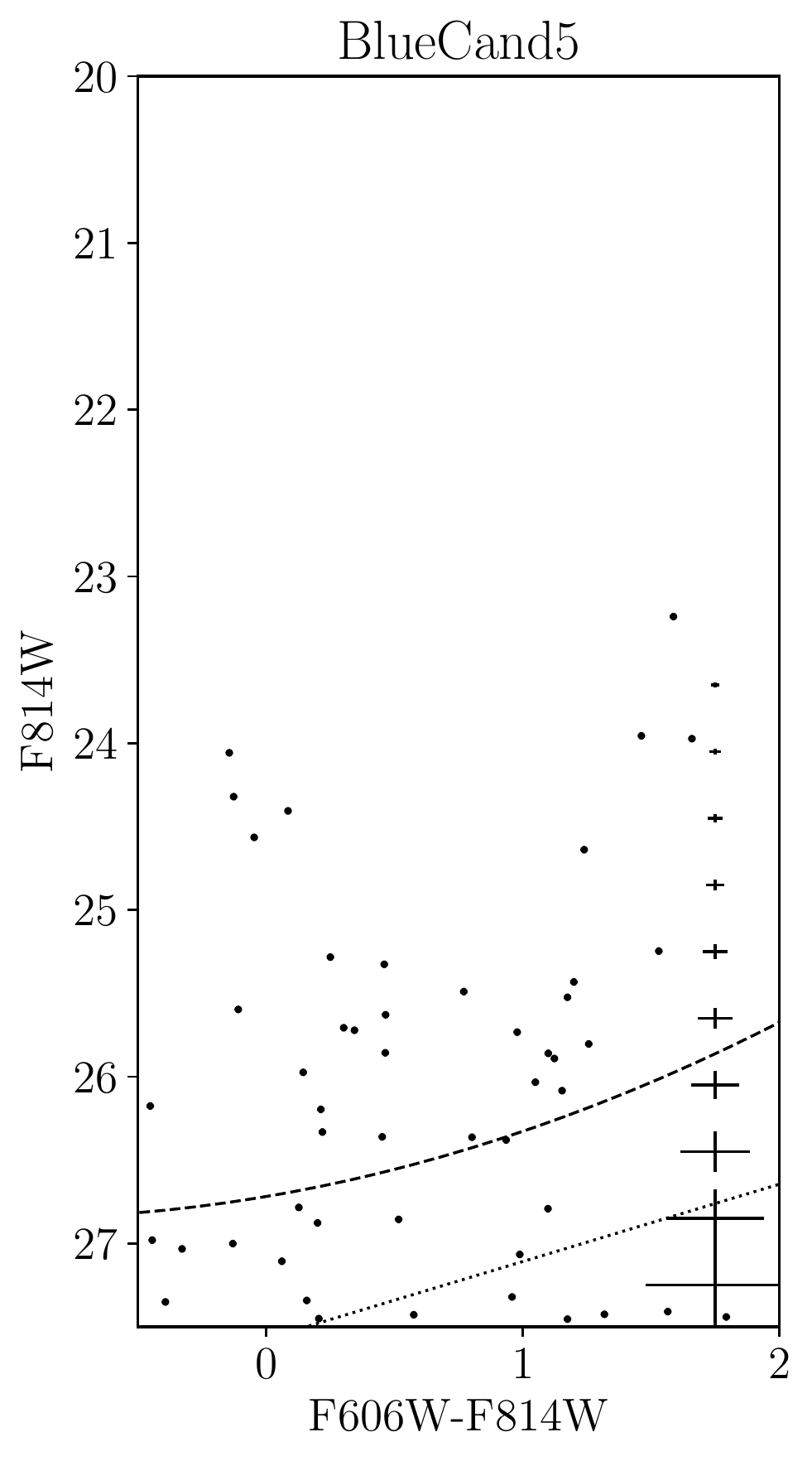}
    \includegraphics[width=0.49\columnwidth]{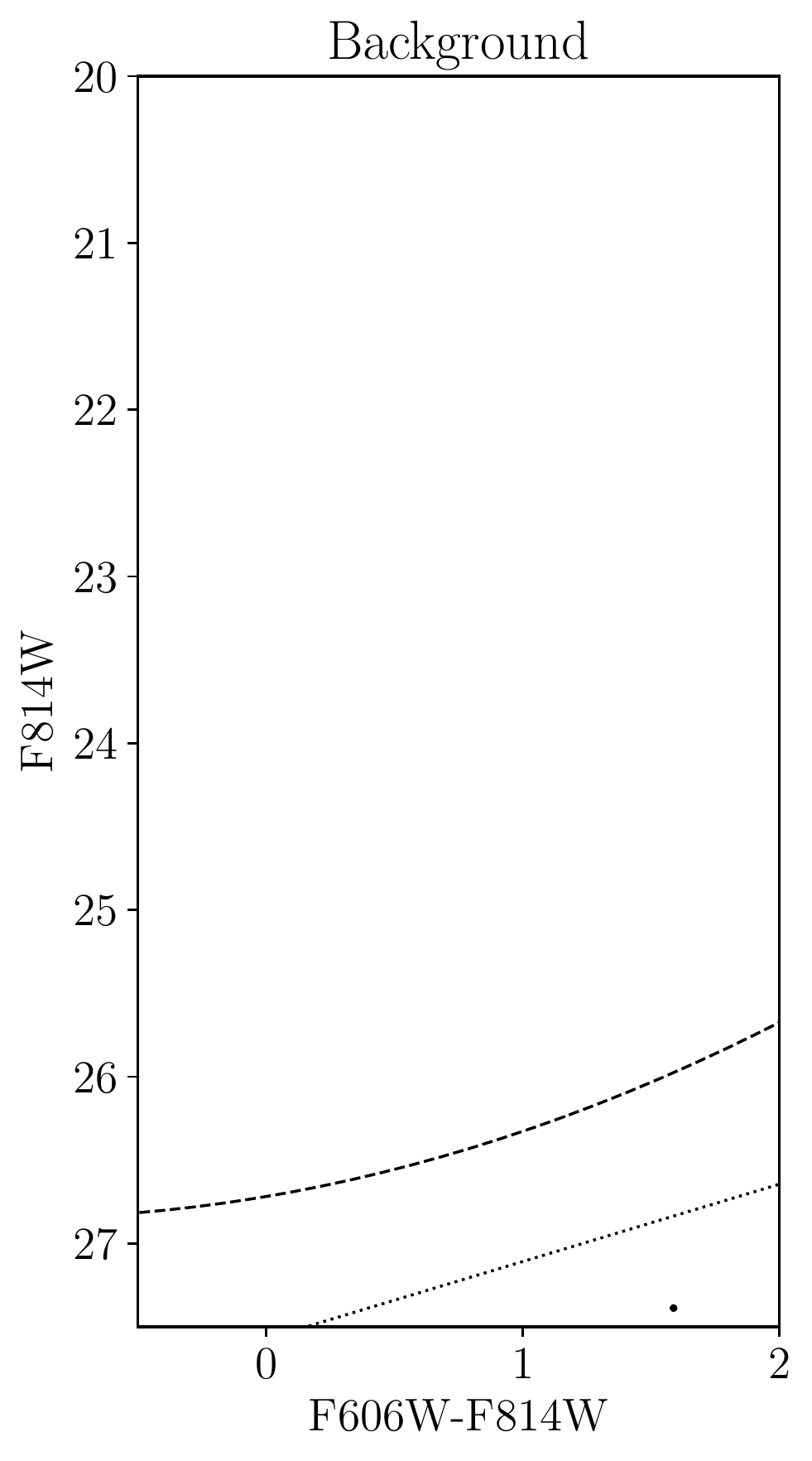}
    \caption{\textit{Top-left}: False color HST F606W+F814W image of BC5. The dashed green ellipse and circle show the regions used to construct the CMD. The component BC5c was identified via H$\alpha$ emission \citepalias{Bellazzini+2022} to be at the same velocity as the main body, but may only be a single cluster of stars. \textit{Top-right}: GALEX NUV+FUV image showing the same field. \textit{Bottom}: CMD within the apertures shown (left) and a blank field aperture (right). See Figure \ref{fig:BC1_HST_GLX} caption for further details.}
    \label{fig:BC5_HST_GLX}
\end{figure*}

\begin{figure}
    \centering
    \includegraphics[width=\columnwidth]{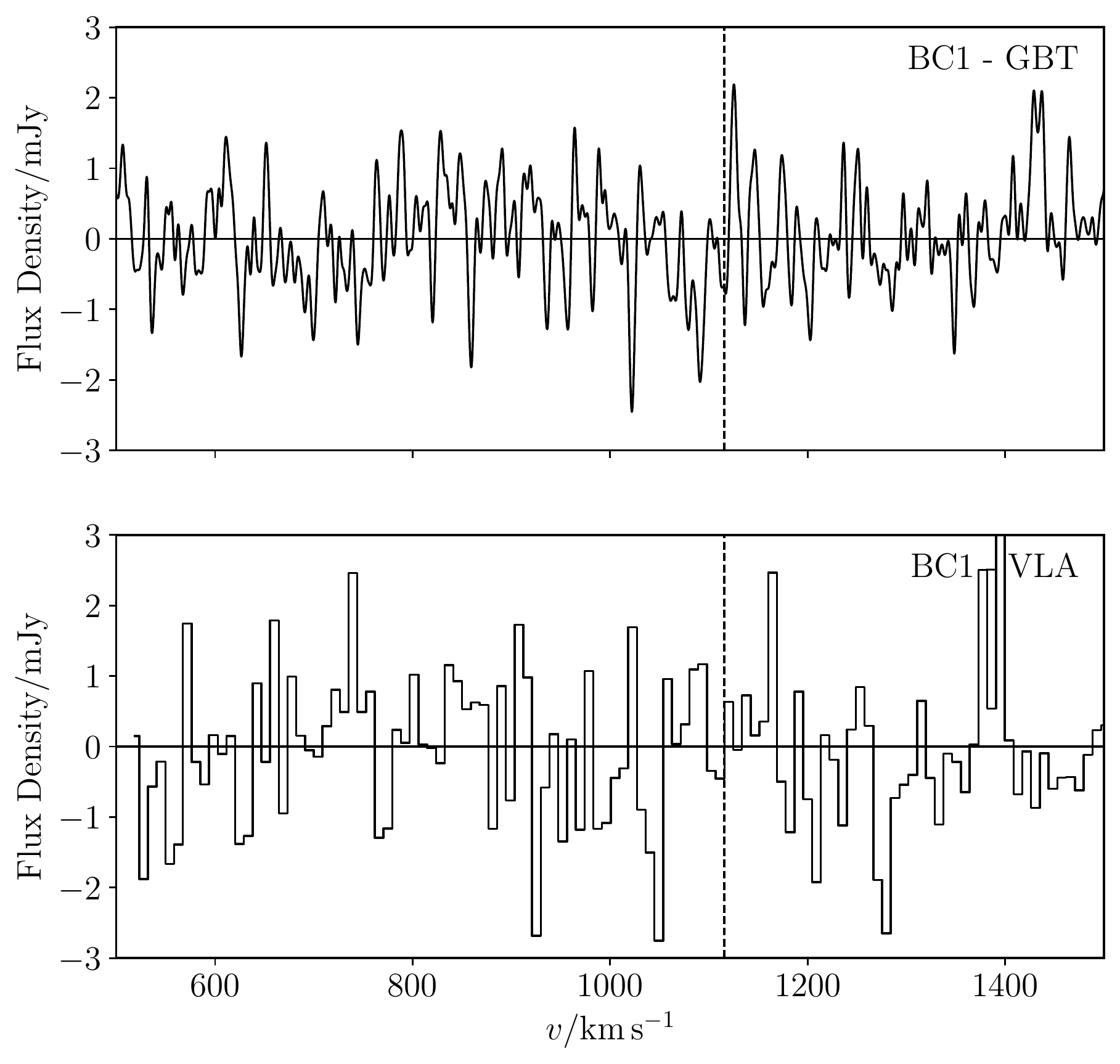}
    \caption{\hi \ spectra of BC1 from the GBT (top) and the VLA (bottom). The VLA spectrum was extracted using an aperture equal in area to the synthesized beam. The vertical dashed lines correspond to the H$\alpha$ velocity measurement from MUSE. No significant signal is detected in either spectrum.}
    \label{fig:BC1_VLA_GBT_specs}
\end{figure}

\begin{figure}
    \centering
    \includegraphics[width=\columnwidth]{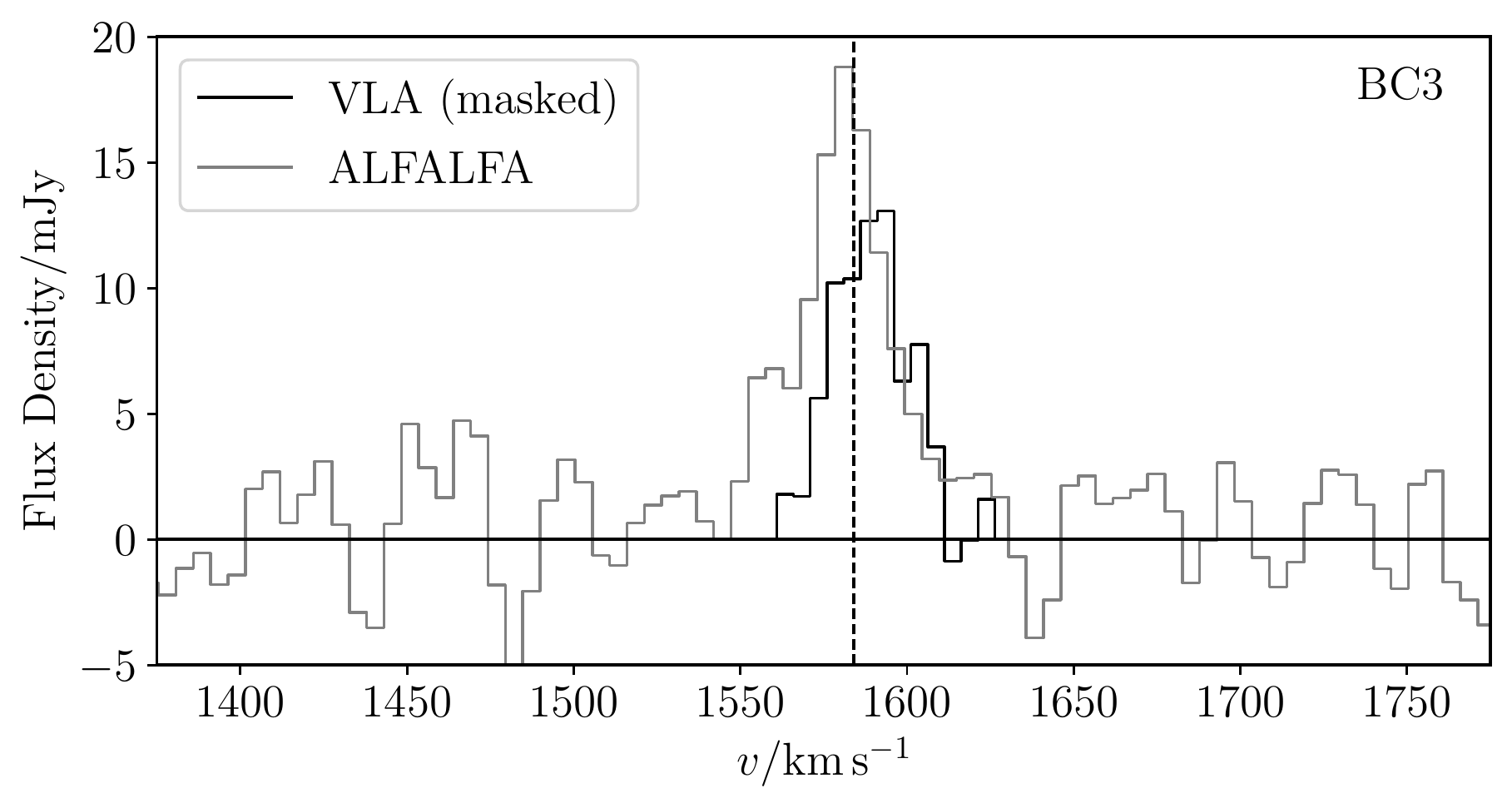}
    \caption{\hi \ spectra of BC3 from ALFALFA and the VLA. The ALFALFA spectrum is the public spectrum from \citet{Haynes+2018} and the VLA spectrum was created using the extended source mask of \citet{Jones+2022}. The vertical dashed line corresponds to the H$\alpha$ velocity measurement from MUSE. BC3 is detected at high signal-to-noise ratio in both spectra and both agree with the H$\alpha$ velocity. However, the VLA measures a somewhat lower flux, with most of the missing emission lying on the approaching side of the line profile. This likely indicates the presence of extended emission below the surface brightness limit of the VLA observations \citep{Cannon+2015,Jones+2022}.}
    \label{fig:BC3_VLA_specs}
\end{figure}

\begin{figure}
    \centering
    \includegraphics[width=\columnwidth]{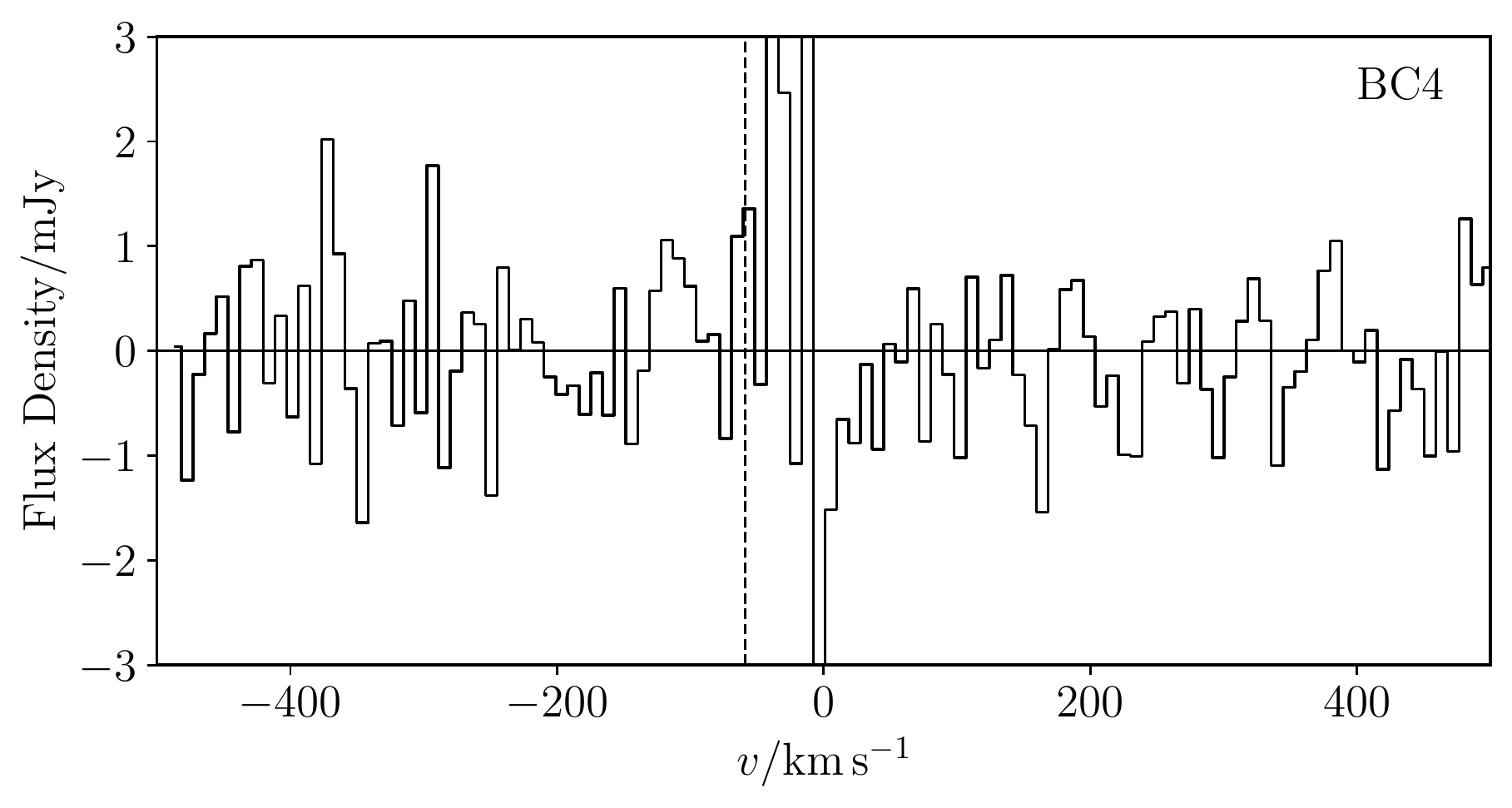}
    \includegraphics[width=\columnwidth]{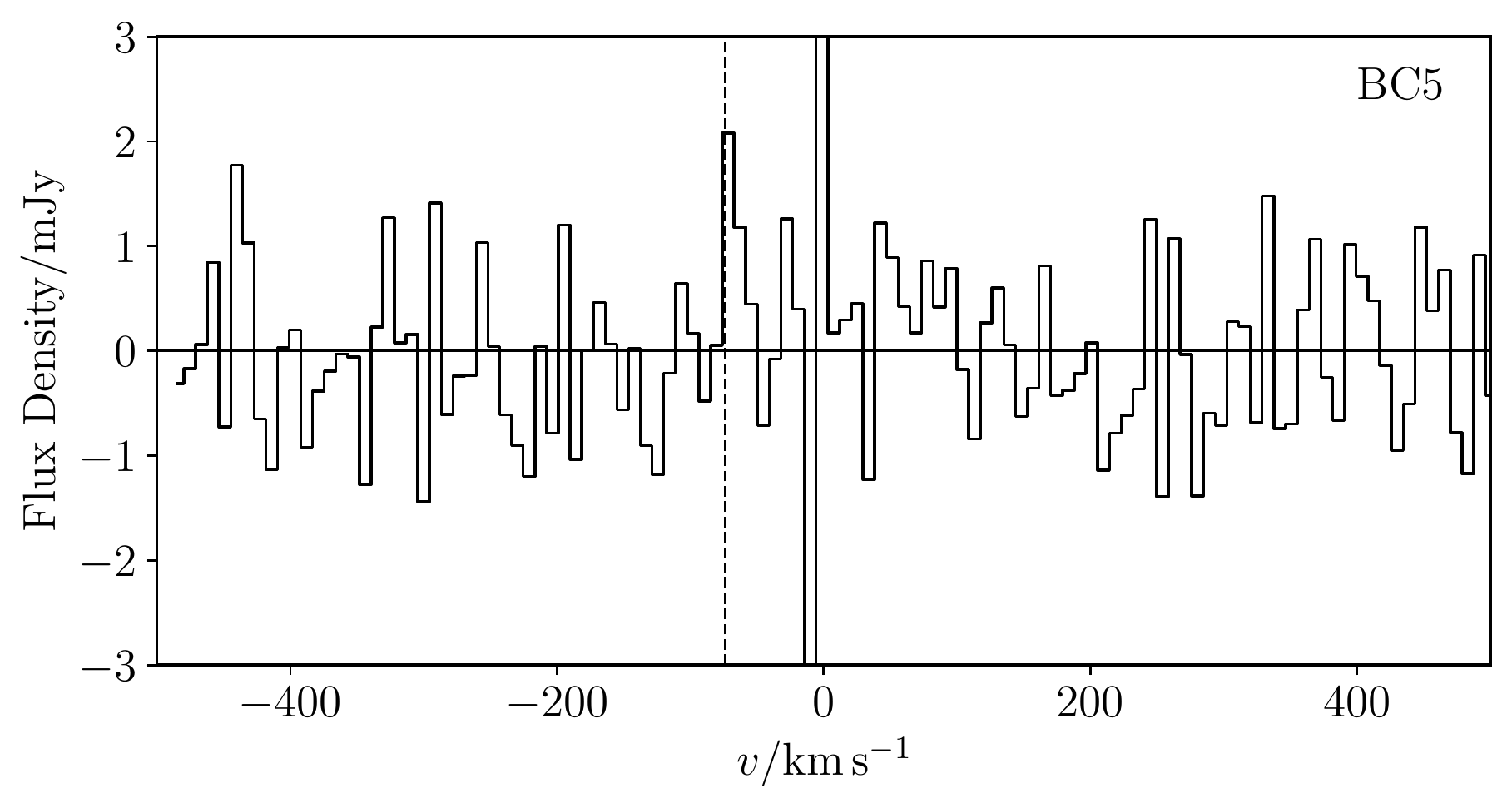}
    \caption{\hi \ spectra in the directions of BC4 (top) and BC5 (bottom) extracted from the VLA \hi \ data cubes within an aperture equal in area to the synthesized beam. The vertical dashed lines correspond to the H$\alpha$ velocity measurements from MUSE.
    Neither shows a significant \hi \ line signal, although they could be contaminated with Milky Way \hi \ emission. The apparent peak coincident with the H$\alpha$ velocity of BC5 is below 3$\sigma$ and is likely a noise spike.}
    \label{fig:BC45_VLA_specs}
\end{figure}

In this section we present the results of our multi-wavelength investigation of BCs, providing a description of their morphology, colors, redshifts, stellar masses, metallicities, and gas content. These physical properties will then be used as the basis for a search for candidate parent objects in the following section and a discussion of potential formation pathways in \S\ref{sec:formation}. 

We include SECCO~1 in this sample throughout and either use quantities measured in previous work or (re)measure them as needed (e.g. to provide equivalent values across the whole sample).

\subsection{Morphology and location}
\label{sec:morph}

The HST images of the BCs are shown in Figures \ref{fig:BC1_HST_GLX}, \ref{fig:BC2_HST_GLX}, \ref{fig:BC3_HST_GLX}, \ref{fig:BC4_HST_GLX}, \ref{fig:BC5_HST_GLX}, while that for SECCO~1 can be found in \citet{Sand+2017}. In all cases, except BC2, the BCs appears to be very blue, highly irregular, and frequently broken up into multiple components. Their stellar populations also appear to be partially resolved with some individual stars discernible. These are almost exclusively blue and likely only represent the youngest, brightest stars, not the underlying stellar population. However, their extremely blue appearance (discussed further in \S\ref{sec:stellarpops} and \S\ref{sec:stellarmasses}) also suggests that any redder, underlying population is likely minimal.

The largest single component of any BC is BC3a (Figure \ref{fig:BC3_HST_GLX}, top-left), which is approximately 30\arcsec \ across its major axis (2.4~kpc at the distance of Virgo). However, what constitutes a single component is quite subjective, for example, BC4a, b, and c could justifiably be considered as a single object \citepalias{Bellazzini+2022}. The smallest components (e.g. BC3c, BC5b, BC5c) are less than 5\arcsec \ ($\sim$400~pc) across and may only consist of a single cluster of stars. 

The six components of BC4 are spread over $\sim$1.5\arcmin \ ($\sim$7~kpc) and may indicate that this is either a very young collection of objects formed in a gas-rich stream, or a somewhat older object that has become gravitationally unbound. Even for the other BCs, which are mostly defined by one or two components, their highly irregular and clumpy structure points to them being extremely low-mass and potentially unbound. Although the individual components of the BCs were identified visually, based primarily on the HST images, nearly all of these clumps have corresponding UV and (usually) H$\alpha$ emission \citepalias{Bellazzini+2022}, which indicate ongoing SF. In the case of the latter, the components are all kinematically associated (see \S\ref{sec:MUSE_results}).

In the case of BC2, the HST image (Figure \ref{fig:BC2_HST_GLX}, top-left) is quite distinct from the other BCs and indicates that this is a spurious candidate. It appears to be a distant background group of galaxies rather than a nearby young object. Unlike the other BCs there is also minimal UV emission (particularly FUV) associated with this candidate (Figure \ref{fig:BC2_HST_GLX}, top-right), and it was undetected in H$\alpha$ by MUSE (\S\ref{sec:MUSE_results}). Furthermore, almost no stars were identified in its CMD (Figure \ref{fig:BC2_HST_GLX}, bottom-left), which is consistent with background (Figure \ref{fig:BC2_HST_GLX}, bottom-right). Henceforth, we will not regard BC2 as a genuine blue stellar system and statements regarding the global properties of BCs should be assumed to include SECCO~1, but not BC2.

Figure \ref{fig:BC_locs} shows the locations of the BCs on the sky in relation to Virgo cluster galaxies and the cluster virial radius. BC2 is shown as an unfilled symbol. All of the BCs are within the virial radius of the cluster. However, none are in the very cluster center, within $\sim$2$^\circ$ ($\sim$575~kpc) of M~87 (the central galaxy in the Virgo cluster, Figure \ref{fig:BC_locs}). BC1 is the closest, with a projected separation of approximately 600~kpc. This may indicate that the parent objects of BCs are recent additions to the cluster.

\subsection{H$\alpha$ velocities and metallicities}
\label{sec:MUSE_results}

\begin{table}
\centering
\caption{Metallicities of BCs}
\begin{tabular}{ccccc}
\hline \hline
Object & $v_{\mathrm{H}\alpha}/\mathrm{km\,s^{-1}}$ & $N_{\mathrm{H}\alpha}$ & $N_\mathrm{O/H}$ & $\langle 12 + \log \mathrm{O/H} \rangle$\\
\hline
BC1    & $1117 \pm 6$ & 18 & 2 & $8.35 \pm 0.15$ \\
BC3    & $1584 \pm 4$ & 15 & 5 & $8.29 \pm 0.17$ \\
BC4    & $-60 \pm 19$ & 16 & 6 & $8.73 \pm 0.15$ \\
BC5    & $-74 \pm 5$ & 4  & 2 & $8.70 \pm 0.14$ \\ 
SECCO1$^\dagger$ & $-153.2 \pm 1.4$ & 33 & 9 & $8.38 \pm 0.11$ \\ \hline
\end{tabular}
\tablenotetext{}{H$\alpha$ redshift and metallicity measurements from \citetalias{Bellazzini+2022}. Columns: (1) object name; (2) mean velocity (and standard deviation) of H$\alpha$ clumps detected with MUSE; (3) number of clumps detected in H$\alpha$; (4) number of clumps detected in H$\alpha$, H$\beta$, [N{\sc ii}], and [O{\sc iii}] (suitable for deriving an O/H estimate); (5) mean oxygen abundance and uncertainties (standard deviation of clumps and scatter in O3N2 calibration, 0.14~dex). 
$^\dagger$Values from \citet{Beccari+2017}.}
\label{tab:metallicity}
\end{table}

MUSE detected H$\alpha$ emission in all BCs, identifying between 4 and 18 distinct clumps of emission in each object \citepalias{Bellazzini+2022}. The mean velocity (and standard deviation) of these clumps in each source is shown in Table \ref{tab:metallicity}. Only BC3 (and SECCO~1) has a prior velocity from an \hi \ detection (Table \ref{tab:BCs}), which matches closely with the H$\alpha$ velocity for that object. All the objects have velocities which are consistent with Virgo cluster membership \citep[$-500 < cz_\odot/\mathrm{km\,s^{-1}} < 3000$, e.g.][]{Mei+2007}, and all are (projected) within the virial radius of the cluster (Figure \ref{fig:BC_locs}). We note that although BC4, BC5, and SECCO~1 all have negative radial velocities, they are in the vicinity of M~86 ($cz_\odot = -224$~\kms), a region of the Virgo cluster where negative radial velocities are common.

As described in \citetalias{Bellazzini+2022}, the average oxygen abundance of each BC was estimated based on N2 and O3N2 \citep[following][]{Pettini+2004}, which were corrected for extinction based on the relative strengths of H$\alpha$ and H$\beta$. The resulting metallicity estimates are shown in Table \ref{tab:metallicity}. All the BCs are extremely high metallicity given their very low stellar masses (\S\ref{sec:stellarpops}), which suggests that they formed from gas pre-enriched in more massive objects. Of particular note are BC4 and 5, both of which are found to be marginally super-solar in metallicity \citep[$12 + \log (\mathrm{O/H})_\odot = 8.69$,][]{Asplund+2009}. The details of the kinematics and metallicity spreads of the clumps within the BCs are discussed in \citetalias{Bellazzini+2022}.

\subsection{\hi \ mass \& limits}
\label{sec:hi_mass}

\begin{table}
\centering
\caption{\hi \ masses of BCs}
\begin{tabular}{ccc}
\hline \hline
Object & $M_\mathrm{HI}/\mathrm{M_\odot}$ & Telescope\\
\hline
BC1    & $<1.6 \times 10^6$ & GBT \\
BC3    & $4.0 \times 10^7$ & Arecibo$^\dag$ \\
BC4    & $<2.9 \times 10^6$ & VLA \\
BC5    & $<3.2 \times 10^6$ & VLA \\ 
SECCO1 & $1.5 \times 10^7$ & Arecibo$^\ddag$ \\ \hline
\end{tabular}
\tablenotetext{}{Columns: (1) object name; (2) \hi \ mass or 3$\sigma$ upper limit; (3) telescope for the stated value. 
$^\dag$\citet{Haynes+2011}.
$^\ddag$\citet{Adams+2015}.}
\label{tab:HImasses}
\end{table}

SECCO~1 is the prototype BC, first detected via its \hi \ emission \citep{Adams+2013}, having a total \hi \ mass of $1.5\times10^{7}$~\Msol \ \citep{Adams+2015}. The low resolution of \hi \ observations means that the main and secondary body of SECCO~1 appear as one source in \hi. However, \citet{Adams+2015} also identified an additional \hi-only component slightly to the north as well as another potential optical component, also to the north, but not coincident with any \hi. Owing to the similar optical/UV appearance of BCs 1-5 it was anticipated that they would also be \hi-rich, which motivated our VLA follow-up program.

In Figures \ref{fig:BC1_VLA_GBT_specs}, \ref{fig:BC3_VLA_specs}, and \ref{fig:BC45_VLA_specs} we present the VLA (and GBT) \hi \ spectra of BCs 1, 3, 4, and 5 (the spectrum of BC2 is discussed in Appendix \ref{sec:BC2spec}). The VLA \hi \ spectra were extracted from the data cubes using an aperture equal to the synethesized beam size, centered on the location of the main body of each BC (Table \ref{tab:BCs}). In addition to these spectra the data cubes were visually inspected channel by channel and \texttt{SoFiA} was run to search for significant emission features that might be extended spatially or spectrally.

Like SECCO~1, BC3 was known a priori to contain a significant \hi \ reservoir as it was originally identified in the ALFALFA survey \citep{Haynes+2011}. However, among BCs 1-5 this is the only object that was detected in our VLA observations. Based on the VLA spectrum (Figure \ref{fig:BC3_VLA_specs}, extracted using the \texttt{SoFiA} source mask), and an assumed distance of 16.5~Mpc, BC3 has an \hi \ mass of $\log M_\mathrm{HI}/\mathrm{M_\odot} = 7.3$. This value is 0.3~dex lower than that measured by ALFALFA \citep{Haynes+2011} suggesting that the VLA has not recovered all the extended flux \citep[this was also noted by][]{Cannon+2015,Jones+2022}. \citet{Jones+2022} show that when viewed in the ALFALFA data cube (which has better column density sensitivity for extended emission than the VLA observations) the \hi \ emission coincident with BC3 appears to connect to the galaxy VCC~2034, approximately 70~kpc to the SW. This galaxy is almost certainly the source of the gas that formed BC3 \citep[discussed further in \S\ref{sec:discuss}, and][]{Jones+2022}.

If the other BCs had comparable \hi \ masses to BC3 and SECCO~1, then they would be detected with the VLA observations, but none were (Figures \ref{fig:BC1_VLA_GBT_specs} \& \ref{fig:BC45_VLA_specs}). The slight caveat is that, because of their low radial velocities, BC4 and BC5 might be blended with MW \hi \ emission. The spectrum of BC5 (Figure \ref{fig:BC45_VLA_specs}, bottom) also appears to have a peak coincident with the H$\alpha$ velocity of BC5. However, this peak is below 3$\sigma$ and extremely narrow, and is likely a noise spike. 

All the BCs that are undetected in \hi \ have optical redshift measurements from MUSE H$\alpha$ observations (BC3 and SECCO~1 do also) and the available \hi \ data can therefore be used with confidence to set upper limits on their \hi \ masses. For BC4 and BC5 the deepest data are those from the VLA, which have rms noise values of 0.9 and 1.0~mJy/beam (at 8.8~\kms \ resolution), respectively, at the velocities of the H$\alpha$ emission. Assuming that any \hi \ emission would fit within one synthesized beam (Table \ref{tab:VLAobs}) and would have a velocity width of 30 \kms, then these equate to 3$\sigma$ upper limits of $\log M_\mathrm{HI}/\mathrm{M_\odot} < 6.46$ and 6.51, respectively, assuming a fiducial distance of 16.5~Mpc in both cases. For BC1 the GBT follow up spectrum is by far the more sensitive. With an rms of 0.28~mJy (at 30~\kms \ resolution) this gives the 3$\sigma$ upper limit as $\log M_\mathrm{HI}/\mathrm{M_\odot} < 6.2$, again assuming a fiducial distance of 16.5~Mpc. These limits are listed in Table \ref{tab:HImasses}.

\subsection{Stellar populations}
\label{sec:stellarpops}

\begin{table}
\centering
\caption{Magnitudes and stellar mass estimates}
\begin{tabular}{cccc}
\hline \hline
Object & F814W & F606W-F814W & $M_\ast/\mathrm{M_\odot}$ \\ \hline
BC1 & $20.29 \pm 0.38$ & $0.08 \pm 0.41$ & $\sim5 \times 10^{4}$ \\ 
BC3 & $20.23 \pm 0.15$ & $-0.23 \pm 0.17$ & $\sim5 \times 10^{4}$ \\
BC4 & $19.86 \pm 0.26$ & $-0.26 \pm 0.29$ & $\sim1 \times 10^{5}$ \\
BC5 & $20.56 \pm 0.10$ & $0.06 \pm 0.12$ & $\sim5 \times 10^{4}$ \\
SECCO1 & $20.39 \pm 0.41$ & $-0.23 \pm 0.46$ & $\sim4 \times 10^{4}$ \\ \hline
\end{tabular}
\tablenotetext{}{Columns: (1) object name; (2) F814W magnitude (extinction corrected); (3) F606W-F814W color (extinction corrected); (4) stellar mass estimate (\S\ref{sec:stellarpops}).}
\label{tab:Mstar}
\end{table}

\begin{figure*}
    \centering
    \includegraphics[width=0.49\columnwidth]{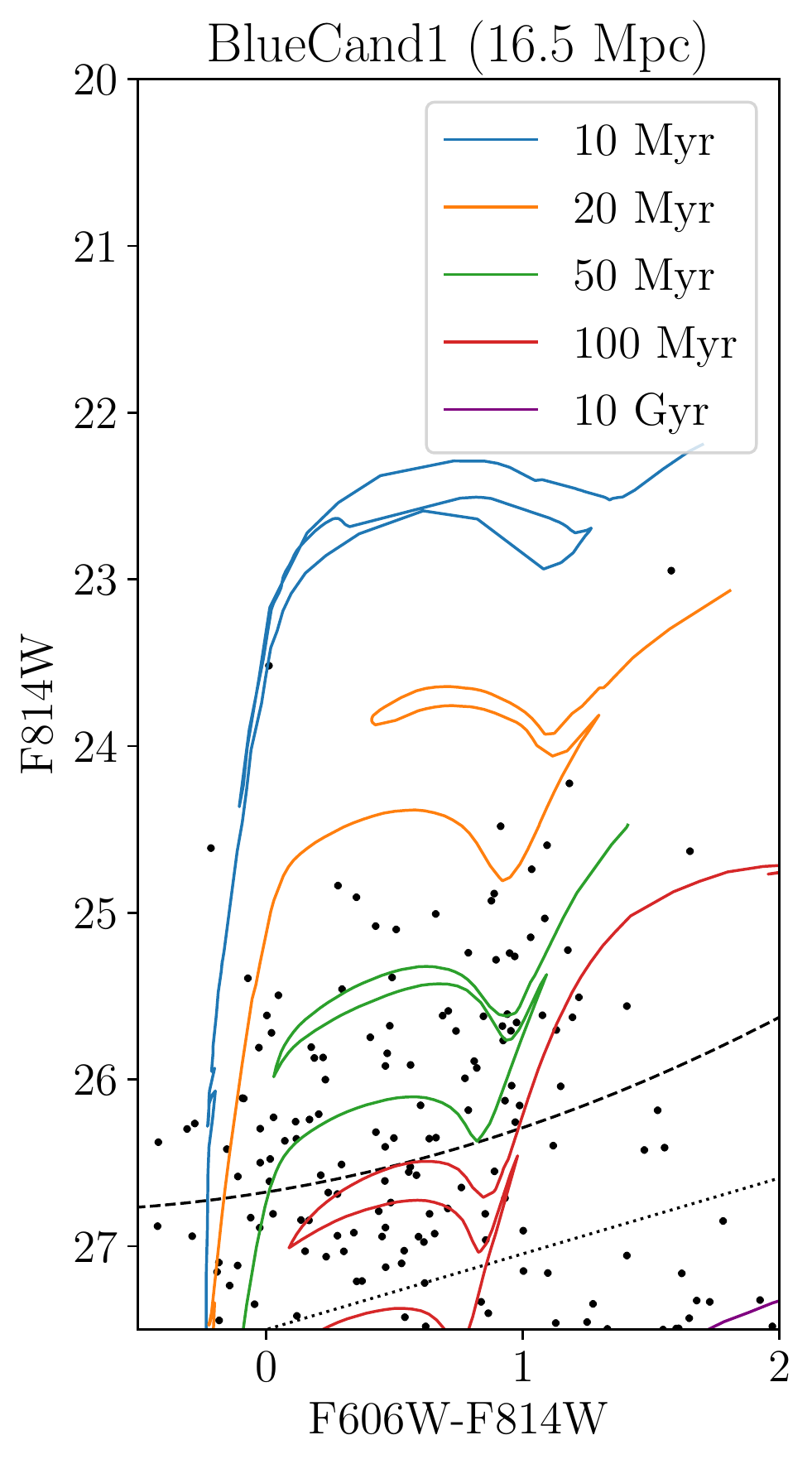}
    \includegraphics[width=0.49\columnwidth]{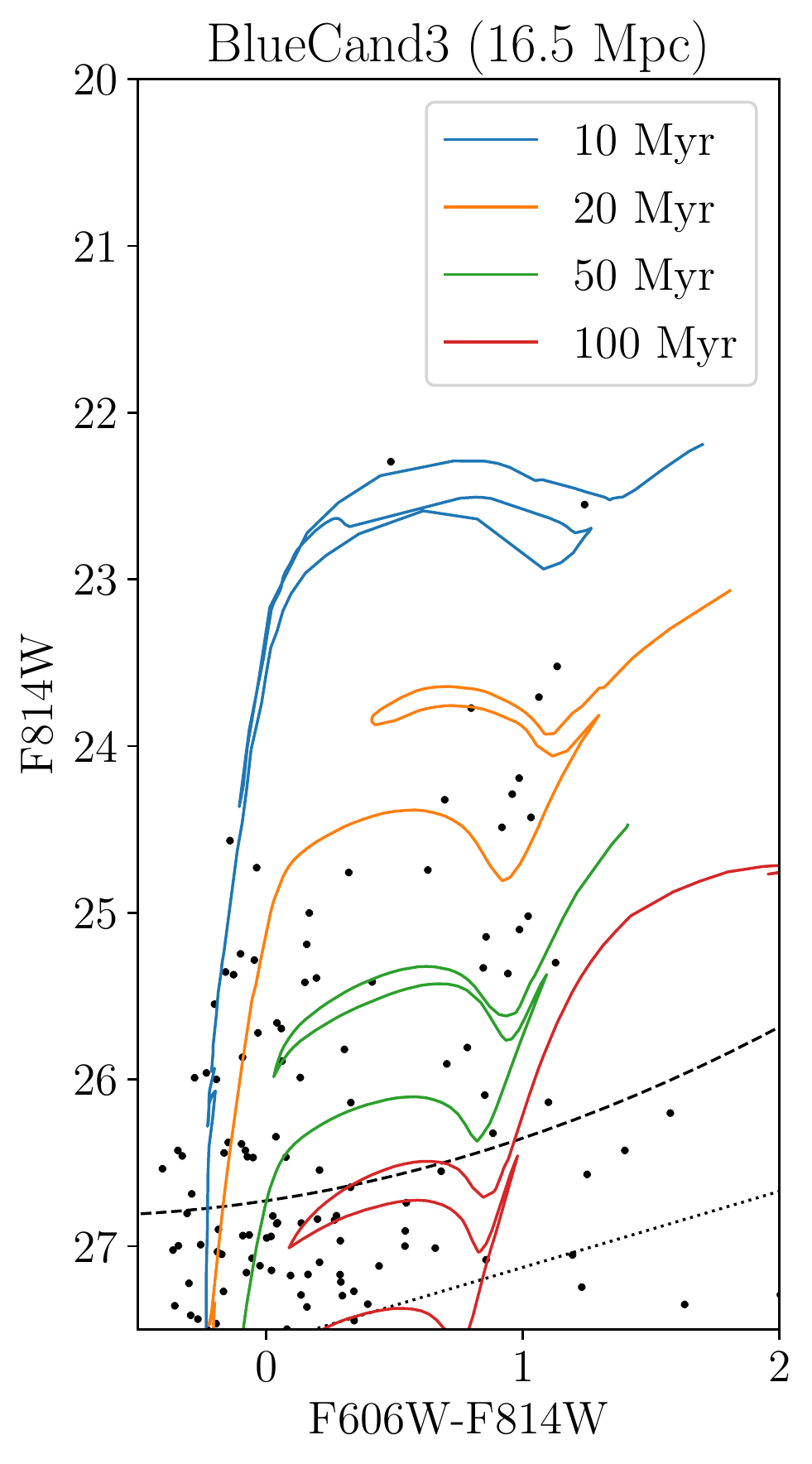}
    \includegraphics[width=0.49\columnwidth]{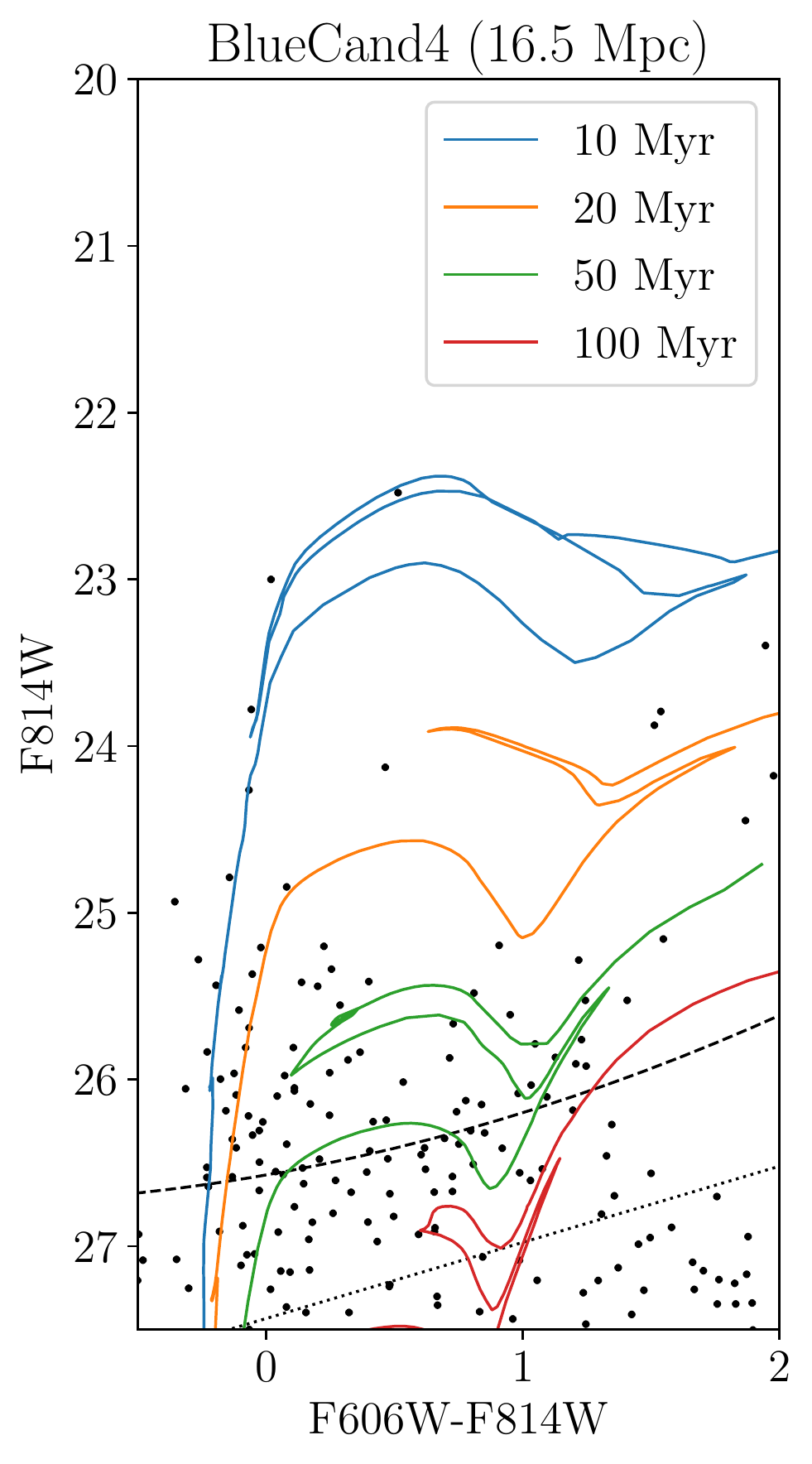}
    \includegraphics[width=0.49\columnwidth]{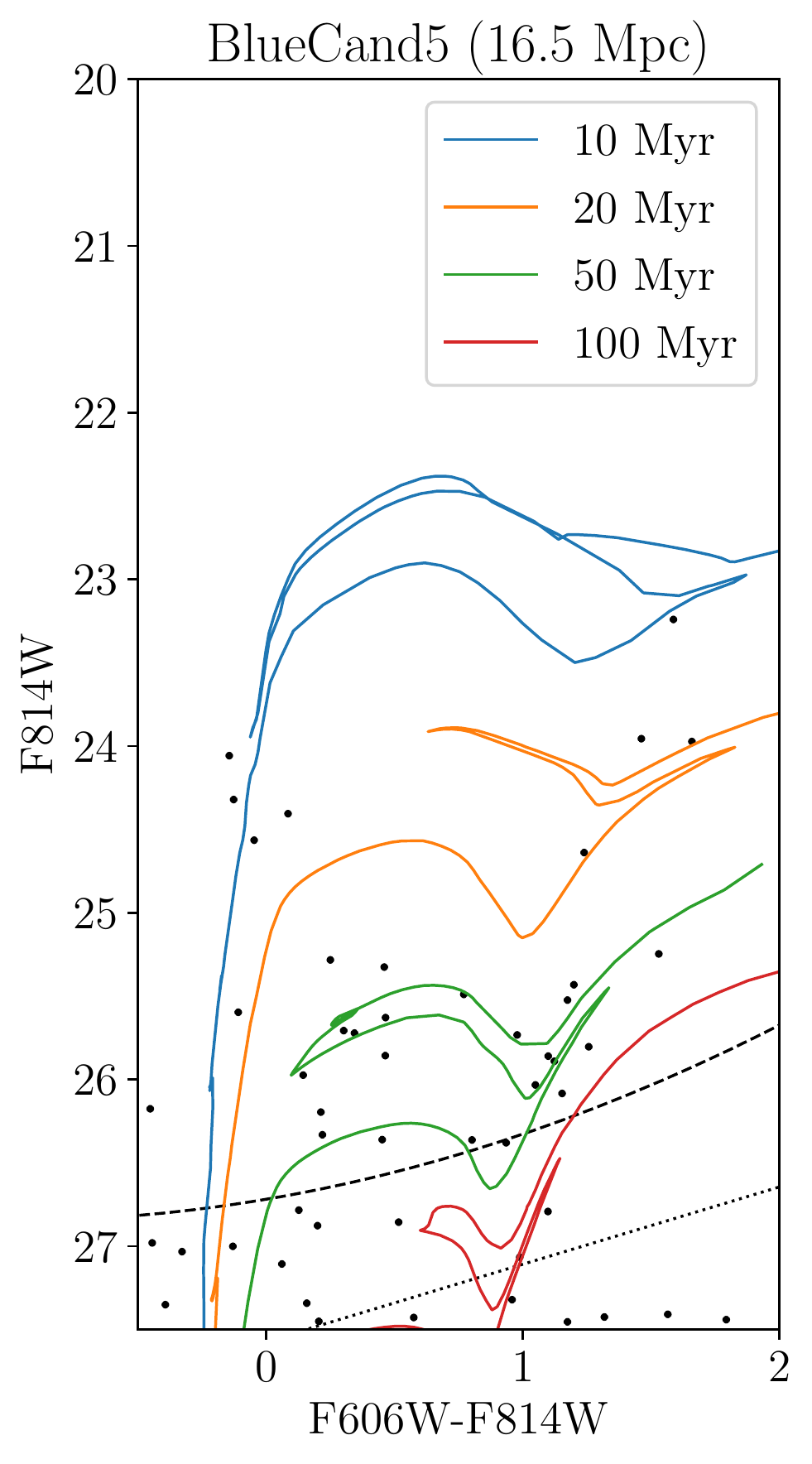}
    \caption{Reproduced CMDs of BCs 1, 3, 4, and 5, with \texttt{PARSEC} isochrones for different stellar population ages overlaid, assuming a distance of 16.5~Mpc to all objects. The isochrones for BC1 and 3 use a metallicity of $[M/H] = -0.35$, which is approximately the value for both objects. For BC4 and 5 the value is $[M/H] = 0.05$. In the latter case it is possible that the two objects are associated, while the similar metallicities of BC1 and 3 are likely by chance. In the leftmost panel we also plot an isochrone indicating where an old (10~Gyr) RGB population would reside in these diagrams, below the completeness limit in the lower right corner and barely visible.}
    \label{fig:isochrones}
\end{figure*}

The HST (and GALEX) images and associated CMDs for all BCs are shown in Figures \ref{fig:BC1_HST_GLX}, \ref{fig:BC2_HST_GLX}, \ref{fig:BC3_HST_GLX}, \ref{fig:BC4_HST_GLX}, and \ref{fig:BC5_HST_GLX}. The blue optical colors and UV emission indicate that the BCs have predominantly young, blue stellar populations. Furthermore, the detection of H$\alpha$ in all BCs indicates that the youngest stars must be $\leq$10~Myr old.

As discussed by \citet{Jones+2022}, the CMD of BC3 (Figure \ref{fig:BC3_HST_GLX}, bottom) is most similar to that of SECCO~1 \citep{Sand+2017}, apparently made up almost entirely of blue main sequence (MS) and helium burning stars (F814W~$\gtrsim$~24.5, F606W-F814W~$\lesssim$~0) and red helium burning (RHeB) stars (23.5~$\lesssim$~F814W~$\lesssim$~26.5 and F606W-F814W~$\gtrsim$~0.6~mag), with almost no candidates for red giant branch (RGB) stars, highlighting the young age of the population. 

BC1's CMD (Figure \ref{fig:BC1_HST_GLX}, bottom) is similar, but the brightest RHeB stars are more numerous and fainter than in BC3. These slight differences likely indicate that BC1 is somewhat older than BC3 \citep[as RHeB peak brightness is a function of age, e.g.][]{McQuinn+2011}, which would be consistent with its non-detection in \hi, if sufficient time has passed for its neutral gas to have been evaporated or stripped.

The CMDs of BC4 and BC5 (Figures \ref{fig:BC4_HST_GLX} \& \ref{fig:BC5_HST_GLX}, bottom panels) are again similar, but the RHeB stars are even fainter and continue to the completeness limit. This likely indicates that BC4 and BC5 are the oldest objects in the sample. The color spread between the blue and RHeB stars is also wider for BC4 and BC5 than for any of the other BCs.

In Figure \ref{fig:isochrones} we overplot \texttt{PARSEC} isochrones \citep[PAdova and TRieste Stellar Evolution Code,][]{Bressan+2012} on the CMDs of each object for a variety of stellar population ages. As pointed out by \citet{Jones+2022} the faintest RHeB stars in BC3 appear consistent with the 50~Myr isochrone, likely indicating that the stellar population in this object cannot be much older than 50~Myr. In the case of the other BCs, as mentioned above, their CMDs imply that their oldest stars are somewhat older (although they must still have formed young stars within the past 10~Myr, as they contain \hii \ regions), but the proximity of the RHeB stars to the completeness limit prevents them from being used to estimate ages. The isochrones also explain the different color gap between the bluest and reddest stars in the CMDs of BC4 and BC5 versus BC1 and BC3. This spread is approximately reproduced in the isochrones and is a function of the higher metallicity of these two objects, which despite their feeble appearance is marginally super-solar (Table \ref{tab:metallicity}). 

The CMD of SECCO~1 was presented and discussed in \citet{Sand+2017} and \citet{Bellazzini+2018}. The general appearance is similar to the other BCs. The spread between the reddest and bluest stars is most similar to BC1 and BC3, again a reflection of the similar metallicities of these objects. \citet{Sand+2017} also simulate a mock stellar population and argue that SECCO~1 must be younger than $\sim$50~Myr based on the luminosity of the RHeB stars. This is roughly consistent with our estimate of 60~Myr in \S\ref{sec:stellarmasses} based on the integrated F814W magnitude and SFR of SECCO~1.

If we compare the CMDs of the BCs to the low-mass, gas-rich dwarf Leo~P \citep{McQuinn+2015b}, then we see a striking difference. In addition to the young blue stars in Leo~P there is also a clear, well-populated RGB at a similar magnitude that is entirely absent from the BCs CMDs. This clear RGB is the result of the old underlying population in Leo~P, but for extremely young stellar populations (which BCs appear to be) no RGB population exists. Furthermore, any RGB stars would be significantly less luminous than the young stars that dominate the CMDs of the BCs. However, the proximity of Leo~P ($D=1.6$~Mpc) means that the depth of its CMD is a mis-match for those of BCs, making it an unfair comparison, despite it being one of the most similar objects known in terms of SFR and stellar mass (but notably not metallicity).

A fairer comparison can be made by considering a blue, irregular dwarf at the distance of Virgo, in this case VCC~1816 (KDG~177, $M_V = -15.2$), which has similar depth HST observations as the BCs. The CMD of this galaxy \citep[Figure 4 of][]{Karachentsev+2014} shows both a blue population (at F606W-F814W$\sim$0) and a red population (at F606W-F814W$\sim$1), similar to BCs. The former is likely made up of blue helium burning stars and young MS stars, as in the BCs, while the latter is likely made up of a combination of asymptotic giant branch and RHeB stars. The number of stars in the CMD increases towards fainter F814W magnitudes (near F606W-F814W$\sim$1) probably indicating the presence of a well-populated RGB near the completeness limit, which is lacking in the BC CMDs. A similar lack of evidence for any RGB was noted for SECCO~1 by \citet{Sand+2017} and \citet{Bellazzini+2018}, but in comparison to a red dwarf spheroidal in Virgo, rather than a star-forming dwarf more in line with the appearance of BCs.

At the distance of the Virgo cluster the tip of the red giant branch (TRGB) is expected to be at F814W~$\sim$~27~mag \citep[e.g.][]{Jiang+2019}, which would be borderline detectable with our HST observations. However, at high metallicities the TRGB becomes less defined and RGB stars become redder, both of which would impede the detectability of an RGB in our observations (Figure \ref{fig:isochrones}, leftmost panel). Thus, it is not possible to conclusively rule out there being an RGB based on the CMDs of BCs. Despite this we still view the existence of an underlying old population as extremely unlikely in these objects. They are extremely blue, to the point where stellar population models struggle to reproduce their colors, even when assuming very young ages (\S\ref{sec:stellarmasses}). In addition, BCs were specifically selected (\S\ref{sec:sample}) to be lacking any visible diffuse red component in the deep NGVS images. Together these points make it highly unlikely that there could be any significant underlying old population of stars, even though the CMDs themselves are insufficiently deep to reach any potential RGB.

Overall the CMDs of the BCs can be characterized as having a population of stars made up exclusively of young blue main sequence and (blue and red) helium burning stars, with no evidence of a RGB. The luminosities of the RHeB stars suggest that the youngest BCs (BC3 \& SECCO~1) are around 50~Myr old. Finally, the remarkably high metallicities measured with MUSE (\S\ref{sec:MUSE_results}) appear to be consistent with the color difference between the reddest and bluest helium burning branch stars in the CMDs.

\subsection{Star formation rates}
\label{sec:SFRs}

SFRs (Table \ref{tab:SFRs}) were estimated for each candidate by measuring the NUV and FUV fluxes within the same apertures used to produce their CMDs (\S \ref{sec:stellarpops}), as described in \S\ref{sec:galex_data}. An additional uncertainty of 15\% was added to the error budget as this is the stated accuracy of the conversion in \citet{Iglesias-Paramo+2006}. 

The SFRs of all BCs fall in the range $-3.5 < \log \mathrm{SFR/M_\odot \, yr^{-1}} < -3$ and are generally quite consistent between NUV and FUV (where both images are available), likely indicating that their SFRs have not varied strongly over the past $\sim$100~Myr (or that they are younger than this). Although matching NUV and FUV SFR estimates could be the result of a bursty SF histories over the past $\sim$100~Myr, with an average rate that equals that of the past $\sim$10~Myr, it seems highly unlikely that this could be the case for all BCs, and a constant SFR is a more natural explanation for this finding. The UV-based SFRs are also roughly consistent with the SFRs estimated from the integrated H$\alpha$ fluxes \citepalias{Bellazzini+2022}, with the slight exception of BC1 (for which the SFR may be beginning to decline), again supporting the assertion that the SFRs appear to have been relatively constant in the recent past. We note that had we adopted a different conversion scheme for our UV-based SFR estimates \citep[e.g.][]{McQuinn+2015a} then our SFR$_\mathrm{FUV}$ values could be up to 0.6~dex higher. However, given the general consistency between the H$\alpha$ and UV-based SFR estimates, the conversion scheme we originally selected appears appropriate for these objects.

This range of SFRs is similar to the faintest dwarf irregular galaxies in the Local Volume \citep{Lee+2009}. However, the extremely low stellar masses of BCs (\S\ref{sec:stellarmasses}) make it difficult to directly compare to equivalent star-forming dwarf galaxies, as almost none are known at these masses. For example, even Leo~P \citep{Giovanelli+2013} is almost an order of magnitude higher stellar mass than most BCs, but its SFR is around an order of magnitude lower \citep[$\log \mathrm{SFR/M_\odot \, yr^{-1}} = -4.4$][]{McQuinn+2015b}. Leo~T \citep{Irwin+2007} is of comparable stellar mass to BCs \citep[$M_\ast = 1.4\times10^5$~\Msol,][]{Weisz+2014}, but is apparently no longer forming stars, or is between episodes \citep{Kennicutt+2008}. If we put the SFRs of BCs in terms of their specific SFRs (sSFR) then they fall in the range $-8.2 < \log (\mathrm{SFR}/M_\ast)/\mathrm{yr^{-1}} < -7.7$, which would place them significantly higher than average, but within the scatter, of sSFR for low-mass, gas-rich, field galaxies \citep{Huang+2012,James+2015}.

\begin{table*}
\centering
\caption{UV fluxes and SFR estimates}
\begin{tabular}{cccccccccc}
\hline \hline
Object &  & $\mathrm{SNR}_\mathrm{NUV}$ & NUV flux                          & $\log \frac{\mathrm{SFR_{NUV}}}{\mathrm{M_\odot\,yr^{-1}}}$          & $\mathrm{SNR}_\mathrm{FUV}$ & FUV flux                          & $\log \frac{\mathrm{SFR_{FUV}}}{\mathrm{M_\odot\,yr^{-1}}}$ & $\log \frac{\mathrm{SFR_{H\alpha}}}{\mathrm{M_\odot\,yr^{-1}}}$ & $\log \frac{M_\mathrm{HI}/\mathrm{SFR}}{\mathrm{yr}}$ \\
\hline
BC1    &            & 10.1    & $9.10 \pm 0.90$ & $-3.25 \pm 0.08$ & 13.7    & $1.98 \pm 0.14$ & $-3.42 \pm 0.07$ & $-$3.9 & $<9.5$ \\
BC3    &            &         & $17.4 \pm 0.7$  & $-3.03 \pm 0.07$ &         & $4.02 \pm 0.09$ & $-3.18 \pm 0.07$ & $-$3.1 & 10.3   \\
       & a          & 23.8    & $14.1 \pm 0.6$  & $-3.13 \pm 0.07$ & 43.9    & $3.36 \pm 0.08$ & $-3.26 \pm 0.07$ &  & \\
       & b          & 9.3     & $2.78 \pm 0.30$ & $-3.83 \pm 0.08$ & 13.6    & $0.57 \pm 0.04$ & $-4.03 \pm 0.07$ &  & \\
       & c          & 4.6     & $0.57 \pm 0.12$ & $-4.52 \pm 0.11$ & 5.6     & $0.10 \pm 0.02$ & $-4.77 \pm 0.10$ & &  \\
BC4    &            &         & $13.1 \pm 0.7$  & $-3.10 \pm 0.07$ &         &                 &                  & $-$3.2 & $<9.6$  \\
       & a          & 16.5    & $4.07 \pm 0.25$ & $-3.60 \pm 0.07$ &         &                 &                  & &  \\
       & b          & 4.7     & $1.04 \pm 0.22$ & $-4.20 \pm 0.11$ &         &                 &                  & &  \\
       & c          & 11.8    & $2.37 \pm 0.20$ & $-3.84 \pm 0.08$ &         &                 &                  & &  \\
       & d          & 6.1     & $1.49 \pm 0.24$ & $-4.04 \pm 0.10$ &         &                 &                  & &  \\
       & e          & 7.4     & $3.56 \pm 0.48$ & $-3.66 \pm 0.09$ &         &                 &                  & &  \\
       & f          & 4.1     & $0.55 \pm 0.13$ & $-4.47 \pm 0.13$ &         &                 &                  & &  \\
BC5    &            &         & $6.13 \pm 0.31$ & $-3.48 \pm 0.07$ &         & $1.20 \pm 0.05$ & $-3.69 \pm 0.07$ & $-$3.8 &  $<10.0$  \\
       & a          & 20.1    & $5.57 \pm 0.28$ & $-3.52 \pm 0.07$ & 27.4    & $1.13 \pm 0.04$ & $-3.72 \pm 0.07$ &  & \\
       & b          & 2.8     & $0.38 \pm 0.14$ & $-4.68 \pm 0.16$ & 2.8     & $0.06 \pm 0.02$ & $-4.96 \pm 0.16$ & &  \\
       & c          & 3.7     & $0.18 \pm 0.05$ & $-4.99 \pm 0.12$ & 1.3     & $0.01 \pm 0.01$ & $-5.67 \pm 0.33$ & &  \\
SECCO1 &            &         & $10.4 \pm 0.8$  & $-3.14 \pm 0.07$ &         &                 &                              & $-$3.2$^\dagger$ & $10.3$  \\
       & MB         & 11.1    & $6.72 \pm 0.06$ & $-3.33 \pm 0.08$ &         &                 &                       & &  \\
       & SB         & 7.6     & $3.63 \pm 0.48$ & $-3.60 \pm 0.09$ &         &                 &                       & &  \\
       \hline
\end{tabular}
\tablenotetext{}{Columns: (1) object name and sub-component (where relevant); (2) SNR of NUV emission (see Section \ref{sec:SFRs} for details); (3) NUV flux in units of $10^{-17} \; \mathrm{erg\,s^{-1}\,cm^{-2}\,\AA^{-1}}$; (4) NUV-based SFR estimate; (5) SNR of FUV emission; (6) FUV flux in units of $10^{-16} \; \mathrm{erg\,s^{-1}\,cm^{-2}\,\AA^{-1}}$; (7) FUV-based SFR estimate; (8) H$\alpha$ SFR estimates from the integrated H$\alpha$ flux of each object in MUSE \citepalias{Bellazzini+2022} following the conversion of \citet{Kennicutt1998}; (9) gas consumption timescale using the larger of the NUV and FUV SFR estimates (we note that this quantity is distance independent). For uniformity, all objects are assumed to be at 16.5~Mpc. 
$^\dagger$H$\alpha$-based SFR estimate from \citet{Beccari+2017,Beccari+2017err}.}
\label{tab:SFRs}
\end{table*}

\subsection{Stellar masses}
\label{sec:stellarmasses}

The integrated F606W and F814W magnitudes of the BCs were measured from the co-added images in each filter. The same apertures indicated in Figures \ref{fig:BC1_HST_GLX}, \ref{fig:BC3_HST_GLX}, \ref{fig:BC4_HST_GLX}, and \ref{fig:BC5_HST_GLX} were used to measure the total magnitude of each source after masking the few clear background galaxies contained within these apertures. Galactic extinction corrections were made using the dust maps of \citet{Schlegel+1998} and the reddening $R_\nu$ values of \citet{Schlafly+2011}. The final magnitudes are listed in Table \ref{tab:Mstar}. The uncertainties were estimated by placing 10 apertures across the full ACS FoV (avoiding bright stars and background galaxies) and using the standard deviation of the counts to approximate the uncertainty in the counts of each BC.

In young stellar populations the emitted light is dominated by the youngest stars, but the mass is generally dominated by the oldest, most numerous stars.
As BCs are apparently such young objects, the correct mass-to-light ratio to use is highly uncertain, and would depend strongly on the assumed age of each object.
Thus, widely used mass-to-light ratio prescriptions \citep[e.g.][]{Zibetti+2009,Taylor+2011} cannot be used with confidence for such a young, irregular, low-mass, and metal-rich stellar population. We therefore adopt an unconventional strategy for estimating the stellar masses of the BCs. If the current SFRs are assumed to be reasonable representations of the SFRs over the (short) lifetimes of the BCs then the total stellar mass is simply the age of each object times its SFR. In order to estimate the age we build up the integrated F814W magnitude of a stellar population forming stars at a constant rate (in 10~Myr steps), based on the \texttt{PARSEC} \citep{Bressan+2012} population models. When the artificial F814W magnitude equals the measured magnitude, we obtain an age estimate for the BC in question (to the nearest 10~Myr). 

To estimate the ages, and subsequently the stellar masses (age $\times$ SFR), we used the NUV SFR measurements (Table \ref{tab:SFRs}) for each object, as these are available for all objects and reflect a slightly longer SF timescale. For BC1, BC3, and SECCO1 a metallicity of $[M/H] = -0.35$ was used, and $[M/H] = 0.05$ for BC4 and BC5. These values approximately correspond to their observed O/H values (Table \ref{tab:metallicity}). The age estimates\footnote{We note that these age estimates should be treated with caution. Ideally the full SF histories of the BCs would be calculated, but the currently existing data are inadequate to do this. These ages represent an approximation to the age of the oldest stellar population in each BC, based on the assumption of a roughly constant SFR.} for BCs 1, 3, 4, 5, and SECCO~1 are 90, 50, 110, 160, and 60~Myr, and the resulting stellar mass estimates are shown in Table \ref{tab:Mstar}.

A significant caveat to this approach is that the \texttt{PARSEC} models are incapable of correctly reproducing the colors of the BCs \citep[as noted by][]{Sand+2017}. Although this issue is not fully addressed in this work, we chose to rely on the F814W magnitudes as the discrepancy is assumed to be most severe for the youngest, bluest stars. Hence the redder band is expected to be somewhat less impacted. We also note that there are encouraging trends in the values that we obtained, that at least indicate internal consistency. For example, BC3 and SECCO~1 are the only BCs detected in \hi \ and we estimate these are by far the youngest objects --- a finding that the CMDs of the BCs would also seem to support. The estimated ages of BC4 and BC5 are also the oldest and it seems plausible (\S\ref{sec:discuss_BC4}) that the two formed from the same origin. In addition, \citet{Junais+2021} independently estimated the stellar mass of BC3 by fitting the spectral energy distribution (from photometry in $ugriz$, H$\alpha$, NUV, and FUV) of each clump with a single stellar population, via a grid search over metallciity and population age, and found a near identical value ($\sim 5 \times 10^4$~\Msol).

\section{Points of origin}
\label{sec:discuss}

\begin{figure}
    \centering
    \includegraphics[width=\columnwidth]{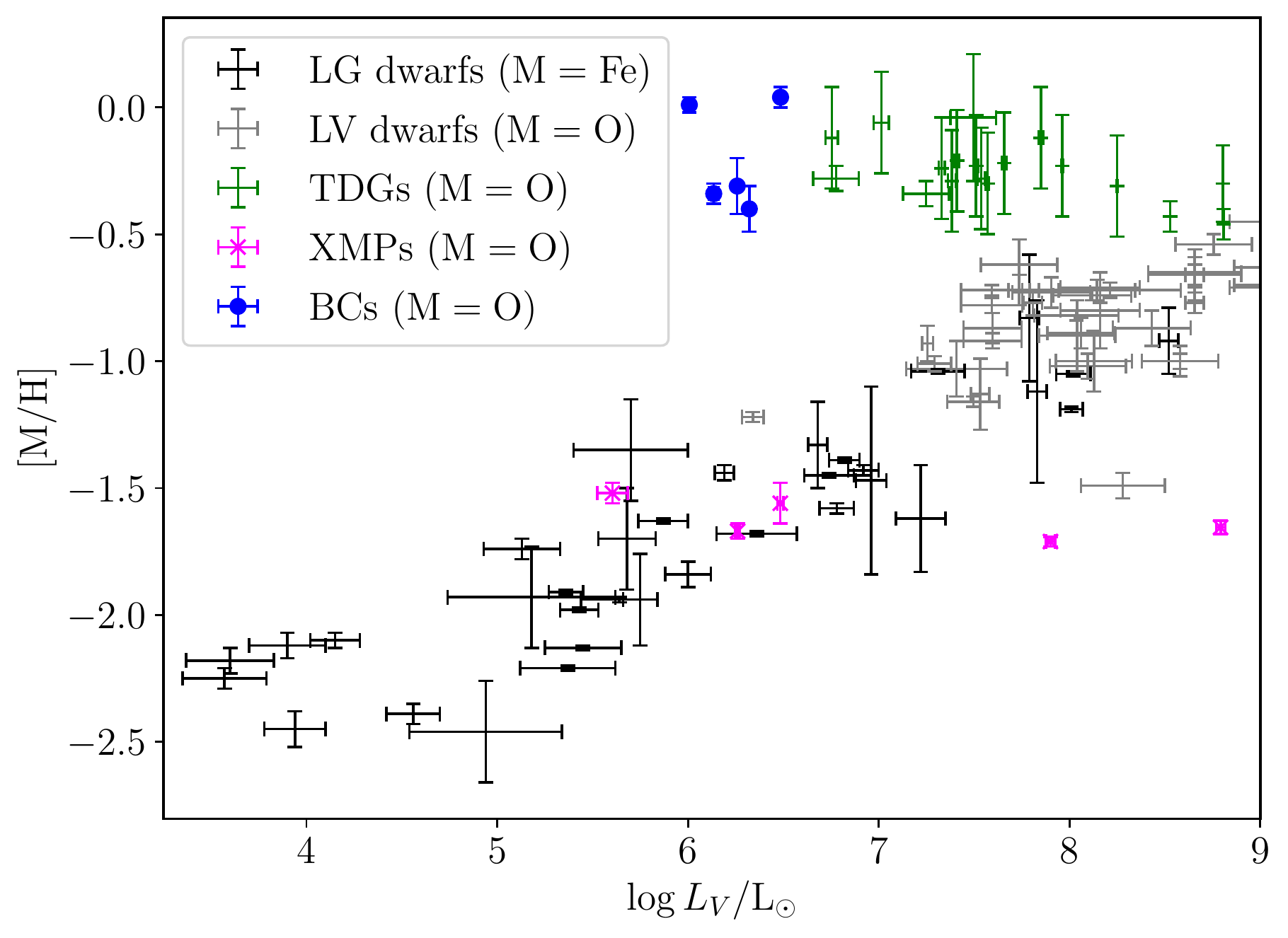}
    \caption{V-band luminosity versus metallicity (relative to solar) for BCs, Local Group dwarfs \citep{Kirby+2013}, Local Volume dwarfs \citep{Berg+2012}, TDGs \citep{Duc+1998,Weilbacher+2003,Duc+2007,Croxall+2009,Lee-Waddell+2018}, and extremely metal-poor galaxies \citep[XMPs,][]{Skillman+2013,McQuinn+2015b,Hirschauer+2016,Hsyu+2017,Izotov+2019,McQuinn+2020}. Metallicity is measured either from Fe/H or O/H as indicated in the legend. Both BCs and TDGs sit well above the luminosity--metallicity relation for dwarf galaxies of equivalent luminosities.}
    \label{fig:LV_Z}
\end{figure}

\begin{figure*}
    \centering
    \includegraphics[width=\columnwidth]{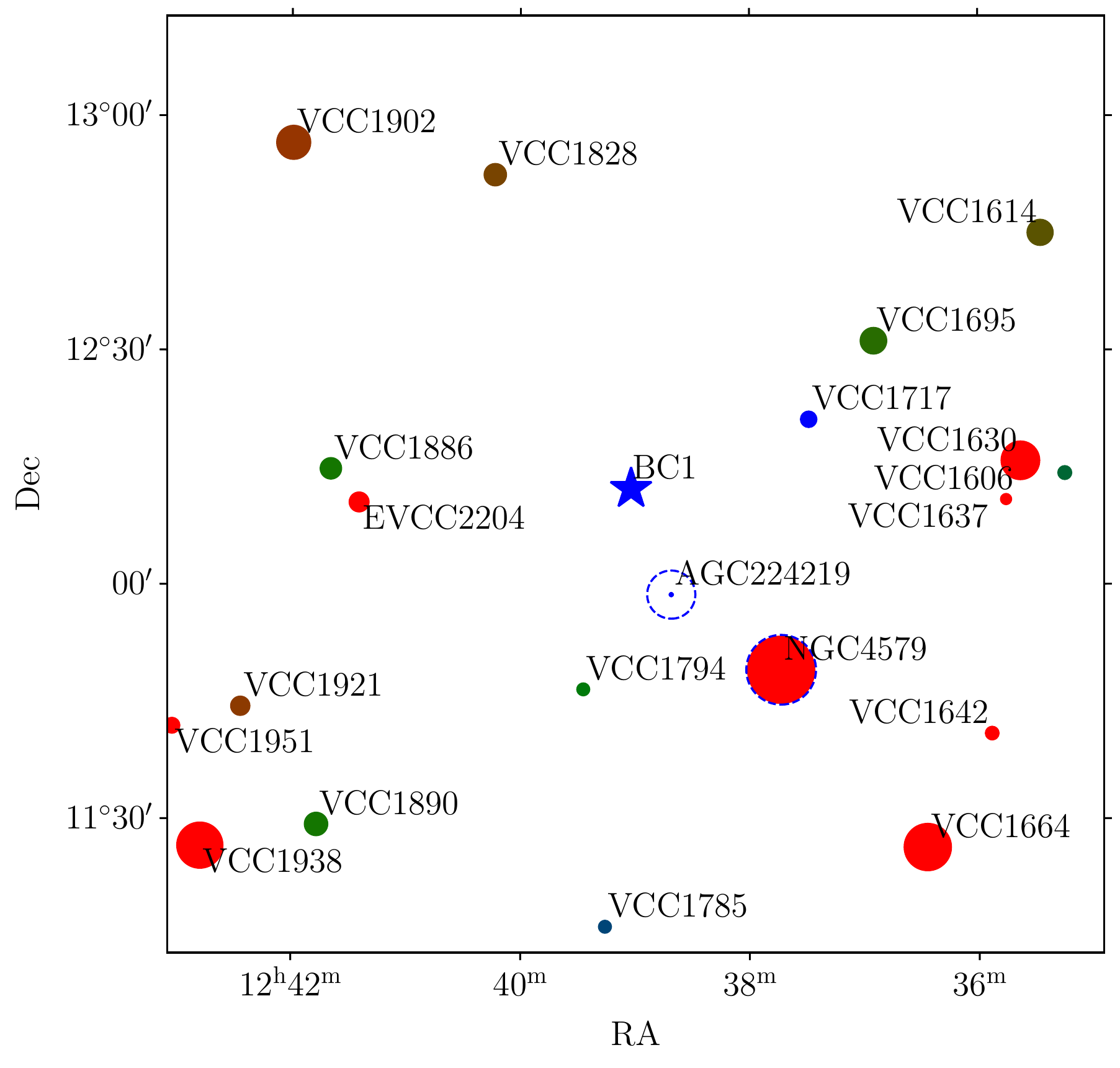}
    \includegraphics[width=\columnwidth]{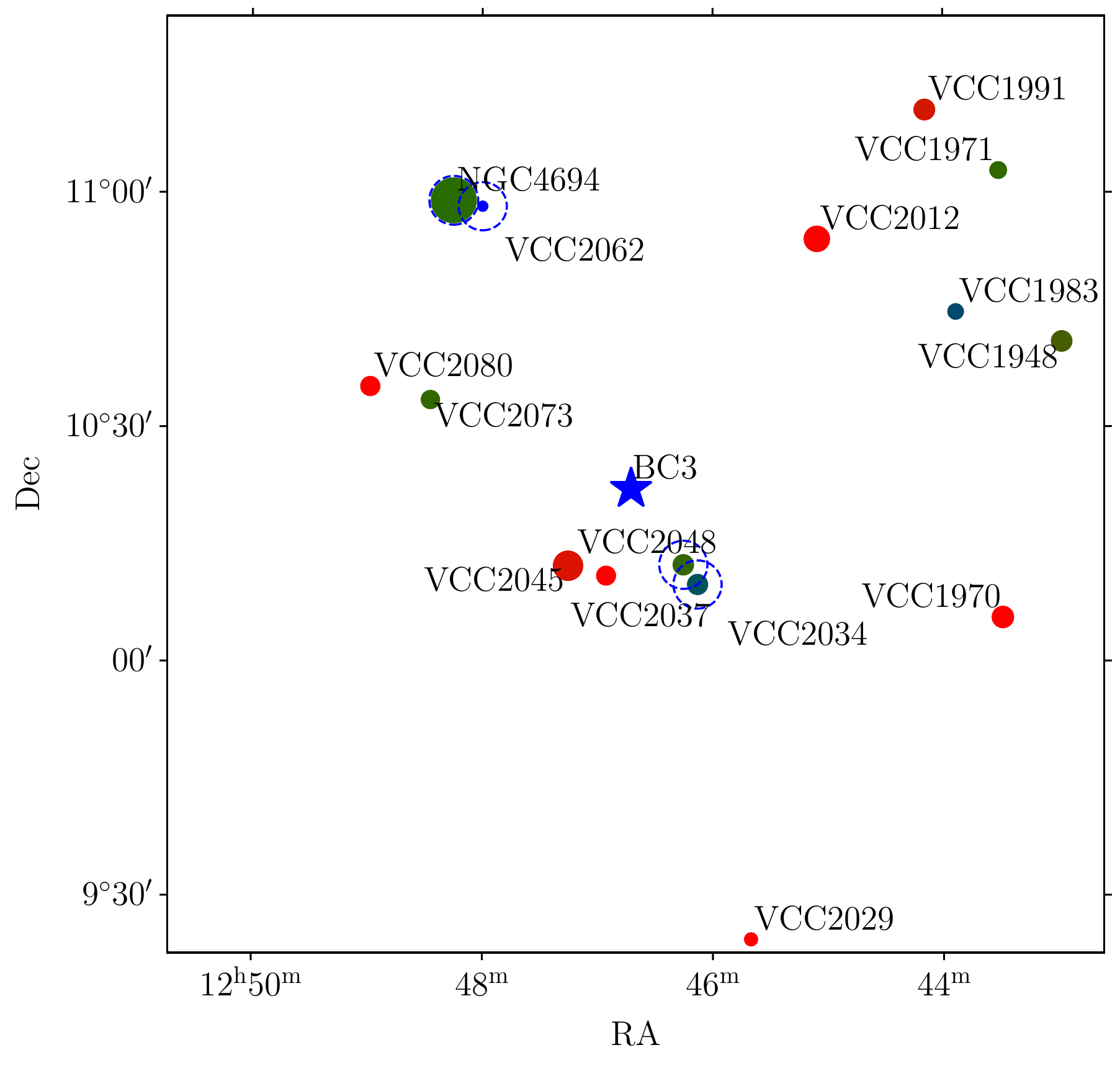}
    \includegraphics[width=\columnwidth]{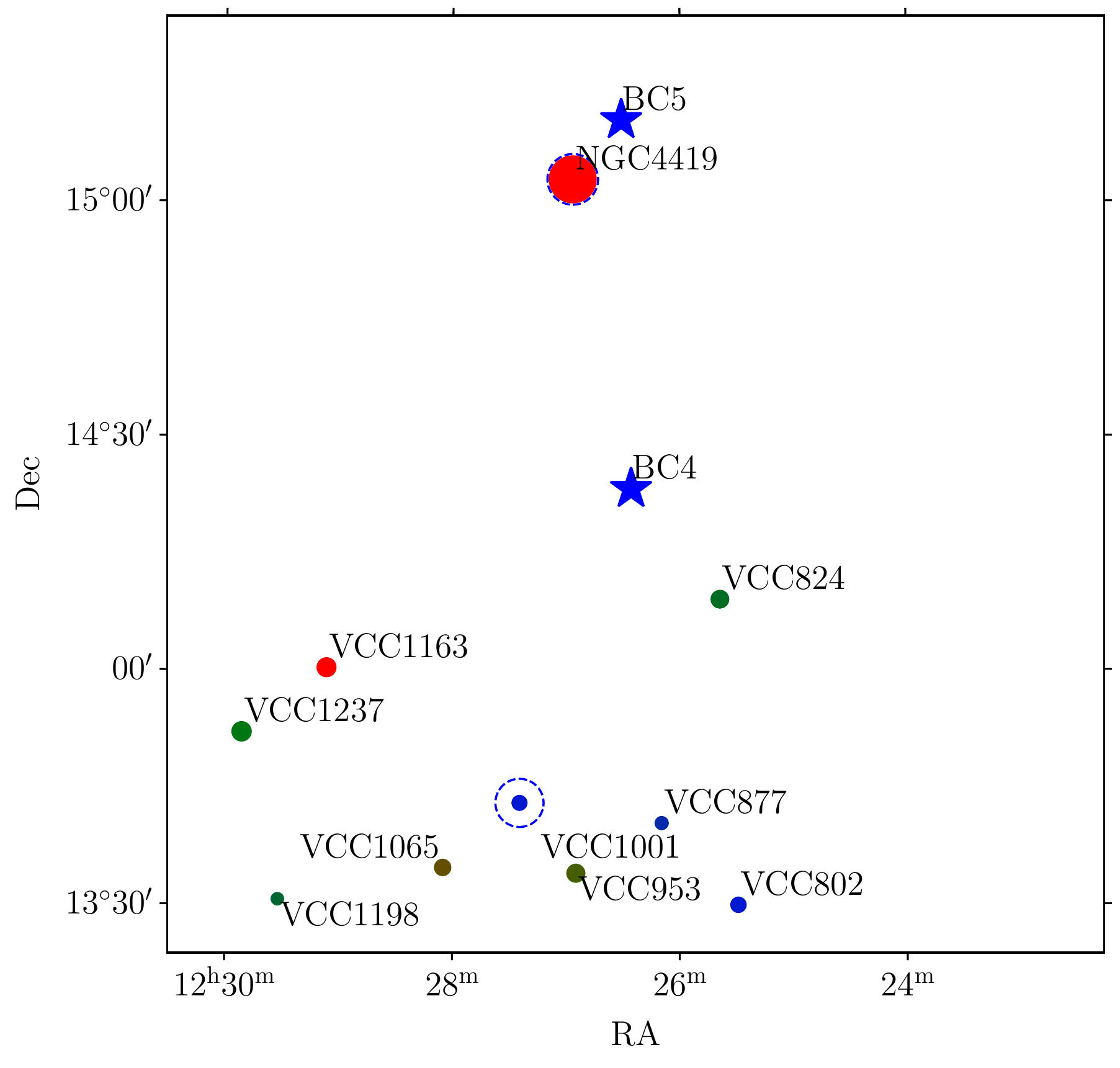}
    \includegraphics[width=\columnwidth]{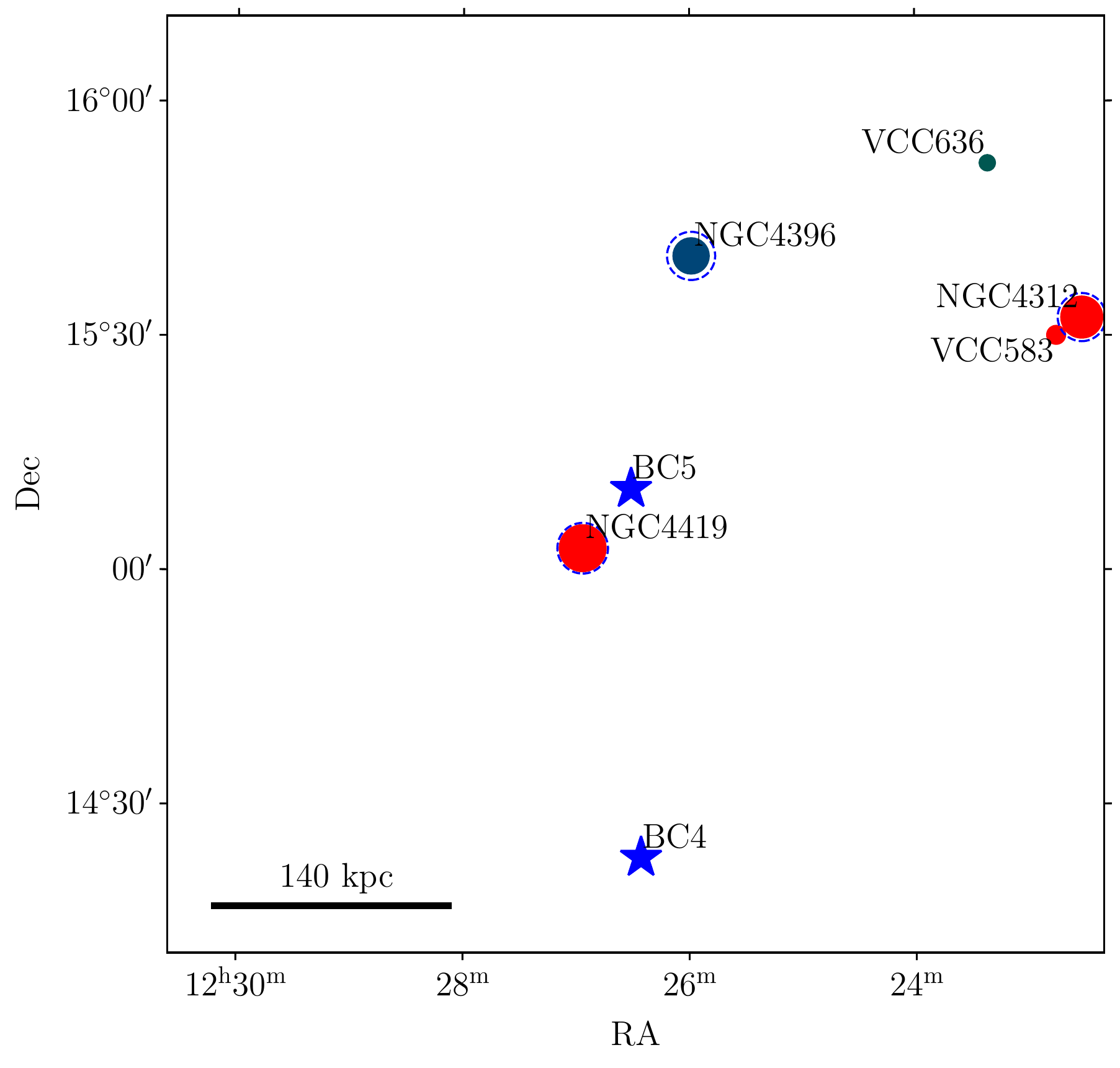}
    \caption{All VCC, EVCC, and ALFALFA neighbors of BC1, 3, 4, and 5 (top-left to bottom-right) in a 4~sq~deg region centered on the BC (blue star in each panel) and within $\pm$500~\kms \ of their H$\alpha$ velocity (Table \ref{tab:metallicity}). The area of each circular marker corresponds to the apparent magnitude (in $g$-band) of the galaxy it represents. The color of the markers corresponds to the galaxy's $g-i$ color, with the narrow transition from blue to red occurring at $g-i=0.9$ (green). Objects circled with a dashed blue line were detected in ALFALFA and thus contain significant quantities of \hi \ (note that BC3 itself was detected in ALFALFA, but is not circled here). At the distance of Virgo 30\arcmin \ corresponds to $\sim$140~kpc. The scale bar in the bottom-right panel applies to all panels.}
    \label{fig:BC_NN}
\end{figure*}

\begin{figure}
    \centering
    \includegraphics[width=\columnwidth]{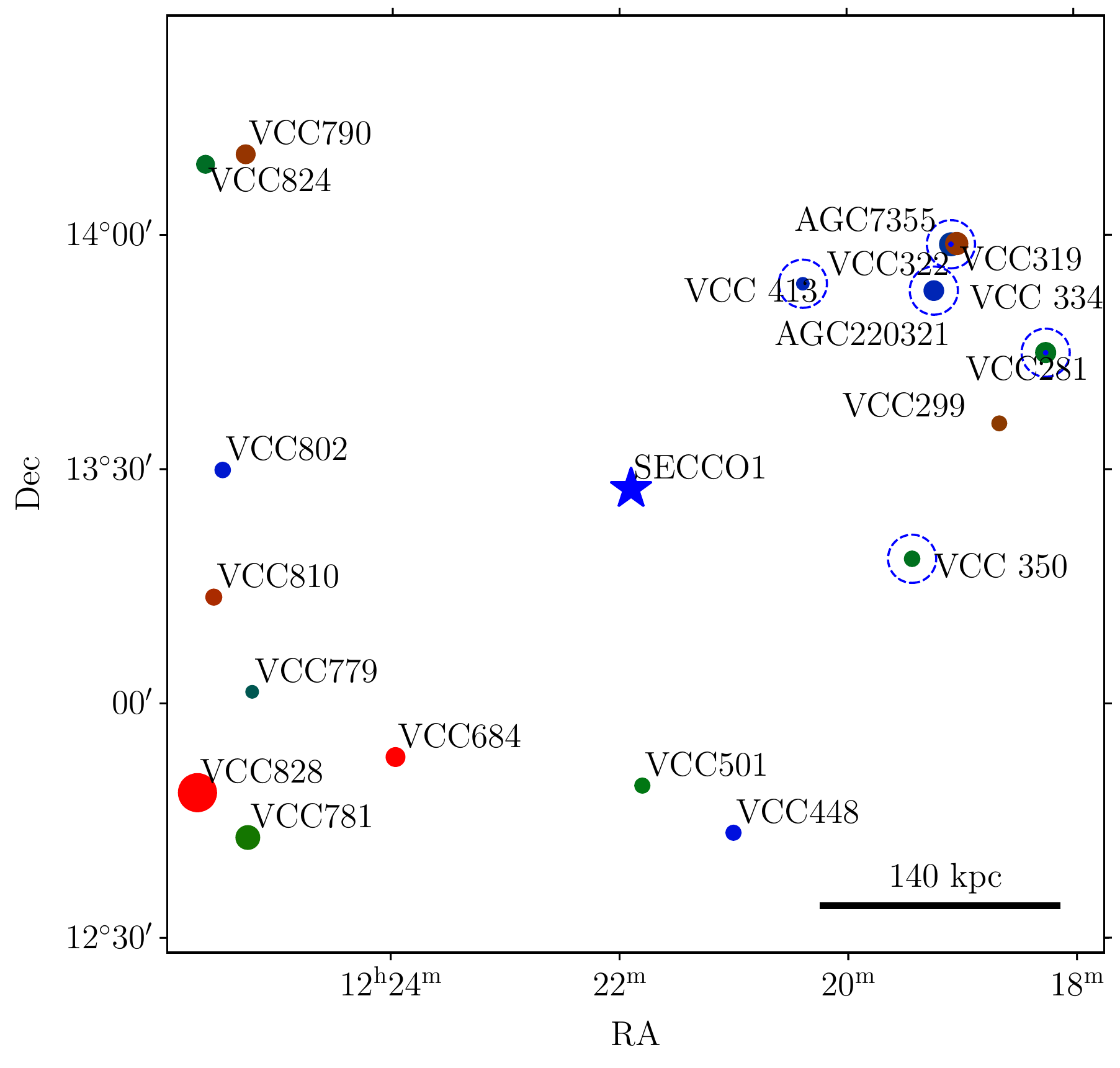}
    \caption{As for Figure \ref{fig:BC_NN}, but for SECCO~1.}
    \label{fig:SECCO1_NN}
\end{figure}

The observations presented above reveal the surprising result that although all the BCs are actively forming stars, only BC3 and SECCO~1 have a detectable quantity of \hi. However, the typical values of the SFR estimates are of the order of $10^{-3} \; \mathrm{M_\odot \, yr^{-1}}$ (Table \ref{tab:SFRs}), which means that even below our \hi \ detection limits (\S\ref{sec:hi_mass}) these objects could still have gas consumption timescales in excess of 1~Gyr (although such long timescales are not uncommon for low-mass galaxies). In addition, the high metallicities of BCs (Table \ref{tab:metallicity}) clearly point to them all having formed from pre-enriched gas that originated in a larger galaxy, as has been shown explicitly to be the case for BC3 \citep{Jones+2022}, where the gas trail can still be traced back to its parent galaxy. 

Figure \ref{fig:LV_Z} compares the metallicities of BCs to other objects of comparable luminosity. As expected, tidal dwarf galaxies (TDGs) are similar to BCs, being of equivalent metallicity, but typically somewhat higher luminosities. This is a point to which we will return in the following section (\S \ref{sec:formation}), but BCs are likely too low-mass to be TDGs. On the opposite end of the metallicity spectrum we compare BCs to a small selection of extremely metal-poor galaxies (XMPs). These objects can appear superficially similar to BCs. Both are usually extremely blue, have clumpy morphologies, and the faintest XMPs are the same luminosity as BCs. However, their metallicities could scarcely be more different and the two populations are clearly distinct in origin.

If we take the metallicities of the BCs and use them to infer a stellar mass from the mass--metallicity relation (MZR) then this should provide a reasonable estimate of the type of galaxies from which they formed. Using the MZR of \citet{Andrews+2013}, the metallicities of the BCs imply that their parent objects could have stellar masses anywhere in the range $8.3 \lesssim \log M_\ast/\mathrm{M_\odot} \lesssim 10.1$ (we note that because the MZR is an asymptotic relation, the lower bound of this range is much better constrained than the upper bound). This covers a broad range from dwarf galaxies almost to Milky Way-like galaxies \citep[$\log M_{\ast,\mathrm{MW}}/\mathrm{M_\odot} = 10.8$,][]{Licquia+2015}, but all are massive enough that, unless particularly low surface brightness (LSB), they should be mostly included in existing catalogs of Virgo cluster galaxies. Furthermore, it is also reasonable to assume that the parent objects are gas-bearing (or were in the recent past), as they must have been able to supply the gas that formed the young stellar populations of the BCs. The quoted range is for the average metallicity and does not account for metallicity variations within the parent galaxies, which could potentially expand the range if the material that formed a BC originated from a region that strongly deviated from the average metallicity.

\subsection{Search for points of origin}

We performed a detailed search considering all known Virgo members in the vicinity (and at a similar velocity to) each BC, paying particular attention to gas-bearing galaxies detected in ALFALFA \citep{Haynes+2018}. Even though most BCs are undetected in \hi, they appear to have formed from stripped gas and must have contained gas in the recent past as they have all formed stars recently. Thus, nearby, gas-rich galaxies are good candidate progenitor systems. The galaxies neighboring each BC are shown in Figures \ref{fig:BC_NN} \& \ref{fig:SECCO1_NN}. Here we present the conclusions of our search, but the full details can be found in Appendix \ref{sec:origin_search}.

Within the entire 4~deg$^2$ region shown around BC1 in Figure \ref{fig:BC_NN} (top-left), there is only one galaxy that contains \hi \ gas and is sufficiently massive to have formed a BC, NGC~4579. This galaxy is approximately 140~kpc (30\arcmin) to the SW of BC1 and separated from it in velocity by $\sim$400~\kms. Thus, BC1 would have needed a large ejection velocity ($>$500~\kms \ in total) for NGC~4579 to be its parent object. Furthermore, other than a slightly \hi-deficient disk\footnote{\citet{Chung+2009} define the \hi-deficiency of a galaxy's disk as the logarithmic decrement between the observed and expected mean \hi \ surface density within the optical disk, where for the latter they use the average value for isolated galaxies from \citet{Haynes+1984}.}, NGC~4579 shows little sign of recent disturbance in either its \hi \ or CO morphology and kinematics \citep{Chung+2009,Brown+2021}. Finally,  NGC~4579 appears to be too metal-rich \citep[$12+\log(\mathrm{O/H}) = 8.87 \pm 0.05$,][]{DeVis+2019} to match the metallicity of BC1 ($\langle 12+\log(\mathrm{O/H}) \rangle = 8.35 \pm 0.15$). Thus, NGC~4579 does not seem to be a viable candidate point of origin for BC1, and the genuine point of origin must presumably be beyond the region shown in Figure \ref{fig:BC_NN} ($>280$~kpc away), but we are unable to identify any strong candidates this far away.

BC3 (also called AGC~226178) was discussed in detail by \citet{Jones+2022}. This is an extremely complicated field with multiple foreground systems projected on it. BC3's nearest apparent neighbor is NGVS~3543 (also called AGC~229166), which \citet{Junais+2021} argued was a LSB galaxy at the same distance as BC3. However, based on the CMDs produced from HST imaging, \citet{Jones+2022} demonstrated that NGVS~3543 is a foreground object at $\sim$10~Mpc, while BC3 is consistent with being in Virgo at 16.5~Mpc. \hi \ observations with the VLA \citep{Cannon+2015} and Arecibo \citep{Giovanelli+2005,Haynes+2011,Minchin+2019} indicate a possible bridge between BC3 and a pair of galaxies, VCC~2034 and 2037. However, the closer (in projection) of these, VCC~2037, is actually another foreground object at approximately 10~Mpc \citep{Karachentsev+2014}. Thus, VCC~2034 ($cz_\odot = 1507$ \kms), $\sim$70~kpc to the SW, is the likely source of BC3's \hi \ gas. However, \citet{Jones+2022} were unable to determine whether ram pressure or tidal stripping was responsible for removing the gas from VCC~2034.

BC4 and BC5 likely formed from the same parent object as they are only separated by 45\arcmin \ on the sky, are at almost the same velocity (Table \ref{tab:metallicity}), have nearly identical metallicity measurements (Table \ref{tab:metallicity}), and have similar age estimates (\S\ref{sec:stellarmasses}). NGC~4419 is in fairly close proximity to both BCs and based on its estimated stellar mass and metallicity, it is likely a close match for the metallicity of these BCs. In addition, there is strong evidence in both \hi \ \citep{Chung+2009} and CO \citep{Brown+2021} that this galaxy is being ram pressure stripped. However, as ram pressure tails only extend in one general direction (in the wake of a galaxy's motion through the ICM) and BC4 is to the south of NGC~4419 and BC5 is to the north, it is extremely unlikely that this is the point of origin of these BCs. The extension of the molecular gas distribution of NGC~4419 is roughly towards the south \citep{Brown+2021}, in the direction of BC4, but away from BC5. If NGC~4419 were simultaneously undergoing ram pressure and tidal stripping then it could plausibly have formed both BCs, but there is no evidence of this in the optical, \hi, or CO images.

There are are a few other gas-bearing galaxies within 1~deg of either BC4 or BC5, but these were all discounted due to a mismatch in properties or because of evidence showing ram pressure stripping in the wrong direction. Upon searching further afield we immediately identified UGC~7695 (VCC~1450, IC~3476) as a strong candidate. This galaxy is a well-studied example of ram pressure stripping in action \citep{Boselli+2021}, and has a prominent bow-shaped wake extending in the approximate direction of BC4 and BC5. Existing measurements of the metallicity of UGC~7695 \citep{Hughes+2012,Boselli+2021} also approximately match those of BC4 and BC5, making this a promising candidate for their point of origin. As the separation between the BCs and UGC~7695 is approximately 450~kpc in projection they would presumably require a very large (perhaps over 1000~\kms) ejection velocity, depending on when the stripping episode began.

Finally, we consider SECCO~1. Figure \ref{fig:SECCO1_NN} shows the neighbors of SECCO~1 within a 4~deg$^2$ field and $\pm$500~\kms, and demonstrates the extraordinary isolation of this system given that it is within the virial radius of a cluster.
The potential points of origin for SECCO~1 have already been discussed extensively by previous works \citep{Adams+2015,Sand+2017,Bellazzini+2018} and we will only review these briefly here. 

If formed by a stripping event then the most likely point of origin is either the M~86 subgroup of Virgo, about 350~kpc to the SE, which exhibits an enormous complex of stripped gas visible in X-rays and H$\alpha$ \citep[][and references therein]{Sand+2017}, or the group of dwarf galaxies $\sim$200~kpc to the NW \citep{Bellazzini+2018}. The proximity of VCC~322, 334, and 319 (compared to the M~86 sub-group) might favor this possibility. However, as we have discussed above, in some cases the separation between parent and BC may be quite large. What is a stronger argument is that the metallicities of VCC~322 and 334 are a close match to that of SECCO~1 \citep{Bellazzini+2018}. VCC~322 also has a stellar tail that extends in the general direction of SECCO~1. Although these galaxies are less massive than some of the others considered, we note that the apparent parent object of BC3 is also a dwarf galaxy and only a few times more massive than BC3 itself.

However, the complex of stripped gas \citep[e.g.][]{Boselli+2018} in the M~86 sub-group is also a good candidate point of origin, for example, if NGC~4438 (VCC~1043, beyond the FoV shown in Figure \ref{fig:SECCO1_NN}) fell towards this sub-group via the location of SECCO~1. In this case, a combination of ram pressure and tidal forces could be responsible for SECCO~1 and the complex of stripped gas. We also noted the blue dwarf irregular IC~3355 near NGC~4438 (in the approximate direction of SECCO~1). However, the lower metallicity of this object \citep[$12 + \log (\mathrm{O/H}) \approx 8.0$,][]{DeVis+2019} suggests that it did not form from stripped gas.

\subsection{Other origin scenarios}

In the above discussion we considered that BCs were likely formed from a gas-bearing galaxy sufficiently massive to be included in existing catalogues of Virgo cluster galaxies. However, there are a few other scenarios that we briefly consider here.

\citet{Junais+2021} suggested that BC3 might have formed from gas stripped from a LSB galaxy. Although this scenario is ruled out for BC3 itself \citep[as the LSB galaxy in question is actually a foreground object,][]{Jones+2022} it is possible that LSB galaxies have been missed in our search above, as they are frequently absent from established catalogs of cluster members. In general this mechanism would imply that the LSB galaxy being stripped would be relatively close by to the BC, as it would presumably have a smaller gas reservoir (that would evaporate more rapidly when stripped) than a larger galaxy. Therefore, even though LSB galaxies can be challenging to detect, it seems unlikely that a close neighbor would have been overlooked in multiple cases, and we do not consider this a likely formation pathway, but note that it is difficult to entirely exclude.

An additional scenario that we considered is the possibility that the BCs could be dark objects that contained neutral gas for an extended period, but formed essentially no stars until very recently \citep[c.f.][]{Kent+2009,Minchin+2019}. This scenario is highly unlikely for two main reasons. Firstly, the search for bona fide dark galaxies that cannot be explained as tidal or spurious objects has turned up few convincing results to date \citep[e.g.][]{Taylor+2013,Cannon+2015}, calling into question whether this scenario is valid. Secondly, the high metallicity of the BCs indicates that there have been multiple prior SF episodes that have enriched their gas, thus ruling out that they could be primordial dark objects \citep[c.f.][]{Corbelli+2021}.

\section{Formation mechanism}
\label{sec:formation}

As shown in Figure \ref{fig:LV_Z}, the universally high metallicities of BCs (in relation to their luminosities or stellar masses) means that the only plausible mechanism for their formation is that they formed from material stripped from a larger galaxy. Their metallicities are a full order of magnitude higher than those of galaxies of the same V-band luminosity, and owing to their extremely young stellar populations, this discrepancy would be even larger if the samples were compared in terms of their stellar masses. Figure \ref{fig:LV_Z} also indicates that BCs are of slightly lower luminosity than TDGs, but of similar metallicity. Their stellar and \hi \ masses indicate that BCs are considerably less massive than long-lived TDGs.

Despite the strong evidence that BCs formed from stripped material, it is unclear whether they formed through tidal or ram pressure stripping. As is the case for BC3, even when the \hi \ connection to the parent galaxy is still detectable \citep{Jones+2022}, it may not be possible to distinguish between ram pressure or tidal forces as the dominant mechanism stripping the gas. Indeed it is possible that both are valid mechanisms.

Regardless of the mechanism by which gas is stripped to form BCs, the parent objects must be new cluster members. \citet{Oman+2016} and \citet{Oman+2021} simulated the stripping and quenching of galaxies falling into clusters and found that essentially all new members are stripped of their gas and quenched during their first orbit, usually around pericenter passage \citep[see also][for reviews]{Cortese+2021,Boselli+2021b}. Thus to have sufficient gas to form a BC, the parent galaxy must likely be on its first infall into the cluster. Although we only have a sample of five objects, this would also appear to agree with their spatial distribution which is inside the virial radius, where significant stripping is expected, but avoids the cluster center.

In the remainder of this section we discuss the evidence for and against tidal and ram pressure formation scenarios, compare BCs to other classes of objects known to form from stripped gas, and give an overview of related simulation results.

\subsection{Comparison to TDGs and the need for ram pressure stripping}

The typical masses of long-lived TDGs are expected to be over $10^8$~\Msol \ \citep{Bournaud+2006}, as below this mass they generally cannot resist the tidal field of their parent galaxies for long enough to escape as bound objects. This threshold mass is considerably larger than any of the BCs, disfavoring a tidal formation pathway, as the most massive BCs (BC3 and SECCO~1) are a few times 10$^7$~\Msol. However, we note that if a lower mass TDG were to be ejected at particularly high speed it may be able to survive, as it would more rapidly escape the tidal field of its parent galaxy. We also note that the simulations of \citet{Bournaud+2006} assume that the parent object of TDGs are loosely MW-like, however, VCC~2034 \citep[the apparent parent object of BC3,][]{Jones+2022} has a stellar mass of only $10^{8.2}$~\Msol. If BCs are tidal in origin then perhaps they formed from lower mass progenitors and are correspondingly lower mass than typical long-lived TDGs. However, such a mechanism could presumably only apply to those BCs (BC1, BC3, and SECCO~1) with slightly lower metallicities that correspond to similarly low-mass progenitors (via the MZR), unless the more metal-rich BCs (BC4 and BC5) formed from recently enriched gas that was stripped before it had sufficient time to mix with the rest of the interstellar medium in the parent galaxy.

TDGs are typically ejected at around the circular velocity of the galaxy they originate from \citep{Bournaud+2006}. For a relatively massive galaxy this might mean an ejection velocity of $\sim$300~\kms. If this were preferentially aligned along the direction perpendicular to the line-of-sight, it would still take a TDG $\sim$1~Gyr to traverse 300~kpc in projection. Thus the isolation of BCs, coupled with their very young stellar populations, is difficult to explain via a tidal formation mechanism. In contrast, in the case of ram pressure stripping the velocity of the galaxies relative to the cluster can exceed 1000~\kms, and galaxies with the largest tails are generally found to be traveling at highest speeds \citep{Jaffe+2018}. The fact that BCs have been identified in a cluster also points to ram pressure stripping as the most likely formation pathway. All gas-rich galaxies falling with sufficient velocity into a cluster are expected to undergo some degree of ram pressure stripping, and while tidal interactions are certainly commonplace in clusters, these most frequently take the form of brief, high speed encounters (e.g. ``galaxy harassment"), which are less likely to strip large quantities of gas \citep[e.g.][]{Smith+2010} than the strong, drawn out interactions in galaxy groups (where TDGs are typically found).

However, although the relative velocity between an infalling galaxy and the ICM can easily exceed 1000~\kms, this does not necessarily translate into an equivalent velocity for the stripped gas, as once stripped it does not immediately become stationary relative to the ICM. The ram pressure stripping simulations of \citet{Kapferer+2009} consider gas-rich galaxies falling at 1000~\kms \ relative to an ICM of varying densities (from $1\times10^{-28}$ to $5\times10^{-27}\;\mathrm{g\,cm^{-3}}$). They show that after 500~Myr of stripping the length of the plume of stripped gas in the wake of the parent galaxy is strongly dependent on the density of the surrounding ICM (e.g. their Figure 20). In this case the most distant gas clouds (for $\rho_\mathrm{ICM} \geq 1\times10^{-27}\;\mathrm{g\,cm^{-3}}$) are $\sim$400~kpc from their parent galaxy, indicating that their average relative velocity over the past 500~Myr has been $\sim$800~\kms. This is somewhat slower than the velocity of the parent galaxy relative to the ICM (and would be considerably slower still for lower ICM densities), but is still several times greater than the relative velocities typically expected for TDGs. 
For comparison, the electron number density of the ICM in Virgo \citep{Nulsen+1995} exceeds $10^{-2}\;\mathrm{cm^{-3}}$ ($\sim2\times10^{-26}\;\mathrm{g\,cm^{-3}}$) near M~87 and at a distance of 230~kpc has decreased to $6\times10^{-4}\;\mathrm{cm^{-3}}$ ($\sim1\times10^{-27}\;\mathrm{g\,cm^{-3}}$).
Thus ram pressure stripping (in Virgo) provides a more viable mechanism for rapidly achieving large separations between stripped material and its parent object. This is especially true within a few hundred kpc of the cluster center, but could be true almost anywhere within the cluster should an infalling galaxy collide with a dense pocket in the ICM \citep[which are known to exist in other clusters, e.g. ][]{Morandi+2014,Eckert+2015}.

The most similar known objects to BCs are ``fireballs" \citep[e.g.][]{Cortese+2007,Yoshida+2008,Hester+2010}, clumps of SF seen in the wake of galaxies being actively ram pressure stripped. Indeed, as discussed by \citet{Bellazzini+2018}, many of the properties of BCs match well with those of fireballs \citep[e.g.][]{Fumagalli+2011}, including their metallicities \citep[e.g. fireballs in the wake of IC~3418 have $8.22 < 12 + \log (\mathrm{O/H}) < 8.38$,][]{Kenney+2014}. Other related objects include SF clumps in filamentary structures in the vicinity of NGC~1275 in the Perseus cluster \citep{Conselice+2001,Canning+2014} and in stripped material in Stephan's Quintet \citep{MendesdeOliveira+2004}.

However, BCs are distinct from fireballs and similar objects, in that they are remarkably isolated (e.g. Figures \ref{fig:BC_NN} \& \ref{fig:SECCO1_NN}). Fireballs are found within a few 10s of kpc of their parent galaxy, where there can be little doubt over their point of origin, and where they may still eventually fall back onto their parent galaxy \citep[e.g.][]{Vollmer+2001,Tonnesen+2012}. To form BCs requires a mechanism which can carry neutral gas several 100s of kpc from a galaxy within the hostile environment of a cluster.

\subsection{Properties of ram pressure stripped gas clumps in simulations}

\citet{Lee+2022} argue that many of the molecular gas clouds seen in the tail of ram pressure stripped galaxies \citep[e.g.][]{Moretti+2018,Jachym+2019} could form in situ, by rapid cooling of warm ionized gas \citep[][also suggest a similar mechanism]{Tonneson+2012,Moretti+2020}. The metal-rich gas and absence of young stars in close proximity (unlike within the disks of most gas-rich galaxies) make conditions favorable for radiative cooling. Furthermore, \citet{Muller+2021} argue that magnetic sheathing could help to protect ram pressure stripped gas from evaporation in a cluster environment.

In the radiative hydrodynamical simulations of \citet{Lee+2022} SF clumps are seen out to $\sim$100~kpc from the parent galaxy. In their models this SF in the distant tail occurs $\sim$200~Myr after the initial onset of ram pressure stripping, suggesting that the very young ages (50-150~Myr) of the stellar populations of BCs may underestimate how long ago their progenitor gas was stripped (although large velocities $>$500~\kms \ would likely still be required to explain their isolation). Finally they note that although bright H$\alpha$ clumps will only track SF activity, more diffuse H$\alpha$ emission (fainter than $6\times10^{38} \; \mathrm{erg\,s^{-1}\,kpc^{-2}}$) is expected throughout the ram pressure tail due to recombinations in the warm ionized gas. Therefore, H$\alpha$ observations significantly more sensitive than this threshold might be capable of robustly identifying the points of origin of BCs. The nominal 1$\sigma$ surface brightness sensitivity of the VESTIGE survey is $2\times10^{-18}\;\mathrm{ergs\,s^{-1}\,cm^{-2}\,arcsec^{-2}}$ \citep{Boselli+2018}, which for a distance of 16.5~Mpc equates to $3.4\times10^{36} \; \mathrm{erg\,s^{-1}\,kpc^{-2}}$. Thus, such features, should they exist, would be  detectable in VESTIGE.

The hydrodynamic simulations of \citet{Kapferer+2009} also produce numerous gas clumps in the wakes of ram pressure stripped galaxies, but out to much greater distances ($\sim$400~kpc). They find that SF is only induced in these clumps if the wind speed exceeds 500~\kms \ and that it is enhanced by yet stronger ram pressure \citep[but note that][find the opposite trend]{Tonneson+2012}. Ram pressure therefore appears to be a promising candidate for producing clumps of star forming gas far from their parent galaxies, but do the physical properties of these systems match with those observed in BCs?

In the case of the above mentioned ram pressure simulations the gas clumps in the wakes of the stripped galaxies are generally presented in terms of gas density rather than masses of distinct clumps. However, in the simulations of \citet{Tonneson+2021} the masses of such clumps are found to be on the order of $10^5$~\Msol, with the most massive distinct clouds being $\sim$10$^6$~\Msol. This matches quite well with the masses of gas clumps typically found in the immediate wakes of ram pressure stripped galaxies \citep[e.g.][]{Poggianti+2019}, but is more than an order of magnitude less massive than BC3 and SECCO~1. However, the earlier ram pressure stripping simulations of \citet{Kronberger+2008} do form bound objects, analogous to TDGs, with total masses of $\sim$10$^7$~\Msol, but these simulations are now thought to oversimplify fluid instabilities \citep[e.g.][]{Sijacki+2012}, calling into question the details of these results.

It terms of metallicity it is generally assumed that, in either the tidal or ram pressure stripping scenario, the BC formed will exhibit the same metallicity as its parent galaxy. However, \citet{Tonneson+2021} also find that all their simulated ram pressure stripped clouds rapidly mix with the ICM. Thus, they predict that the metallicity of ram pressure stripped clouds should decrease with distance from their parent galaxy. This seems to be directly contradicted by the high metallicities of BCs, given their relative isolation and large separations (e.g. $>$300~kpc in some cases) from their apparent points of origin. We also note that \citet{Calura+2020} find that more massive \hi \ clouds (similar to SECCO~1 and BC3) can survive intact for on the order of a Gyr, while moving rapidly through the ICM. It may be that the gas clouds from which BCs form are exceptional objects and not typical of the underlying population of gas clouds that are stripped in ram pressure events. For example, these could be some of the most loosely bound gas that is the first to be stripped, or they could be stripped by a denser clump of the ICM. This could explain the lack of similar objects in the simulations of \citet{Tonneson+2021}.

As a closing remark for this discussion we also note that despite the apparently simple requirement for ram pressure stripping (i.e. sufficient ram pressure to overcome the gravitational attraction of the gas disk) and extensive efforts to simulate this process in increasing detail, there remain systems that are challenging to explain. In particular, the recent discovery of an enormous (apparently) ram pressure stripped \hi \ tail in a system outside of a cluster, where no significant inter-galatic medium could be detected \citep{Scott+2022}, poses difficult questions regarding its origin, and could even suggest that some BC-like objects might exist outside of clusters.

\subsection{Summary}

In summary, BCs appear to be distinct from both TDGs and fireballs, being too low-mass to be the former, too high-mass to be the latter, and too isolated for either. The isolation of some BCs is their property that is hardest to explain, and would seem to necessitate the large velocities expected in ram pressure stripping events, but not for strong tidal interactions. We therefore favor ram pressure stripping as the most likely formation mechanism of BCs.
If ram pressure stripping is confirmed to be the formation pathway then BCs can be thought of as ``ram pressure dwarfs", analogous to tidal dwarfs, but unlikely to survive as bound structures on long timescales. Simulations provide a somewhat conflicting picture of how such objects might form via ram pressure stripping, however, this may be because BCs represent atypical objects that, unlike fireballs, are not formed in large numbers during stripping episodes. Regardless of their formation mechanism the properties of BCs appear to be distinct from any other stellar systems of which we are aware.

\section{Fate and production rate}
\label{sec:fate}

Based on their morphologies and stellar mass estimates, BCs are unlikely to be gravitationally bound. In the case of BC3 and SECCO~1, their \hi \ content might be sufficient for them to remain bound in the short term \citep[e.g.][]{Calura+2020}, but this neutral gas (the majority of their total mass) will eventually be lost to the ICM. It is challenging to accurately assess the stability of BCs due to their irregular morphologies and because their velocity dispersions are not well resolved by the MUSE observations. However, based on the stellar mass estimates in Table \ref{tab:Mstar} and their apparent sizes, we estimate that if they are extremely dynamically cold (e.g. $\sigma_v < 1$~\kms) then they may be bound, but for $\sigma_v > 2$~\kms \ they would certainly be unbound \citep[using Equation 8 of][]{Calura+2020}. The stability of BCs is considered further in \citetalias{Bellazzini+2022}, however, the most likely scenario is that each BC as a whole is unbound, but individual component clumps or star clusters may be bound, if sufficiently dynamically cold. Thus, in the long term BCs will likely disperse (either as individual stars or star clusters) into the intracluster light. However, even if BCs were to remain bound, without sustained SF they would quickly become essentially undetectable. Currently they are only identifiable at all (in optical/UV) because of their young, blue stars.

In \citet{Jones+2022} we argued that as BCs are only expected to be visible for a short period, they must be continually produced in the cluster. We estimated that an object such as BC3 might be detectable for at most 500~Myr, meaning that for five BCs to be visible today, they must be being produced at a rate on the order of 1 per 100~Myr. However, given that all five of the known BCs appear to have ages of less than 200~Myr, this might be a more reasonable estimate, making the production rate closer to 1 per 50~Myr. We speculate that this might be a common phenomenon with many newly infalling galaxies producing such objects.

With this in mind we also note that, in hindsight, the metallicities of BCs are perhaps not surprising. As mentioned in \S\ref{sec:discuss}, the metallicities of BCs correspond to a stellar mass range $8.3 \lesssim \log M_\ast/\mathrm{M_\odot} \lesssim 10.1$. However, this wide range likely encompasses most galaxies that could possibly form BCs (suggesting it is a common occurrence). Galaxies significantly less massive than $\log M_\ast/\mathrm{M_\odot} = 8.3$ would have \hi \ reservoirs scarcely larger than those of SECCO~1 and BC3, and are thus probably too small to form a BC themselves. Whereas galaxies significantly more massive than $\log M_\ast/\mathrm{M_\odot} = 10.1$ are increasingly uncommon and increasingly likely to be gas-poor.

\section{Future directions}
\label{sec:future}

Although the faintness and peculiar properties of BCs make them challenging objects to study, we suggest a few directions where progress could likely be made.

We experienced significant difficulties in attempting to identify the point of origin of most of the BCs, likely because it has been several hundred Myr since some of them were first stripped from their parent galaxy. However, if ram pressure stripping is the formation pathway of these objects then it is possible that extremely faint H$\alpha$ trails still connect the BCs to their parent objects \citep[e.g.][]{Lee+2022}. A deep H$\alpha$ search around the BCs (or others identified in the future), is a promising approach to robustly identifying their parent objects, which in turn will allow for more detailed study of their formation mechanism. X-ray emission is also frequently found to accompany H$\alpha$ tails of ram pressure stripped galaxies \citep[e.g.]{Sun+2007,Sun+2021}, and may represent another means to characterize the properties of the stripping events that formed the BCs in cases where the parent object can be identified.

CO observations with the Atacama Large Millimeter/submillimeter Array (ALMA) have successfully made detections of individual clumps of molecular gas in the wakes of a ram pressure stripped galaxies \citep[e.g.][]{Jachym+2019} as well as individual giant molecular clouds in TDGs \citep{Querejeta+2021}. If BCs still contain significant quantities of molecular gas (as would be expected based on their recent SF) then they should be readily detectable with ALMA, especially as their high metallicity measurements imply a favorable CO-to-H$_2$ conversion factor in comparison to other low-mass objects \citep{Bolatto+2013}. 

\citet{Lee+2022} find that ram pressure stripped gas clouds may travel for over a hundred Myr before SF occurs in them. Although we are limited by a very small sample size, all of the BCs with slightly older stellar populations estimates ($\gtrsim100$~Myr, rather than $\sim$50~Myr) are undetected in \hi. The long gas consumption (by SF) timescales in Table \ref{tab:SFRs} indicate that the gas in these systems is not (for the most part) consumed by SF. BCs 1, 4, and 5 must have contained significant cold gas reservoirs within the past 200~Myr to permit the formation of their observed stellar populations, and they likely still contain some molecular gas as they have all formed new stars in the past 10~Myr, yet today we find no evidence of any \hi \ content. If BCs have traveled through the ICM for significantly longer than the current age of their stellar populations in order to reach their current state of isolation, perhaps this indicates that it is the SF episode itself that triggers the evaporation of the neutral gas. For example, it seems plausible that an object like SECCO~1 is essentially an earlier stage of an object like BC4 and BC5. Both are broken into two main components, but SECCO~1 has yet to lose its gas, and BC4 appears as though it may be disintegrating (Figure \ref{fig:BC4_HST_GLX}). If this were the case then SECCO~1 would likely be on the verge of losing its \hi \ gas. This disagrees somewhat with the findings of \citet{Calura+2020}. However, those authors note that the details of the SF episode, particularly when it began, are quite uncertain, and we suggest that this possibility might warrant further investigation.

Along similar lines, with the HST observations it has only been possible to characterize the stellar populations of BCs from the stars formed in the past $\sim$50~Myr. While some BCs may genuinely contain no stars that are older than this, others might. To detect or rule out this older stellar population, and therefore to constrain the full SF histories of BCs, is possible with the James Webb Space Telescope observations (JWST). With a moderate investment of observing time ($\sim$10~h) JWST is capable of detecting stars several magnitudes below the TRGB, should an RGB exist, at the distance of the Virgo cluster. Such observations would not only determine the age of the oldest stellar component of BCs, but (if RGB stars exist in BCs) would also be capable of conclusively demonstrating Virgo membership via TRGB distance measurements.

Finally, we note that if our hypothesis is correct and BCs are commonly produced when new member galaxies fall into a cluster, then they should exist in other clusters as well as Virgo. Unfortunately, due to how faint BCs are, they would be undetectable in any galaxy clusters significantly farther away than Virgo. We therefore suggest that the Fornax cluster could be a suitable location to extend the search and would represent an independent environment where our findings could be cross-checked. The distance modulus for Fornax is only $\sim$0.5~mag greater than for Virgo, thus the brightest BCs (Table \ref{tab:Mstar}) would likely still be detectable and slightly longer HST observations could provide similar quality CMDs.

The ongoing MeerKAT Fornax survey \citep{Serra+2016,Kleiner+2021} aims to map 12~deg$^2$ of the cluster in \hi. These observations will be approximately 5 times deeper than our pointed VLA observations of BCs, and will have a factor of $\sim$3 times better angular resolution. Thus this survey will be ideal for identifying ``dark", dense \hi \ clouds analogous to those in \citet{Adams+2013} and \citet{Cannon+2015} that led to the discovery of BCs. Furthermore, the improved column density sensitivity will exceed that of ALFALFA and will be readily capable of detecting residual \hi \ streams that might still connect young BCs to their parent objects, as is the case for BC3 \citep{Jones+2022}. The Fornax cluster is also the target of both the Next Generation Fornax Survey \citep{Munoz+2015}, $ugi$ imaging with the Dark Energy Camera on the Blanco telescope, and the Fornax Deep Survey with the Very Large Survey Telescope \citep{Peletier+2020}, which is imaging the Fornax cluster in $ugri$ at comparable depth to the NGVS in Virgo. Together these surveys will provide the means to identify BCs both through their young blue stellar populations and, where it exists, their \hi \ gas.

\section{Conclusions}
\label{sec:conclude}

We have presented follow up HST, VLT/MUSE, VLA (and GBT) observations of five candidate young, blue, faint, stellar systems in the direction of the Virgo cluster that are analogous to SECCO~1. With the exception of one spurious object, we find that these are all comparable to SECCO~1 in terms of their extremely low stellar masses, blue stellar populations, and high metallicities, leading us to conclude that they must have formed from gas stripped from more massive galaxies. However, only one is detected in \hi, suggesting that the others have likely survived sufficiently long to lose much of their initial gas content. Some of these objects are also surprisingly isolated, residing several hundred kpc from the nearest potential source of gas, which poses a challenge for robustly identifying their points of origin.

We considered both tidal and ram pressure stripping scenarios as the potential formation mechanism of these stellar systems. Although we cannot confidently exclude either of these mechanisms, and indeed there may not be one single mechanism responsible for all BCs, ram pressure stripping is most consistent with the observed properties. In particular, the isolation of some BCs is difficult to explain with the low velocities ($\leq$300~\kms) expected for ejected TDGs, but can more naturally be explained by ram pressure stripping proceeding at $>$1000~\kms. In addition, BCs are likely too low mass to be long-lived TDGs. However, gas clumps formed in ram pressure stripping simulations are typically much lower mass than BCs (based on the \hi \ masses of BC3 and SECCO~1), and we suggest that these objects may be atypical and form from the first loosely bound gas to be stripped, or as a result of stripping in a clumpy ICM. These massive clumps ($\sim$10$^7$~\Msol) of stripped gas moving at high speed can likely survive sufficiently long in the ICM to form the stellar populations observed and to become relatively isolated. However, they will ultimately lose their gas content (the majority of their total mass) and likely become unbound. 

BCs therefore represent a new class of stellar system that form from large ($\sim$10$^7$~\Msol) clumps of pre-enriched, stripped gas, are (assumed to be) dark matter free, and are capable of surviving sufficiently long in the hostile ICM to become isolated ($>$100~kpc away) from their parent galaxies.

A further census of this class of object in the Virgo cluster, and potentially the Fornax cluster, will allow for improved constraints on their lifetimes and how frequently they are produced. However, robust identification of their parent objects will remain challenging, owing to their isolation. Deep, wide-field H$\alpha$ imaging, to identify diffuse emission, is a potential approach for systems where the majority of the neutral gas has already been evaporated.


\begin{acknowledgments}
The authors thank Kyle Artkop for assistance in identifying blue candidates in Virgo. We also thank Toby Brown and co-authors for providing their X-ray mosaic from archival ROSAT observations of the Virgo cluster.
This work is based on observations made with the NASA/ESA Hubble Space Telescope, obtained at the Space Telescope Science Institute, which is operated by the Association of Universities for Research in Astronomy, Inc., under NASA contract NAS5-26555.  These observations are associated with program \# HST-GO-15183.  Support for program \# HST-GO-15183 was provided by NASA through a grant from the Space Telescope Science Institute, which is operated by the Association of Universities for Research in Astronomy, Inc., under NASA contract NAS5-26555.
It is also based on observations collected at the European Organisation for Astronomical Research in the Southern Hemisphere under ESO programme 0101.B-0376A.
This work used both previously unpublished and archival data from the Karl G. Jansky Very Large Array. The National Radio Astronomy Observatory is a facility of the National Science Foundation operated under cooperative agreement by Associated Universities, Inc. The data were observed as part of programs 13A-028 (PI: J.~Cannon) and 18A-185 (PI: K.~Spekkens). 
The work used images from the Dark Energy Camera Legacy Survey (DECaLS; Proposal ID 2014B-0404; PIs: David Schlegel and Arjun Dey). Full acknowledgment at \url{https://www.legacysurvey.org/acknowledgment/}.

DJS acknowledges support from NSF grants AST-1821967 and 1813708.
MB acknowledges the financial support to this research from the INAF Main Stream Grant 1.05.01.86.28 assigned to the program {\em The Smallest Scale of the Hierarchy (SSH)}
KS acknowledges support from the Natural Sciences and Engineering Research Council of Canada (NSERC).
BMP is supported by an NSF Astronomy and Astrophysics Postdoctoral Fellowship under award AST-2001663.
EAKA is supported by the WISE research programme, which is financed by the Dutch Research Council (NWO).
GB acknowledges financial support through the grant (AEI/FEDER, UE) AYA2017-89076-P, as well as by the Ministerio de Ciencia, Innovación y Universidades, through the State Budget and by the Consejería de Economía, Industria, Comercio y Conocimiento of the Canary Islands Autonomous Community, through the Regional Budget.
JS acknowledges support from the Packard Foundation.
MPH acknowledges support from NSF/AST-1714828 and grants from the Brinson Foundation.
JMC, JF, and JLI are supported by NST/AST 2009894.
R.~R.~M. gratefully acknowledges support by the ANID BASAL project FB210003. 
Research by DC is supported by NSF grant AST-1814208.
AK acknowledges financial support from the State Agency for Research of the Spanish Ministry of Science, Innovation and Universities through the "Center of Excellence Severo Ochoa" awarded to the Instituto de Astrof\'{i}sica de Andaluc\'{i}a (SEV-2017-0709) and through the grant POSTDOC\_21\_00845 financed from the budgetary program 54a Scientific Research and Innovation of the Economic Transformation, Industry, Knowledge and Universities Council of the Regional Government of Andalusia.

\end{acknowledgments}

\facilities{Blanco, GALEX, GBT, HST (ACS), ROSAT, VLA, VLT:Yepun (MUSE)}
\software{\href{http://americano.dolphinsim.com/dolphot/}{\texttt{DOLPHOT}} \citep{Dolphin2000}, \href{https://casa.nrao.edu/}{\texttt{CASA}} \citep{CASA}, \href{https://www.astropy.org/index.html}{\texttt{astropy}} \citep{astropy2013,astropy2018}, \href{https://aplpy.github.io/}{\texttt{APLpy}} \citep{aplpy2012,aplpy2019}, \href{https://photutils.readthedocs.io/en/stable/}{\texttt{Photutils}} \citep{photutils}, \href{https://reproject.readthedocs.io/en/stable/}{\texttt{reproject}} \citep{reproject}, \href{https://acstools.readthedocs.io/en/latest/}{\texttt{acstools}} \citep{acstools}, \href{https://sites.google.com/cfa.harvard.edu/saoimageds9}{\texttt{DS9}} \citep{DS9}, \href{https://dustmaps.readthedocs.io/en/latest/}{\texttt{dustmaps}} \citep{Green2018}, \href{https://astroalign.readthedocs.io/en/latest/}{\texttt{Astroalign}} \citep{astroalign}, \href{https://www.astromatic.net/software/sextractor/}{\texttt{Sextractor}} \citep{Bertin+1996}, \href{https://aladin.u-strasbg.fr/}{\texttt{Aladin}} \citep{Aladin2000,Aladin2014}}

\bibliography{refs}{}
\bibliographystyle{aasjournal}



\appendix

\section{VLA spectrum of BC2}
\label{sec:BC2spec}

\begin{figure}
    \centering
    \includegraphics[width=0.5\textwidth]{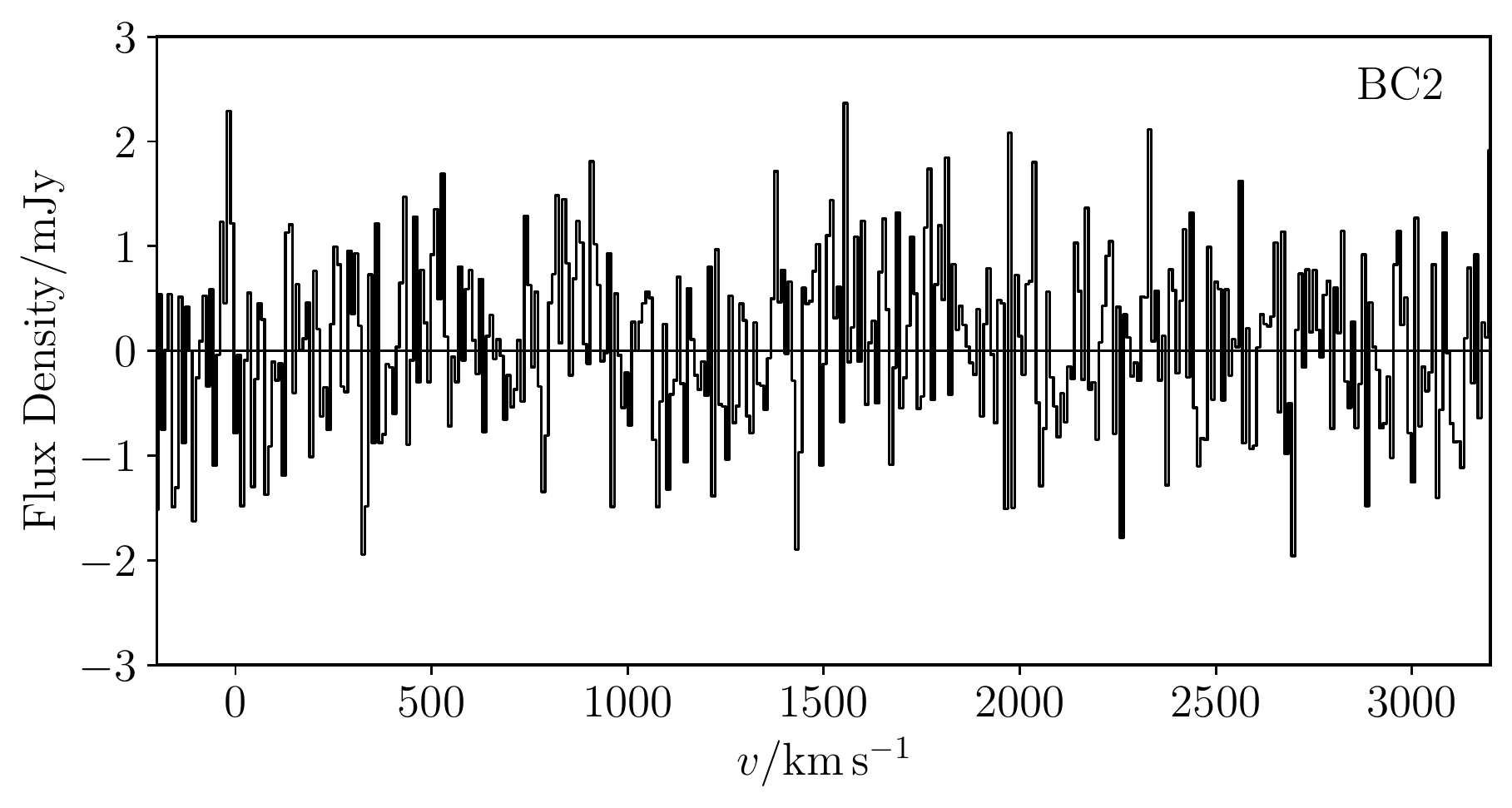}
    \caption{\hi \ spectrum in the directions of BC2 extracted from the VLA \hi \ data cube within an aperture equal in area to the synthesized beam.}
    \label{fig:BC2VLAspec}
\end{figure}

Although the HST imaging of BC2 (Figure \ref{fig:BC2_HST_GLX}) indicates that it is a background group of galaxies rather than a blue stellar system at the distance of Virgo, we have included an \hi \ spectrum extracted from the VLA data cube for completeness (Figure \ref{fig:BC2VLAspec}). As with the other BCs the spectrum is extracted with a beam-sized aperture centered on the optical position of BC2. However, unlike the other VLA spectra this spectrum covers a broader range of velocities as BC2 has no known redshift from H$\alpha$ emission. As expected, no significant \hi \ signal could be identified.

\section{Search for points of origin}
\label{sec:origin_search}

In the following subsections we search for neighboring galaxies that could have supplied the gas from which the BCs formed. In particular, we consider all ALFALFA \citep{Haynes+2018}, Virgo Cluster Catalog \citep[VCC,][]{Binggeli+1985}, and Extended Virgo Cluster Catalog \citep[EVCC,][]{Kim+2014} galaxies in the vicinity of each BC, in an attempt to identify the most probable point of origin in each case.

\subsection{BC1}

As shown in Figure \ref{fig:BC_NN} (top left), the nearest galaxy (in projection) with a redshift within 500~\kms \ of BC1 ($cz_\odot = 1118$ \kms), is AGC~224219 (LSBVCC~79), 14.5\arcmin \ to the SW, which was detected in \hi \ by ALFALFA and AGES \citep[Arecibo Galaxy Environment Survey,][AGESVC2\_30]{Taylor+2013}, and marginally detected near the primary beam edge of our VLA observation. After this there is the large spiral NGC~4579 (VCC~1727) further to the SW (30\arcmin \ in total), which was also detected in \hi \ by ALFALFA. 

AGC~224219 ($cz_\odot = 1019$ \kms) is a very LSB dwarf with a central surface brightness of 24.8~mag~arcsec$^{-2}$ in $g$-band \citep{Davies+2016}. Its stellar population is also clearly redder than BC1, which comes as a surprise given that it has at least an order of magnitude more \hi. The other progenitor possibility, NGC~4579 ($cz_\odot = 1517$ \kms) is a much larger, gas-bearing, spiral galaxy. This galaxy was imaged by the VIVA \citep[VLA Imaging of Virgo spirals in Atomic gas,][]{Chung+2009} survey and they concluded that it is slightly \hi-deficient and its \hi \ distribution is somewhat truncated (presumably by ram pressure stripping). However, other than this, the \hi \ distribution and kinematics are regular and there is no sign of an ongoing or recent interaction. ALMA (Atacama Large Millimeter Array) observations of its CO molecular gas from VERTICO \citep[Virgo Environment Traced in CO survey;][]{Brown+2021} present a similarly regular morphology. We also note that if BC1 originated from NGC~4579 then it must have had a very large ejection velocity, at least 400~\kms \ along the line-of-sight (LoS). If we also assume BC1 is on the order of 200~Myr in age \citep[c.f. BC3,][]{Jones+2022} then the component perpendicular to the LoS would need to be on the order of 500~\kms, giving a total ejection velocity of $\sim$650~\kms, a large, but not impossible value.

Perhaps both BC1 and AGC~224219 originated from a past stripping event of NGC~4579. They are both along the same direction from NGC~4579 and both have very similar radial velocities. This scenario would also be consistent with the observation that AGC~224219 has more \hi \ gas than BC1, as the latter would have traveled twice as far from its parent object ($\sim$150~Mpc, projected, in total). However, this scenario would not naturally explain why AGC~224219 appears to have an older, redder stellar population, while BC1 is dominated by young, blue stars (Figure \ref{fig:BC1_HST_GLX}) and H\,{\sc ii} regions. Furthermore, if a large quantity of gas had been stripped from NGC~4579 in the recent past then its \hi \ and CO distributions would probably show more signs of disturbance and other clumps of in situ SF would be expected in the immediate vicinity.

The discussion above exhausts the possible sources of gas, within 1$^\circ$ ($\sim$300~kpc) and $\pm$500 \kms, from which BC1 could have formed. Assuming that we have not erroneously rejected the true source of the gas that formed BC1, we are left with three possibilities: a) the origin of the gas lies farther away than 1$^\circ$, b) the gas originally belonged to a faint or LSB object that we have missed or excluded due to a lack of a redshift, or c) that BC1 was a (long-lived) dark object that contained gas but no stars prior to its recent SF episode. 

Based on the simulations of \citet{Kapferer+2009}, neutral gas being ram pressure stripped can be found several hundred kpc from the source galaxy. However, increasing the search radius gives a large number of potential candidates owing to the density of the Virgo region. If we narrow these down to those that are blue or contain \hi, and are very close in velocity to BC1 then VCC~1686 (IC~3583, $cz_\odot = 1122$~\kms, 71\arcmin \ away), VCC~1992 (IC~3710, $cz_\odot = 1012$~\kms, 75\arcmin \ away), AGC~225847 ($cz_\odot = 982$~\kms, 81\arcmin \ away), and VCC~1931 ($cz_\odot = 1183$~\kms, 83\arcmin \ away) are the most promising candidates (note that these are beyond the FoV shown in Figure \ref{fig:BC_NN}). These are all irregular dwarf galaxies that appear to be forming stars and are likely interacting with other galaxies in the cluster. 

VCC~1686 can be discounted as a foreground galaxy based on a TRGB distance measurement \citep{Karachentsev+2014}. Estimating the stellar masses of the remaining 3 candidates (assuming $D=16.5$~Mpc) with the mass-to-light ratio of \citet{Taylor+2011} and photometry of \citet{Kim+2014}, then applying the MZR of \citet{Andrews+2013}, reveals that all have expected metallicities at least 0.2~dex lower than BC1 (Table \ref{tab:metallicity}). Thus we are left with no strong candidate that matches the properties of BC1.

\subsection{BC4 \& BC5}
\label{sec:discuss_BC4}

There are no galaxies with significant \hi \ reservoirs (detectable in ALFALFA at the distance of Virgo) within 30\arcmin \ and 500~\kms \ of BC4 (Figure \ref{fig:BC_NN}, bottom left). However, the only (possible) Virgo member in the VCC or EVCC within 30\arcmin \ is VCC~824 ($cz_\odot = 392$ \kms), a red ($g-i = 0.84$), nucleated, LSB dwarf. The nearest galaxy detected in \hi \ is NGC~4419 (VCC~958, $cz_\odot = -275$ \kms), approximately 40\arcmin \ to the north of BC4. The \hi \ distribution of this galaxy has been greatly truncated \citep{Chung+2009}, indicating that it has undergone significant ram pressure stripping. The VERTICO CO map of NGC~4419 \citep{Brown+2021} indicates that the molecular gas may extend roughly to the south. This could indicate that this galaxy has already fallen past the cluster center and is now moving to the north, in which case this could be a good candidate origin of the gas that formed BC4. The \citet{Taylor+2011} relation for stellar mass (based on $M_i$ and $g-i$) gives an estimate of $\log M_\ast/\mathrm{M_\odot} = 10.25 \pm 0.10$ for NGC~4419 (assuming $D=16.5$~Mpc). Using this stellar mass estimate in the MZR of \citet{Andrews+2013} gives a metallicity estimate of $12 + \log(\mathrm{O/H}) = 8.74 \pm 0.18$, which matches remarkably well with the observed metallicity of the H\,{\sc ii} regions in BC4, $\langle 12 + \log(\mathrm{O/H}) \rangle = 8.73 \pm 0.15$ (\S \ref{sec:MUSE_results}).

A little farther to the north there is another gas-bearing galaxy, NGC~4396 (VCC~865, $cz_\odot = -122$~\kms, 77\arcmin \ away). This galaxy was also studied in the VIVA project, which found that it is in the process of being stripped by ram pressure. However, in this case the gas clearly extends to the NW, indicating that NGC~4396 is unlikely to be the source of the gas that formed BC4.

Skipping over a few low-mass \hi \ detections (to which we shall return), the next nearest gas-bearing galaxies are to the south, NGC~4402 (VCC~873) and NGC~4438 (VCC~1043), both of which are $\sim$1.3$^\circ$ from BC4 (beyond the FoV in Figure \ref{fig:BC_NN}). NGC~4402 was imaged in \hi \ by the VIVA survey, revealing that the galaxy is deficient in \hi, exhibits a truncated disk and clear signs of ongoing ram pressure stripping, with the current gas tail emanating in a NW direction \citep{Crowl+2005}. However, NGC~4402 has $cz_\odot = 230$~\kms, which places it $\sim$300~\kms \ away from BC4. Although this velocity separation does not rule out NGC~4402 as the origin of BC4,  it means it is less favored than other candidates. NGC~4438 is just to the east of NGC~4402, and unfortunately does not have any existing \hi \ imaging. However, the DECaLS images of NGC~4438 display a highly irregular morphology and the galaxy appears partially disrupted. The central velocity of the \hi \ emission \citep{Haynes+2018} is $cz_\odot = 104$~\kms, placing it significantly closer to BC4 in velocity. Again using the \citet{Taylor+2011} stellar mass-to-light relation and the photometry from the EVCC, these two galaxies have stellar mass estimates of $\log M_\ast/\mathrm{M_\odot} = 9.98 \pm 0.10$ and $10.57 \pm 0.10$, respectively. Thus neither would match with the metallicity of BC4 (via the MZR) quite as well as NGC~4419, but would agree comfortably within the relation's scatter.

We also note that BC4 is in the vicinity of SECCO~1 and \citet{Sand+2017} suggested the same sub-group of Virgo (containing NGC~4402 and NGC~4438) as the potential source of the gas in that object. The extension of NGC~4402 to the NW \citep{Crowl+2005,Chung+2009} and the highly disturbed H$\alpha$ emission \citep{Kenney+2008,Sand+2017} in the vicinity of both galaxies makes them both plausible candidates for the origin of SECCO~1. However, this does not preclude them also being candidates for BC4. It is possible that several such objects may be formed from the ram pressure stripping of a single large galaxy, but the significantly lower metallicity of SECCO~1 \citep[though still high for such a low mass object, $12 + \log(\mathrm{O/H}) = 8.38$,][]{Beccari+2017} suggests that it was likely formed from a different parent object than BC4.

Finally, we return to the two blue, irregular, dwarf galaxies to the south of BC4 that we skipped over above, VCC~1001 ($cz_\odot = 340$~\kms, 43\arcmin \ away) and VCC~945 (IC~3355, $cz_\odot = -16$~\kms, 73\arcmin \ away). These are likely too small and far away from BC4 to have been the source of the gas it must have had recently to support its SF. However, they may be related to BC4, or indeed to NGC~4402 and NGC~4438. It is also worth noting that three very similar objects were seen in the vicinity of BC1 on the other side of the cluster.

BC5 is approximately 45\arcmin \ north of BC4, raising the possibility that perhaps it is part of the same extended structure, indeed the two objects have almost identical metallicity measurements (Table \ref{tab:metallicity}) and radial velocities (Table \ref{tab:BCs}). The closest galaxy detected in \hi \ (and within 500 \kms) is NGC~4419 (VCC~958, $\sim$10\arcmin \ to the SE), as it was for BC4 (Figure \ref{fig:BC_NN}, bottom right). Again this object would be an excellent match for BC5's metallicity (based on the MZR). However, if we assume these objects formed from ram pressured stripped material then it seems unlikely that NGC~4419 could be the origin of both BC4 and BC5, as BC4 is 40\arcmin \ to its south and BC5 is 10\arcmin \ to its NW, while ram pressure tails usually extend in one general direction (i.e. in the wake of the galaxy's motion through the ICM). If instead they were formed from tidally stripped gas then this objection is removed, as they may have formed from tidal tails on opposite sides of the galaxy. 

The next nearest galaxy is NGC~4396, 30\arcmin to the NW. As discussed above, this galaxy is also being stripped, however, as with BC4, the \hi \ tail extends in the wrong direction. 

These two galaxies are the only \hi-bearing galaxies within 1$^\circ$ and $\pm 500$~\kms. The only other galaxy in the VCC or EVCC within this range is VCC~583, a red, LSB dwarf elliptical approximately 58\arcmin \ to the west. This dwarf is next to the \hi-bearing galaxy NGC~4312, which might have been a good candidate for the origin of both BC4 and BC5 via ram pressure stripping, except that multiple TFR distance estimates place it well in the foreground of the cluster \citep{Yasuda+1997,Russell2002,Sorce+2014}.

Thus if we wish to look for better-matched candidates then we must look further afield. Here we come across a promising candidate, UGC~7695 (VCC~1450, IC~3476; $cz_\odot = -159$~\kms, $\sim$1.5$^\circ$ to the SE of BC4, beyond the FoV shown in Figure \ref{fig:BC_NN}). This galaxy was extensively studied by \citet{Boselli+2021}. It displays a striking bow-shaped region bright in H$\alpha$ and UV, a clear sign of ram pressure stripping. \citet{Boselli+2021} argue that this galaxy is likely being stripped almost edge-on and that the stripped gas is forced up, over and around the disk. The metallicity of UGC~7695 is also consistent with that of BC4 and BC5, with the highest values measured by \citet{Boselli+2021} being $12 + \log (\mathrm{O/H}) \approx 8.7$ and the average being $\left< 12 + \log (\mathrm{O/H})\right> = 8.60 \pm 0.12$ \citep{Hughes+2012}.

The orientation of UGC~7695 and its velocity relative to the cluster center suggest that the transverse velocity through the cluster is likely on the order of 1500~\kms \ \citep{Boselli+2021}. Based on this velocity, the material that formed BC4 and BC5 would have needed to have been stripped approximately 250-300~Myr ago. This is consistent with the approximate ages of the stellar populations of the BCs that we estimated in \S\ref{sec:stellarpops}, in that those age estimates are less ($\sim$200~Myr). But, \citet{Boselli+2021} conclude that the structures seen in UGC~7695 are considerably younger, 50-150~Myr. However, their simulations of the stripping of UGC~7695 assume constant properties for the ICM, whereas if it has traversed over 450~kpc through the cluster, then the surrounding density will have changed significantly during the course of its interaction. It is also possible that the gas that formed BC4 and BC5 was the first material to be stripped, before the ram pressure was sufficient to strip the main disk. We note that the metallicities of BC4 and BC5 correspond to the most metal-rich measurements of UGC~7695. This may be an indication that their seed gas originated in a particularly metal-rich region, such as gas that has been enriched by recent supernovae and pushed far from the plane of the galaxy. We therefore suggest that despite the young age estimate of the ram pressure stripping event in \citet{Boselli+2021}, UGC~7695 is still consistent with being the origin of these two BCs.

Finally, we note that in the ALFALFA data cube the \hi \ emission appears to extend slightly from UGC~7695 in the approximate direction of BC4 and BC5, but unfortunately this emission cannot be reliably traced as it becomes blended with Galactic \hi \ emission at the same velocity. Given the morphology, orientation, and location of UGC~7695, along with its estimated metallicity, this is by far the best candidate for the origin of that material that formed BC4 and BC5.

\end{document}